\documentclass[aps, prd, superscriptaddress, reprint, twocolumn, floatfix]{revtex4-2}

\usepackage{tikz}
\usetikzlibrary{decorations.markings, patterns}
\usepackage{latexsym, amsmath, amsfonts, amssymb}
\usepackage{mathrsfs}
\usepackage{mathtools}
\usepackage[american]{babel}
\usepackage{bbm}
\usepackage[pdfencoding=auto]{hyperref}
\hypersetup{colorlinks, citecolor=[rgb]{.7,0,0}, linkcolor=[rgb]{0,0,0.7}, urlcolor=[rgb]{0,0,0.5}}
\usepackage{booktabs}

\newcommand{\du}[2]{_{#1}^{\phantom{#1}#2}}

\newcommand{\ket}[1]{\ensuremath{| #1 \rangle}}

\newcommand{\braketmatrix}[3]{\left \langle #1 \middle| #2 \middle| #3 \right \rangle}
\newcommand{\wt}{\widetilde}

\newcommand{\wb}{\overline}
\newcommand{\matht}[1]{\ensuremath{\boldsymbol{#1}}}
\newcommand{\tps}[2]{\texorpdfstring{#1}{#2}}
\newcommand{\ts}{\textstyle}
\newcommand{\ds}{\displaystyle}

\newcommand{\eg}{\textit{e.g.}}

\newcommand{\ie}{\textit{i.e.}}

\newcommand{\nn}{\nonumber}

\newcommand{\smat}[1]{\big( \begin{smallmatrix} #1 \end{smallmatrix} \big)}
\newcommand{\be}{\begin{equation}} \newcommand{\ee}{\end{equation}}
\newcommand{\bea}{\begin{equation} \begin{aligned}} \newcommand{\eea}{\end{aligned} \end{equation}}
\newcommand{\ba}{\begin{array}} \newcommand{\ea}{\end{array}}

\newcommand{\cC}{\mathcal{C}}
\newcommand{\cD}{\mathcal{D}}
\newcommand{\cE}{\mathcal{E}}
\newcommand{\cF}{\mathcal{F}}
\newcommand{\cG}{\mathcal{G}}
\newcommand{\cH}{\mathcal{H}}
\newcommand{\cI}{\mathcal{I}}
\newcommand{\cJ}{\mathcal{J}}
\newcommand{\cK}{\mathcal{K}}
\newcommand{\cL}{\mathcal{L}}

\newcommand{\cN}{\mathcal{N}}
\newcommand{\cO}{\mathcal{O}}
\newcommand{\cP}{\mathcal{P}}
\newcommand{\cQ}{\mathcal{Q}}

\newcommand{\cS}{\mathcal{S}}
\newcommand{\cT}{\mathcal{T}}

\newcommand{\cY}{\mathcal{Y}}

\newcommand{\bN}{\mathbb{N}}

\newcommand{\bR}{\mathbb{R}}

\newcommand{\bZ}{\mathbb{Z}}

\newcommand{\fq}{\mathfrak{q}}

\newcommand{\rB}{\mathrm{B}}
\newcommand{\rF}{\mathrm{F}}
\newcommand{\rR}{\mathrm{R}}

\newcommand{\sT}{{\sf{T}}}

\newcommand{\unit}{\mathbbm{1}}

\newcommand{\rSU}{\mathrm{SU}}
\newcommand{\rU}{\mathrm{U}}
\newcommand{\rSL}{\mathrm{SL}}
\newcommand{\rPSL}{\mathrm{PSL}}

\newcommand{\su}{\mathfrak{su}}

\DeclareMathOperator{\Tr}{Tr}

\DeclareMathOperator{\Ber}{Ber}

\DeclareMathOperator{\sign}{sign}
\DeclareMathOperator{\sgn}{sgn}

\DeclareMathOperator{\re}{\mathbb{R}e}
\DeclareMathOperator{\im}{\mathbb{I}m}

\DeclareMathOperator{\diag}{diag}

\DeclareMathOperator{\arccosh}{arccosh}

\begin{document}

\preprint{SISSA 02/2024/FISI}

\title{\tps{\matht{\cN=2}}{N=2} SYK models with dynamical bosons and fermions}

\newcommand{\SISSA}{\affiliation{SISSA and INFN, via Bonomea 265, 34136 Trieste, Italy}}

\newcommand{\ICTP}{\affiliation{International Centre for Theoretical Physics (ICTP), Strada Costiera 11, 34151 Trieste, Italy}}

\newcommand{\KITS}{\affiliation{Kavli Institute for Theoretical Sciences (KITS), University of the Chinese Academy of Sciences, Beijing 100190, China}}

\author{Francesco Benini}
\SISSA
\ICTP

\author{Tom\'as Reis}
\SISSA

\author{Saman Soltani}
\SISSA

\author{Ziruo Zhang}
\SISSA
\KITS

\begin{abstract}
\noindent
We study a class of SYK models with $\cN=2$ supersymmetry, described by $N$ fermions in chiral Fermi multiplets, as well as $\alpha N$ first-order bosons in chiral multiplets. The interactions are characterized by two integers $(p,q)$. We focus on the large $N$ and low energy limit of these models. Despite the presence of dynamical bosons, we find conformal behavior akin to the standard SYK model. We use $\cI$-extremization of a Witten index to study the supersymmetric solutions. In particular, we find an exact expression for the entropy, which matches the numerical solutions to the Schwinger--Dyson equations. We further solve the model both in the large $p$ and large $p,q$ limits. Numerically, we verify our analytical results and obtain estimates for the Schwarzian coupling in the near zero-temperature limit. We also study the low-lying spectrum of operators to determine the parameter ranges where the Schwarzian mode dominates the IR dynamics. Lastly, we study out-of-time-ordered correlators to show that the model is maximally chaotic.
\end{abstract}

\maketitle


\section{Introduction and summary}

The quantum-mechanical models introduced by Sachdev, Ye, and Kitaev (SYK) \cite{Sachdev:1992fk, *Kitaevtalk:2015} have the properties of being solvable at strong coupling in the large $N$ limit, exhibiting maximally chaotic behavior, and developing an emergent conformal symmetry in the infrared (IR) \cite{Sachdev:1992fk, Parcollet:1999itf, Sachdev:2010um, Kitaevtalk:2015, Polchinski:2016xgd, Maldacena:2016hyu, Maldacena:2016upp}. Their IR physics is dominated by a pseudo Goldstone boson, the Schwarzian mode, which is dual to two-dimensional JT dilaton gravity \cite{Teitelboim:1983ux, *Jackiw:1984je} in AdS$_2$. Given the universal appearance of near-AdS$_2$ throats in the near-horizon region of near-extremal black holes, the SYK model poses itself as a model of quantum black-hole physics: it captures the dynamics of a universal ``breathing'' mode that lives in the near-horizon region \cite{Iliesiu:2020qvm, *Heydeman:2020hhw}. Many variations on the SYK model have been studied, including \cite{Anninos:2016szt, Fu:2016vas, Davison:2016ngz, Murugan:2017eto, Peng:2017spg, Bulycheva:2017uqj, Bulycheva:2018qcp, Marcus:2018tsr, Wang:2019bpd, Gu:2019jub, Pan:2020mqy, Peng:2020euz, Tikhanovskaya:2020elb, Tikhanovskaya:2020zcw, Gates:2021jdm, Gates:2021iff, Heydeman:2022lse, Murugan:2023vhk, Biggs:2023mfn} which are close to the subject of this paper.

An interesting question is whether one can microscopically derive a certain type of SYK-like models directly from gravity or string theory \cite{Anninos:2016szt}. A step in that direction was made in \cite{Benini:2022bwa} in the context of holography and black holes in a higher-dimensional AdS space. The quantum mechanical model found in \cite{Benini:2022bwa} presents many peculiar features, \eg, it has $\cN=2$ supersymmetry, it describes both dynamical fermions and bosons belonging to many species with unequal abundance in a large $N$ limit, it enjoys multiple Abelian symmetries.
It is not a priori clear what the impacts of those features are on low-energy physics. For instance, the $\cN=4$ supersymmetric model constructed in \cite{Anninos:2016szt} --- which also contains both fermions and bosons --- fails to exhibit emergent superconformal symmetry at low energy, and thus it cannot be a good model of near-BPS black holes.

With that motivation in mind, in this paper we study a class of SYK-like models with $\cN=2$ supersymmetry, made of $N$ regular fermions in chiral Fermi multiplets, as well as $\alpha N$ bosons in chiral multiplets with one-derivative kinetic terms (\textit{a.k.a.}\ first-order bosons), in the large $N$ limit. The parameter $\alpha$ controls the relative abundance. The interactions are described by ``superpotential'' J-terms (in the language of \cite{Witten:1993yc}), are controlled by two positive integers $(p,q)$ (where $p$ is odd), and may include --- depending on $p,q$ --- a scalar potential and Yukawa interactions. As in SYK, the interactions are all-to-all and given by random variables that represent quenched disorder.

Our model is distinct from previous supersymmetric SYK models in the literature. On the one hand, it has dynamical bosons unlike \cite{Fu:2016vas, Heydeman:2022lse}. On the other hand, the dynamical bosons have one-derivative kinetic term and lie in a supersymmetric multiplet different from that of the dynamical fermions, making our model distinct from proposals such as those of \cite{Gates:2021jdm, Gates:2021iff, Anninos:2016szt, Biggs:2023mfn} and ultimately leading to different physics.
There are, naturally, similarities as well, for instance with the $\cN=2$ two-fermion model of \cite{Heydeman:2022lse}. Indeed, our model has both a $\rU(1)_R$ R-symmetry and a $\rU(1)_F$ flavor symmetry, and we study the model with chemical potentials turned on, finding similar physical properties. We are particularly interested in the role played by dynamical bosons, which could lead to a variety of phenomena such as spin-glass phases or condensation (see \eg\ \cite{Fu:2016yrv, Gur-Ari:2018okm, Baldwin:2019dki, Christos:2021wno, Swingle:2023nvv, Biggs:2023mfn}) or simply prevent the IR emergence of (super)conformal symmetry \cite{Anninos:2016szt}. In our model, instead, bosons turn out to be tamed and lead to conformal solutions --- both with and without supersymmetry --- similarly to the standard SYK model. This implies that our $(p,q)$ models retain the main features expected from near-BPS black hole horizons, yet accommodating the presence of dynamical bosons, and make us confident that also the more convoluted, but also more realistic, quantum mechanical model of \cite{Benini:2022bwa} might exhibit the same properties.

Let us summarize our results. In Section~\ref{sec: model} we present our models. In Section~\ref{sec: sol of model} we solve the models in the annealed approximation and at leading order in $1/N$, in terms of Schwinger--Dyson equations. In particular, we verify that the systems admit both supersymmetric and non-supersymmetric conformal solutions in the IR. For certain values of the chemical potentials, we also find some non-conformal solutions in which the correlators behave as those of gapped free fields.
Supersymmetric solutions at zero temperature can also be studied through a symmetry-refined Witten index. From it, employing $\cI$-extremization \cite{Benini:2015eyy, Benini:2016rke}, we extract the zero-temperature entropy and the R-symmetry charge of the vacuum. Both are compatible with our numerical calculations and with the Luttinger--Ward relation \cite{Luttinger:1960ua}.
The fact that the zero-temperature entropy computed in the annealed approximation agrees with the exact large $N$ result obtained from the index supports the validity of the annealed approximation and the claim that the models are conformal in the IR, rather than in a spin-glass phase
\footnote{Ruling out a spin-glass phase is a notoriously delicate matter. The existence of a superconformal solution to the Schwinger--Dyson equations in the annealed approximation and its compatibility with the Witten index is only circumstantial evidence that our models are conformal in the IR, and a more direct proof would certainly be welcome.}.
We analyze the systems at large $p$, as well as at large $p$ and $q$. In these limits, we are able to reach an intermediate low-energy regime which is conformal, but not the very low-energy behavior, which is also conformal but with different spectrum. The two agree only in the supersymmetric case.

In Section~\ref{sec: numerics} we perform a thorough numerical analysis of our models, both for supersymmetric and non-supersymmetric values of the chemical potentials, for different values of $(p,q)$. In particular we solve the Schwinger--Dyson equations, confirming the IR conformal behavior found analytically. We test the Luttinger--Ward relations and the computation of the zero-temperature entropy. Going to non-zero temperature allows us to extract the Schwarzian coupling. 

In Section~\ref{sec: spectrum} we study the low-lying spectrum of operator dimensions. Besides observing the presence of the Schwarzian mode, the two currents for the R and flavor symmetries,  and their superpartners, we determine in which window of parameters we expect the Schwarzian mode to dominate the IR dynamics and identify windows in which other modes are expected to dominate. In Section~\ref{sec: chaos exp} we extract the Lyapunov exponents from the out-of-time-ordered correlator (OTOC) 4-point functions, finding that the models are maximally chaotic (they saturate the bound of \cite{Maldacena:2015waa}).

We conclude in Section~\ref{sec: discussion} by presenting some puzzling results that we hope to resolve in future work. We provide a few appendices with technical details of the computations.

\section{\tps{\matht{\cN=2}}{N=2} SYK models with dynamical bosons}
\label{sec: model}

We are interested in quantum mechanical models with $\cN=2$ supersymmetry (one complex supercharge) described by $N$ Fermi multiplets $\cY_{a}$ ($a=1, \dots, N$) and $\alpha N$ chiral multiplets $\Phi_b$ ($b=1, \dots, \alpha N$), where $\alpha$ is a positive real constant that we call the abundance parameter
\footnote{At finite $N$ the parameter $\alpha$ is quantized such that $\alpha N \in \bZ$, but at large $N$, $\alpha$ is essentially continuous.}.
The field content of a Fermi multiplet $\cY$ is a dynamical fermion $\eta$ and an auxiliary scalar $f$, while that of a chiral multiplet $\Phi$ is, in our models, a dynamical boson $\phi$ (with one-derivative kinetic term) and an auxiliary fermion $\psi$. In the Lagrangian formulation and in Lorentzian signature, the models are labelled by two positive integers $p,q$ ($p$ is odd) and have action:
\begin{align}
\label{lorentzian action}
S_\text{L} &= \int\! dt \, \biggl[ i \eta_a^\dag \bigl( \partial_t + i \mu_\eta \bigr) \eta_a + f_a^\dag f_a + i \phi_b^\dag \bigl( \partial_t + i \mu_\phi \bigr) \phi_b \nn \\
&\quad + \psi_b^\dag \psi_b - J_{a_1 \ldots a_p b_1 \ldots b_q} \Bigl( p \, f_{a_1}\eta_{a_2} \cdots \eta_{a_p} \phi_{b_1} \cdots \phi_{b_q} 
\nn \\
&\quad + q \, \eta_{a_1} \cdots \eta_{a_p}\psi_{b_1} \phi_{b_2} \cdots \phi_{b_q} \Bigr) + \text{h.c.} \biggr].
\end{align}
We use the convention that repeated indices are summed over. These so-called J-term interactions in the second and third lines are inspired by the ones obtained in \cite{Benini:2022bwa}, corresponding here to the $p=1$, $q=2$ case. The couplings $J_{a_1\ldots a_pb_1\ldots b_q}$ are complex Gaussian random variables with zero mean and variance
\be
\bigl\langle J^{\phantom{*}}_{a_1\ldots a_p b_1 \ldots b_q} J^*_{c_1 \ldots c_p d_1 \ldots d_q} \bigr\rangle = \frac{J \, \delta_{a_1 \ldots a_p}^{[c_1 \ldots c_p]} \, \delta_{b_1 \ldots b_q}^{( d_1 \ldots d_q)} }{p \, q \, N^{p+q-1}}  \;.
\ee
Here $J$ has dimension of mass, and is kept fixed in the large $N$ limit so that the averaged partition function has uniform scaling. The couplings are antisymmetric in the first $p$ indices and symmetric in the next $q$ indices. Note that $p$ must be odd, while $q$ is any positive integer. In (\ref{lorentzian action}) we have also included the chemical potentials $\mu_\phi$, $\mu_\eta$. Under charge conjugation
\begin{align}
\eta_a &\mapsto \eta_a^\dag \;,\quad& f_a &\mapsto f_a^\dag \;,\;& \phi_b &\mapsto \phi_b^\dag \;,\;& \psi_b &\mapsto \psi_b^\dag \;, \nn \\
\mu_\eta &\mapsto - \mu_\eta \;,\;& \mu_\phi &\mapsto - \mu_\phi \;,
\end{align}
the Fermi multiplet kinetic terms are invariant, while the chiral multiplet kinetic terms are odd and thus they break charge conjugation symmetry.

Integrating out the auxiliary fields, the action becomes
\begin{align}
S_\text{L} &= \int\! dt \biggl[ i \eta_a^\dag \bigl( \partial_t + i \mu_\eta \bigr) \eta_a + i \phi_b^\dag \bigl( \partial_t + i\mu_\phi \bigr) \phi_b \\
&\qquad\quad - \Bigl\lvert p \, J_{a_1\ldots a_pb_1\ldots b_q} \eta_{a_2} \cdots \eta_{a_p}\phi_{b_1} \cdots \phi_{b_q} \Bigr\rvert^2  \nn \\
&\qquad\quad - \Bigl\lvert q \, J_{a_1\ldots a_pb_1\ldots b_q} \eta_{a_1} \cdots \eta_{a_p} \phi_{b_2} \cdots \phi_{b_q} \Bigr\rvert^2 \biggr] \,. \nn
\end{align}
In the Hamiltonian formulation, the bosonic conjugate momentum to $\phi$ is $\Pi_\phi = i \phi^\dag$ and therefore the (anti\nobreakdash-)commutation relations are $\{\eta_a, \eta^\dag_c \} = \delta_{ac}$ and $[\phi_b, \phi^\dag_d] = \delta_{bd}$ (all other ones vanishing). The Hamiltonian is $\tilde H = H + \mu_\eta Q_\eta + \mu_\phi Q_\phi$, where
\bea
\label{interacting Hamiltonian}
H &= \bigl\lvert p \, J_{a_1\ldots a_pb_1\ldots b_q} \eta_{a_2} \cdots \eta_{a_p} \phi_{b_1} \cdots \phi_{b_q} \bigr\rvert^2 \\
&\quad + \bigl\lvert q \, J_{a_1 \ldots a_p b_1 \ldots b_q} \eta_{a_1} \cdots \eta_{a_p} \phi_{b_2} \cdots \phi_{b_q} \bigr\rvert^2
\eea
is the interacting Hamiltonian with chemical potentials turned off, while $Q_\eta = \frac12 [\eta_a^\dag, \eta_a]$ and $Q_\phi = \frac12 \{ \phi_b^\dag, \phi_b\}$ are the charge operators. The ordering ambiguities in $H$ are fixed by supersymmetry by requiring that $H = \{\cQ, \cQ^\dag\}$ in terms of the supercharge
\be
\cQ \,\sim\, J_{a_1 \ldots a_p b_1 \ldots b_q} \eta_{a_1} \cdots \eta_{a_p} \phi_{b_1} \cdots \phi_{b_q} \,.
\ee
Each bosonic degree of freedom $\phi$ describes a particle in a magnetic field (see, \eg, \cite{Dunne:1989hv}). Upon canonical quantization, $Q_\phi \geq \frac12$ and the bosonic Hilbert space is a Fock space with lowest weight state $|0\rangle$ defined by $\phi |0\rangle = 0$ and generated by the creation operator $\phi^\dag$. For $p=1$ the Hamiltonian $H$ includes a positive-definite scalar potential for $\phi$, and $\tilde H$ is bounded from below for any given value of $\mu_\phi$ and finite $N$. However $J_{a_1\ldots a_pb_1\ldots b_q}$ goes to zero at large $N$, and thus one should take $\mu_\phi>0$ in order to avoid instabilities (as we will see, the limit $\mu_\phi \to 0$ is stable after taking the large $N$ limit). For $p>1$, $H$ has flat directions and thus one should restrict to $\mu_\phi>0$ even at finite $N$.

To go to Euclidean signature we set $\tau = it$ and define $i S_\text{L} = - S_\text{E}$ (we also use a bar in place of $\dag$). Then the Euclidean Lagrangian in components reads:
\begin{align}
\cL_\text{E} &= \wb\eta_a \bigl( \partial_\tau + \mu_\eta \bigr) \eta_a - \wb{f}_a f_a + \wb\phi_b \bigl( \partial_\tau + \mu_\phi \bigr) \phi_b - \wb\psi_b \psi_b \nn\\
&\quad + J_{a_1\ldots a_pb_1\ldots b_q} \Bigl( p \, f_{a_1} \eta_{a_2} \cdots \eta_{a_p} \phi_{b_1} \cdots \phi_{b_q}  \nn\\
&\qquad + q \, \eta_{a_1} \cdots \eta_{a_p}\psi_{b_1} \phi_{b_2} \cdots \phi_{b_q} \Bigr) + \text{h.c.} \;.
\end{align}
It will be convenient to work in superspace with coordinates $(\tau, \theta, \wb\theta)$ and use the superfields
\bea
\Phi &= \phi + \theta\psi + \tfrac{1}{2} \theta\wb\theta \, (\partial_\tau+\mu_\phi)\phi \;, \\
\cY &= \eta - \theta f + \tfrac{1}{2} \theta\wb\theta \, (\partial_\tau+\mu_\eta)\eta \;.
\eea
Our conventions for supersymmetry and superspace are in Appendix~\ref{app: superspace}, while we refer to \cite{Benini:2022bwa, Hori:2014tda} for more detailed presentations.
Using superspace integrals, the Euclidean action is
\begin{align}
S_\text{E} &= - \int\! d^3T \, \bigl( \wb\cY_a \cY_a + \wb\Phi_b \Phi_b \bigr) \\
&\quad + J_{a_1\ldots a_pb_1\ldots b_q} \int\! d^2T \; \cY_{a_1} \cdots \cY_{a_p}\Phi_{b_1}\cdots\Phi_{b_q} + \text{h.c.}. \nn
\end{align}
In the following we will mostly work in Euclidean signature and drop the subscript E.

\begin{table}
$\ds
\arraycolsep=3mm
\begin{array}{c | c c | c c}
    \toprule
    & Q_\eta & Q_\phi & Q_F & R \\
    \midrule
    \eta & 1 &  0  &  q  &  R[\eta]  \\
     f  &  1-p  &  -q  &  q  &  R[\eta]-1  \\[.2em]
     \phi  &  0  &  1  &  -p  &  R[\phi]  \\
     \psi  &  -p  &  1-q  &  -p  &  R[\phi]-1 \\
    \bottomrule
\end{array}
$
\caption{\label{tab: charges}%
Charge assignments under the continuous global symmetries $\rU(1)^2$, using two alternative parametrizations. The R-charge assignments satisfy $p \, R[\eta] + q \, R[\phi]=1$.}
\end{table}

The group of continuous global symmetries is $\rU(1)^2$, which we can parametrize by the two charges $Q_\eta$, $Q_\phi$ with charge assignments as in Table~\ref{tab: charges}. An alternative parametrization is in terms of $\rU(1)_F\times \rU(1)_R$ with charges $Q_F = q \, Q_\eta - p \, Q_\phi$ and $R=R[\eta] \, Q_\eta + R[\phi] \, Q_\phi$ that satisfy $p \, R[\eta] + q \, R[\phi]=1$. Here $Q_F$ is a flavor symmetry that commutes with supersymmetry, whilst $R$ is a generic R-symmetry. For future reference, $Q_\eta$ and $Q_\phi$ can be written in terms of $Q_F$ and $R$ as
\be
\label{Qeta Qphi ito QF R}
Q_\eta = p \, R + R[\phi] \, Q_F \;,\qquad Q_\phi = q \, R - R[\eta] \, Q_F \;.
\ee
When $d \equiv \gcd(p,q)$ is bigger than 1, there is also a discrete flavor symmetry $\bZ_d$. Indeed consider the discrete symmetry $g$ that acts as
\be
\label{def g Z_gcd}
g: \qquad \cY_a \;\mapsto\; e^{\frac{2\pi i}p} \; \cY_a \;,\qquad\quad \Phi_b \;\mapsto\; \Phi_b \;.
\ee
One can check
\footnote{Indeed let $\ell$ be the inverse of $q/d$ modulo $p/d$,
then a rotation $\rU(1)_F$ by an angle $\alpha = 2\pi \ell /p$ acts as $g^d$.}
that $g^d \in \rU(1)_F$. Besides, one can write the fermion parity operator as
\be
(-1)^F = g^{\frac{p-1}2} e^{\pi i R_0} \,,
\ee
where $R_0$ is a reference R-symmetry with assignments $R_0[\eta]=\frac{1}{p}$, $R_0[\phi]=0$. When $d=1$ this is $e^{\pi i R}$ for some R-symmetry $R$ that is a linear combination of $R_0$ and $Q_F$, but otherwise it is not.

The relations between chemical potentials, according to Table~\ref{tab: charges}, are
\be
\label{def muF muR}
\mu_F = R[\phi] \, \mu_\eta - R[\eta] \, \mu_\phi \;,\qquad \mu_R = p\, \mu_\eta + q\, \mu_\phi \;.
\ee
In particular, whenever we want to impose $\mu_R = 0$ because of supersymmetry, we also fix $\mu_\eta = -q \mu_\phi/p$ and $\mu_F = - \mu_\phi/p$, and therefore we should restrict to $\mu_\eta < 0$ and $\mu_F < 0$.

\section{Solution in the annealed approximation}
\label{sec: sol of model}

Having established the model, we want to explore its large $N$ dynamics. When averaging over the couplings, particularly when focusing on correlators and 4-point functions, it is in principle preferable to first calculate the observables in one instance of the model, and then average. This is usually called \textit{quenched disorder}, which is analytically very hard to control. The approach we follow here is that of \textit{annealed disorder}, in which we calculate the averaged action first and then derive the observables from it. For quantities such as correlators and the entropy, those two approaches can lead to different results. In the standard SYK model, many observables are found to be ``self-averaging'' which means that the differences between distinct averaging schemes are subleading in $1/N$. We expect a similar behavior in our model. A crucial counterexample would be if the entropy calculation were dominated by a replica-breaking solution, which is not captured by the annealed approximation. The fact that our model contains fermions might be crucial to rule this out, as argued in \cite{Baldwin:2019dki}. In the rest of the paper we use annealed disorder, assuming it approximates the quenched system to leading order in $1/N$.

\subsection{Equations of motion at large \tps{\matht{N}}{N}}

We follow the standard steps performed in \cite{Fu:2016vas, Heydeman:2022lse} to derive the equations of motion (EOMs) at large $N$. For simplicity, we start with no chemical potentials turned on, and perform our manipulations in superspace. The corresponding expressions in components, and including non-vanishing chemical potentials, can be found in Appendix~\ref{app: component expr}.

After averaging over $J_{a_1\ldots a_pb_1\ldots b_q}$, the partition function is 
\begin{align}
\langle Z\rangle &= \int\! \cD\cY_a \, \cD\Phi_b \, \exp\biggl[ \, \int\! d^3T \, \Bigl( \wb\cY_a \cY_a + \wb\Phi_b \Phi_b \Bigr)  \\
&\quad + \frac{J}{p \, q \, N^{p+q-1}} \int\! d^2T_2 \, d^2 \wb{T}_1 \times {} \nn\\
&\qquad \times \wb\cY_{a_1}(T_1) \, \cY_{a_1}(T_2) \, \cdots \, \wb\cY_{a_p}(T_1) \, \cY_{a_p}(T_2) \times {} \nn\\
&\qquad \times \wb\Phi_{b_1}(T_1) \, \Phi_{b_1}(T_2) \, \cdots \, \wb\Phi_{b_q}(T_1) \, \Phi_{b_q}(T_2) \biggr] \,. \nn
\end{align}
We define the bilocal fields
\bea
\label{def bilocal fields}
\cG_{\cY}(T_1,T_2) \,&\equiv\, {\ts \frac{1}{N} \sum\nolimits_a } \, \wb\cY_a(T_1) \, \cY_a(T_2) \;, \\
\cG_{\Phi}(T_1,T_2) \,&\equiv\, {\ts \frac{1}{N} \sum\nolimits_b } \, \wb\Phi_b(T_1) \, \Phi_b(T_2)\;,
\eea
which are chiral in $T_2$ and anti-chiral in $T_1$. They encode the two-point functions of the model. We introduce them --- together with the bilocal Lagrange multipliers $\Sigma_\cY$ and $\Sigma_\Phi$ --- into the action by inserting the following identity in the path integral:
\begin{align}
\label{pi insert bilocal}
1 &= \!\int\! \cD\cG_\cY \cD\Sigma_\cY \cD\cG_\Phi \cD\Sigma_\Phi  \exp \biggl\{ {-}N \!\int\! d^2T_2  \, d^2\wb T_1 \times {} \\
&\quad \times \Bigl[ \Sigma_\cY(T_1,T_2) \Bigl( \cG_\cY(T_1,T_2) - \tfrac{1}{N} \wb\cY_a(T_1) \cY_a(T_2) \Bigr) \nn\\
&\qquad + \Sigma_\Phi(T_1,T_2) \Bigl( \cG_\Phi(T_1,T_2) - \tfrac{1}{N} \wb\Phi_b(T_1) \Phi_b(T_2) \Bigr) \Bigr] \biggr\}  \,. \nn
\end{align}
At this stage, $\cG_{\cY,\Phi}(T_1,T_2)$ and $\Sigma_{\cY,\Phi}(T_1,T_2)$ are generic superfields which are chiral in $T_2$ and anti-chiral in $T_1$. Each of them is specified by four independent functions of $\tau_{1,2}$ according to the expansions in \eqref{bilocal definition component}. The components of $\Sigma_{\cY,\Phi}$ encode the self-energies of the fields in the model.
The kinetic term of $\cY_a$ can be rewritten in a bilocal way using the identity 
\begin{align}
& \int\! d^3T \; \wb\cY_a \cY_a \\
&\qquad = \int\! d^2T_2 \, d^2\wb{T}_1 \; \wb\cY_a(T_1) \, D_1 \overline{D}_2 \, \delta^3(T_1-T_2) \, \cY_a(T_2) \,, \nn
\end{align}
and similarly for $\Phi_b$. Here the superspace delta function is $\delta^3(T_1-T_2) \equiv ( \wb\theta_1 - \wb\theta_2 )(\theta_1-\theta_2)\delta(\tau_1-\tau_2)$, while $D_1$ and $\overline{D}_2$ are (anti-)chiral superspace derivatives with respect to $T_1$ and $T_2$, respectively (see Appendix~\ref{app: superspace}). For concreteness, we have
\begin{align}
& D_1 \wb{D}_2 \delta^3(T_1-T_2) = \delta(\tau_1 - \tau_2)  - \frac{ \theta_1 \theta_2 \wb\theta_1 \wb\theta_2 }4 \partial_{\tau_1}^2 \delta(\tau_1 - \tau_2)
\nn\\
& \hspace{8.5em} - \frac{\theta_1 \wb\theta_1 + \theta_2 \wb\theta_2 - 2 \theta_2 \wb\theta_1 }2 \partial_{\tau_1} \delta (\tau_1 - \tau_2) \nn\\
&\qquad = \delta \Bigl( \tau_1 - \tau_2 -\tfrac{1}{2}\, \theta_1 \wb\theta_1 -\tfrac{1}{2}\,\theta_2 \wb\theta_2 +\theta_2 \wb\theta_1 \Bigr) \;.
\end{align}
The fields $\cY_a$ and $\Phi_b$ only appear quadratically in the action and can be integrated out, producing Berezinian (\textit{a.k.a.} superdeterminant) factors in the path integral. We are left with an expression for the annealed partition function in terms of the bilocal fields:
\begin{align}
\label{bilocal action superspace}
\langle Z\rangle &= \int\! \cD\cG_\cY \, \cD\Sigma_\cY \, \cD\cG_\Phi \, \cD\Sigma_\Phi \, \exp\biggl\{ - N \biggl[ \\
&\;\; - \log \Ber \Bigl( D_1 \overline{D}_2 \delta^3(T_1-T_2) + \Sigma_\cY \Bigr) \nn\\
&\;\;  + \alpha \log \Ber \Bigl( D_1\overline{D}_2 \delta^3 (T_1 - T_2) + \Sigma_\Phi  \Bigr) \nn\\
&\;\; + \!\int\! d^2T_2 \, d^2\wb{T}_1 \biggl( \Sigma_{\cY}\, \cG_{\cY}  + \Sigma_\Phi  \, \cG_\Phi  - \frac{J}{pq} \, \cG_{\cY}^p \, \cG_{\Phi}^q \biggr) \biggr] \biggr\} . \nn
\end{align}
Here Ber is the Berezinian (or superdeterminant), whose definition is given in  \eqref{berezinian}, and bilocal fields should be read as $\cG_\Phi = \cG_\Phi (T_1,T_2)$.

Let us write down the equations of motion for the bilocal fields in terms of superfields. Extremizing \eqref{bilocal action superspace} with respect to $\cG_{\cY,\Phi}$ leads to the algebraic equations:
\bea
\label{superspace algebraic eq}
\Sigma_\cY(T_1, T_2) &= (J/q) \, \cG_\cY(T_1, T_2)^{p-1} \, \cG_\Phi(T_1, T_2)^q \,, \\
\Sigma_\Phi(T_1, T_2) &= (J/p) \, \cG_\cY(T_1, T_2)^p \, \cG_\Phi(T_1, T_2)^{q-1} \,.
\eea
Extremizing with respect to $\Sigma_{\cY, \Phi}$ leads to the integro-differential equations:
\begin{align}
\label{integral eom antichiral}
& D_2\cG_\cY(T_1, T_2) \! + \!\!\int\!\! d^2T_3 \, \Sigma_\cY(T_2,T_3) \, \cG_\cY(T_1,T_3)  = \overline{\delta}^2(T_1,T_2) \nn\\
& D_2\cG_\Phi(T_1, T_2) \! + \!\!\int\!\! d^2T_3 \, \Sigma_\Phi(T_2,T_3) \, \cG_\Phi(T_1,T_3) = \nn\\
&\hspace{13em} = -\alpha \, \overline{\delta}^2(T_1,T_2) \;.
\end{align}
If the derivative terms $D_2\cG_{\cY,\Phi}$ are dropped, one can show that \eqref{superspace algebraic eq} and \eqref{integral eom antichiral} are invariant under super-reparametrizations $\text{Diff}^+(S^{1|2})$ defined in \eqref{general super reparam}, where the superfields transform as
\begin{align}
& \cG_{\cY,\Phi}(T_1, T_2) \;\mapsto\; \bigl( D\theta'_1 \bigr){}^{2\Delta_{\eta,\phi}} \bigl( -\wb{D} \, \wb\theta{}_2' \bigr){}^{2\Delta_{\eta,\phi}} \cG_{\cY,\Phi}(T'_1, T'_2) \nn\\
\label{superdiff trans}
& \Sigma_{\cY,\Phi}(T_1, T_2) \;\mapsto \\
&\qquad\qquad  \bigl( D\theta'_1 \bigr){}^{1-2\Delta_{\eta,\phi}} \bigl( -\wb{D} \, \wb\theta{}_2' \bigr){}^{1-2\Delta_{\eta,\phi}} \Sigma_{\cY,\Phi}(T'_1, T'_2) \nn
\end{align}
with $p \, \Delta_\eta + q \, \Delta_\phi = \frac12$. The chiral measure and the (anti\nobreakdash-)chiral delta functions transform as in \eqref{chiral measure reparam} and \eqref{chiral delta reparam}. This shows that $\eta$ and $\phi$ are chiral primary operators under $\rSU(1,1|1) \subset \text{Diff}^+(S^{1|2})$.
While \eqref{integral eom antichiral} is anti-chiral in $T_{1,2}$, there is an equivalent way to write the equations using chiral superfields.
The equations \eqref{integral eom antichiral} and their chiral counterparts are derived by extremizing Ber with respect to $\Sigma_{\cY,\Phi}$ in \eqref{bilocal action superspace}. Indeed, treating it formally like any other operator determinant, one gets:
\bea
\label{superspace sigma deriv}
\cG_\cY(T_1,T_2) &= \bigl( D \wb{D} \, \delta^3 + \Sigma_\cY \bigr)^{-1}(T_1,T_2) \,,
\\ 
\cG_\Phi(T_1,T_2) &= - \alpha \, \bigl( D \wb{D} \, \delta^3 + \Sigma_\Phi \bigr)^{-1}(T_1,T_2) \,.
\eea
By convoluting with either the chiral or anti-chiral coordinate of $\bigl( D \wb{D} \, \delta^3 + \Sigma_{\cY,\Phi} \bigr)$ and given the property \eqref{delta sign in inverse}, one obtains \eqref{integral eom antichiral} or their chiral counterparts.

\subsection{Conformal solutions}

We now assume that, at low energies, the kinetic terms are negligible and thus that the integro-differential equations \eqref{integral eom antichiral} can be approximated by dropping the first term on the left-hand-side. Besides, assuming that the fermion number symmetry remains unbroken, we search for solutions with vanishing mixed fermionic components in $\cG$ and $\Sigma$, \ie, we impose
\bea
\label{zero mixed cpts}
G_{\bar\eta f} &= G_{\bar{f} \eta} = G_{\bar\phi \psi} = G_{\bar\psi \phi} = 0 \;, \\
\Sigma_{\bar\eta f} &= \Sigma_{\bar{f} \eta} = \Sigma_{\bar\phi \psi} = \Sigma_{\bar\psi \phi} = 0 \;.
\eea
One can check that this is a consistent ansatz. The EOMs for the remaining bosonic components are:
\begin{align}
\Sigma_{\bar\eta \eta} &= J \, \Bigl( \tfrac{p-1}{q} \, G_{\bar\phi \phi} \, G_{\bar{f} f} + G_{\bar\eta \eta} \, G_{\bar\psi \psi} \Bigr) \, G\du{\bar\eta \eta}{p-2} \, G\du{\bar\phi \phi}{q-1} \,, \nn\\
\Sigma_{\bar\phi \phi} &= J \, \Bigl( G_{\bar\phi \phi} \, G_{\bar{f} f} + \tfrac{q-1}{p} \, G_{\bar\eta \eta} \, G_{\bar\psi \psi} \Bigr) \, G\du{\bar\eta \eta}{p-1} \, G\du{\bar\phi \phi}{q-2} \,, \nn\\
\Sigma_{\bar{f} f} &= \tfrac{J}{q} \, G\du{\bar\eta \eta}{p-1} \, G\du{\bar\phi \phi}{q} \,,\quad
\Sigma_{\bar\psi \psi} = \tfrac{J}{p} \, G\du{\bar\eta \eta}{p} \, G\du{\bar\phi \phi}{q-1} \,,
\label{alg eom zero mixed}
\end{align}
where all functions are of $(\tau_1, \tau_2)$, and
\bea
\label{integral eom}
\ts \int\! d\tau_3 \, \Sigma_{\bar\eta \eta}(\tau_2,\tau_3) \, G_{\bar\eta \eta}(\tau_1,\tau_3) &= \delta(\tau_1 {-} \tau_2) \,, \\
\ts \int\! d\tau_3 \, \Sigma_{\bar{f} f}(\tau_2,\tau_3) \, G_{\bar{f} f}(\tau_1,\tau_3) &= - \delta(\tau_1 {-} \tau_2) \,, \\
\ts \int\! d\tau_3 \, \Sigma_{\bar\phi \phi}(\tau_2,\tau_3) \, G_{\bar\phi \phi}(\tau_1,\tau_3) &= - \alpha \, \delta(\tau_1 {-} \tau_2) \,, \\
\ts \int\! d\tau_3 \, \Sigma_{\bar\psi \psi}(\tau_2,\tau_3) \, G_{\bar\psi \psi}(\tau_1,\tau_3) &= \alpha \, \delta(\tau_1 {-} \tau_2) \,. 
\eea
We consider turning on chemical potentials $\mu$ for various $\rU(1)$ global symmetries. This corresponds to turning on a real background gauge field $A_t = - \mu$ in Lorentzian signature, and an imaginary background field $A_\tau = i \mu$ in Euclidean signature. Notice that the super-reparametrization symmetry \eqref{superdiff trans} can be preserved only if $\mu_R=p\mu_\eta+q\mu_\phi=0$. For generic chemical potentials, one is left with the standard reparametrization symmetry $\text{Diff}^+(S^{1})$, under which the fields transform as
\begin{align}
\label{diff trans}
G_{\bar A A}(\tau_1, \tau_2) \;&\mapsto\; \bigl( \partial_{\tau_1} \tau'_1 \bigr){}^{\Delta_{A}} \; \bigl( \partial_{\tau_2} \tau'_2 \bigr){}^{\Delta_{A}} \; G_{\bar A A}(\tau_1, \tau_2) \\
\Sigma_{\bar A A}(\tau_1, \tau_2) \;&\mapsto\; \bigl( \partial_{\tau_1} \tau'_1 \bigr){}^{1-\Delta_{A}}  \bigl( \partial_{\tau_2} \tau'_2 \bigr){}^{1-\Delta_{A}} \, \Sigma_{\bar A A}(\tau_1, \tau_2) \nn
\end{align}
where the label $A$ runs over all field species. As argued in \cite{Gu:2019jub}, shifting the self-energies $\Sigma$ by the chemical potentials, the EOMs in the IR are unchanged. 

In the presence of a chemical potential, we use the following conformal ansatz for the two-point function at zero temperature ($\beta=\infty$) of a boson or a fermion with charge $Q$ and dimension $\Delta$:
\be
\label{G conformal ansatz}
G(\tau) = g\,G_s\left(\tau;\Delta,\cE\right) \equiv g \, \frac{e^{\pi\cE} \, \Theta(\tau) + s \, e^{-\pi\cE} \, \Theta(-\tau) }{ |\tau|^{2\Delta} } \,,
\ee
where $s=1$ for a boson and $s=-1$ for a fermion. We review its derivation in Appendix~\ref{app: IR and UV limits}. This ansatz is invariant under $\rPSL(2,\bR)$ transformations acting as $\tau\mapsto\tau'=\frac{a\tau+b}{c\tau+d}$ with $ad-bc=1$, accompanied by a suitable gauge transformation of $A_\tau$
\footnote{When the transformed time difference $\tau'(\tau)-\tau'(0)$ has the opposite sign with respect to $\tau$, one needs to keep track of the extra factor $s \, e^{2\pi\cE}$ when the operators commute past each other, see Appendix~\ref{app: conformal ansatz with chem}.}.
The constant $g$ is bound to be positive by unitarity. The 2-point function at non-zero temperature is obtained through the reparametrization $\phi(\tau) = \tan \frac{\pi\tau}{\beta}$ with $\tau \in \bigl( -\frac{\beta}{2}, \frac{\beta}{2} \bigr)$:
\begin{align}
\label{conf 2pt finite beta}
& G(\tau) = g\,G_s^\beta(\tau;\Delta,\cE) \\
&\equiv g \, e^{-2\pi\cE\frac{\tau}{\beta}} \, \biggl\lvert \frac{\pi}{ \beta \sin \bigl( \frac{\pi\tau}{\beta} \bigr) }\biggr\rvert^{2\Delta} \Bigl( e^{\pi\cE} \, \Theta(\tau) + s \, e^{-\pi\cE} \, \Theta(-\tau) \Bigr) . \nn
\end{align}
The parameter $\cE$ is called spectral asymmetry and is related to the chemical potential. Then the extension of $G(\tau)$ beyond $\tau \in \bigl( -\frac{\beta}{2}, \frac{\beta}{2} \bigr)$ satisfies $G(\tau+\beta) = s \, G(\tau)$.

We solve the equations of motion using the same strategy as in \cite{Heydeman:2022lse}. The equations \eqref{integral eom} can be uniformly written as $\int\! d\tau_3 \, \Sigma(\tau_2,\tau_3) \, G(\tau_1,\tau_3) = - s \, \delta(\tau_1-\tau_2)$, which in Fourier space simply read
\be
\label{IR integral eqn}
\Sigma(\omega) \, G(-\omega) = - s \,,
\ee
where $\Sigma$ is understood to be $\Sigma_{\bar\eta\eta}$ or $\Sigma_{\bar{f}f}$ for the Fermi multiplet components, and $\alpha^{-1}\Sigma_{\bar{\phi}\phi}$ or $\alpha^{-1}\Sigma_{\bar{\psi}\psi}$ for the chiral multiplet ones. 
The Fourier transform of (\ref{G conformal ansatz}) is
\footnote{In our conventions $G(\omega) = \protect\int \! d\tau \, e^{-i\omega\tau} \, G(\tau)$.}
\begin{align}
\label{G conformal ansatz in omega}
G (\omega) &= -i \, g \sgn(\omega) \, \Gamma(1-2\Delta) \,  |\omega|^{2\Delta-1} \times \\
&\quad \times \Bigl( e^{\pi\cE + \pi i \Delta \sgn(\omega)} - s \, e^{-\pi\cE - \pi i\Delta \sgn(\omega)} \Bigr) \,. \nn
\end{align}
The expressions for $\Sigma$ in Fourier and normal space follow from (\ref{IR integral eqn}):
\begin{align}
\Sigma(\omega) &= \frac{ 1 }{ 2 i g  \, \Gamma(1-2\Delta) \, \bigl( \cosh 2\pi\cE - s \cos 2\pi\Delta \bigr) } \times {} \nn\\
&\quad \times \frac{ e^{ - \pi\cE - \pi i\Delta \sgn(\omega) } - s \, e^{\pi\cE + \pi i\Delta \sgn(\omega) } }{  \sign(\omega) |\omega|^{2\Delta-1} } \,, \nn\\
\label{sigma G FT}
\Sigma(\tau) &= \frac{(1-2\Delta) \sin 2\pi\Delta }{ 2\pi g \, \bigl( \cosh 2\pi\cE-s\cos 2\pi\Delta \bigr) } \times {} \\
&\quad \times \, \Bigl( e^{-\pi\cE} \, \Theta(\tau) + s \, e^{\pi\cE} \, \Theta(-\tau)\Bigr) \, |\tau|^{2\Delta-2} \,. \nn
\end{align}

We denote the parameters appearing in the ansatz for $G_{\bar\phi \phi}$ as $\cE_\phi$, $\Delta_\phi$, $g_\phi$, and similarly for the other fields. Plugging $\Sigma$ from \eqref{sigma G FT} into the algebraic equations in \eqref{alg eom zero mixed} and matching spectral asymmetries and dimensions, one obtains four linear relations:
\bea
\label{aux dim eps constr}
\Delta_f &= 1 - (p-1)\Delta_\eta - q \, \Delta_\phi \;, \\
\cE_f &=  -(p-1) \cE_\eta - q \, \cE_\phi \;, \\
\Delta_\psi &= 1 - p \, \Delta_\eta - (q-1) \Delta_\phi \;, \\
\cE_\psi &= - p \, \cE_\eta - (q-1) \cE_\phi \;.
\eea
Matching the coefficients in \eqref{alg eom zero mixed} and making use of \eqref{aux dim eps constr} gives the equations:
\begin{widetext}
\begin{align}
\label{matching coeff in alg}
& \frac{ (1 - 2 \Delta_\eta) \sin(2 \pi \Delta_\eta) }{ \cosh(2 \pi \cE_\eta) + \cos(2 \pi \Delta_\eta)} = 2\pi J g_\eta^{p-1} g_\phi^{q-1} \biggl( \frac{p-1}{q} g_\phi g_f + g_\eta  g_\psi \biggr) \,, \\
& \frac{ \bigl[ 1-2(p-1) \Delta_\eta - 2q \Delta_\phi \bigr] \sin \bigl[ 2\pi(p-1) \Delta_\eta + 2\pi q \Delta_\phi \bigr] }{ \cosh \bigl[ 2\pi(p-1) \cE_\eta + 2\pi q \cE_\phi \bigr] - \cos \bigl[ 2\pi (p-1) \Delta_\eta + 2\pi q \Delta_\phi \bigr] } = \frac{2\pi J}{q} g_\eta^{p-1} g_\phi^{q} \, g_f \,, \nn \\
& \frac{ \alpha \, (1-2\Delta_\phi) \sin(2\pi\Delta_\phi) }{ \cosh(2\pi\cE_\phi) - \cos(2\pi\Delta_\phi)} = 2\pi J g_\eta^{p-1} g_\phi^{q-1} \biggl( g_\phi g_f + \frac{q-1}{p} g_\eta g_\psi \biggr) \,, \nn \\
& \frac{ \alpha \, \bigl[ 1 - 2p\Delta_\eta - 2(q-1) \Delta_\phi \bigr] \sin \bigl[ 2\pi p\Delta_\eta + 2\pi (q-1) \Delta_\phi \bigr] }{ \cosh \bigl[ 2\pi p \cE_\eta + 2\pi (q-1) \cE_\phi \bigr] + \cos \bigl[ 2\pi p \Delta_\eta + 2\pi (q-1) \Delta_\phi \bigr] } = \frac{2\pi J}{p} g_\eta^p \, g_\phi^{q-1} g_\psi \,. \nn
\end{align}
Substituting the second and fourth equations into the first and third ones one can eliminate the parameters $g_\#$ and $J$ and obtain two equations that determine $\Delta_\eta$ and $\Delta_\phi$ as functions of $\cE_\eta$ and $\cE_\phi$:
\begin{align}
\label{conf consistency cond}
\frac{(1-2 \Delta_\eta) \sin(2\pi\Delta_\eta) }{ \cosh(2\pi\cE_\eta) + \cos(2\pi\Delta_\eta)} &= \frac{(p-1) [1-2(p-1) \Delta_\eta-2q\Delta_\phi] \sin [ 2\pi(p-1) \Delta_\eta + 2\pi q \Delta_\phi] }{ \cosh [2\pi(p-1)\cE_\eta + 2\pi q \cE_\phi] - \cos [2\pi (p-1) \Delta_\eta + 2\pi q\Delta_\phi] } \\
&\quad + \frac{\alpha p[1- 2p\Delta_\eta - 2(q-1)\Delta_\phi] \sin[2\pi p\Delta_\eta + 2\pi(q-1)\Delta_\phi] }{ \cosh[2\pi p\cE_\eta + 2\pi(q-1)\cE_\phi] + \cos[2\pi p\Delta_\eta + 2\pi(q-1)\Delta_\phi] } \,, \nn \\
\frac{\alpha(1 - 2\Delta_\phi) \sin(2\pi\Delta_\phi) }{ \cosh(2\pi\cE_\phi) - \cos(2\pi\Delta_\phi)} &= \frac{q [1-2(p-1)\Delta_\eta - 2q\Delta_\phi] \sin [2\pi(p-1)\Delta_\eta + 2\pi q\Delta_\phi] }{ \cosh [2\pi(p-1)\cE_\eta + 2\pi q\cE_\phi] - \cos [2\pi (p-1)\Delta_\eta + 2\pi q\Delta_\phi] } \nn \\
&\quad + \frac{\alpha (q-1) [1 - 2p\Delta_\eta - 2(q-1)\Delta_\phi] \sin[2\pi p\Delta_\eta + 2\pi(q-1)\Delta_\phi] }{ \cosh[2\pi p\cE_\eta + 2\pi(q-1)\cE_\phi] + \cos[2\pi p\Delta_\eta + 2\pi(q-1)\Delta_\phi] } \nn \,.
\end{align}
\end{widetext}
Substituting the solution back into \eqref{matching coeff in alg} then determines the combinations of coefficients $g_\eta^{p-1} g_\phi^q g_f$ and $g_\eta^pg_\phi^{q-1}g_\psi$.
That only those two combinations can be determined in this way is due to the fact that the IR Schwinger--Dyson equations have an emergent symmetry \cite{Fu:2016vas}.
For solutions to be consistent with the approximation \eqref{integral eom} in which we dropped the kinetic terms, we need $\Delta_{\eta,\phi}>0$ and $\Delta_{f,\psi}> 1/2$ which, due to \eqref{aux dim eps constr}, imply:
\bea
\label{conf limit consistency}
\Delta_\eta &>0 \,,\quad& (p-1) \Delta_\eta + q \, \Delta_\phi &< \tfrac12 \,, \\
\Delta_\phi &> 0 \,,\quad& p \, \Delta_\eta + (q-1) \, \Delta_\phi &< \tfrac12 \,.
\eea

\subsection{Superconformal solutions}

Assuming translational invariance and supersymmetry, the bilocal superfields must be functions of the invariant $T_{12}$ in \eqref{transl chiral anti-chiral inv}. This implies that they are determined by their lowest components as defined in \eqref{bilocal definition component}:
\begin{align}
\label{superconformal solutions}
\cG_{\cY}(T_1,T_2) &= G_{\bar\eta \eta}(T_{12}) \,, & \cG_{\Phi}(T_1,T_2) &= G_{\bar\phi \phi}(T_{12}) \,, \\
\Sigma_{\cY}(T_1,T_2) &= \Sigma_{\bar{f} f}(T_{12}) \,,& \Sigma_{\Phi}(T_1,T_2) &= \Sigma_{\bar\psi \psi}(T_{12}) \,. \nn
\end{align}
Expanding the components of $\cG_{\cY,\Phi}$ one obtains the constraints
\bea
\label{susy constr coeff}
\Delta_f &= \Delta_\eta + \tfrac{1}{2} \,,\quad& \cE_f &= \cE_\eta \,,\quad& g_f &= 2 \Delta_\eta g_\eta \,, \\
\Delta_\psi &= \Delta_\phi + \tfrac{1}{2} \,,\quad& \cE_\psi &= \cE_\phi \,,\quad& g_\psi &= 2 \Delta_\phi g_\phi \,.
\eea
Together with \eqref{aux dim eps constr}, they imply
\be
\label{susy constr dim eps}
p \, \Delta_\eta + q \, \Delta_\phi = \tfrac12 \;,\qquad\quad p \, \cE_\eta + q \, \cE_\phi = 0 \;.
\ee
In accord with (\ref{def muF muR}), it will be convenient to parametrize the spectral asymmetries as
\be
\label{cE solution}
\cE_\phi = - p \, \cE \;,\qquad\qquad \cE_\eta = q \, \cE \;. 
\ee
At low temperature we can identify $2\pi \cE = - \beta \mu_F$ in terms of the chemical potential $\mu_F$ for $\rU(1)_F$ (while a chemical potential for $\rU(1)_R$ explicitly breaks supersymmetry). Note that we should restrict to $\cE \geq 0$.
The consistency conditions in \eqref{matching coeff in alg} reduce to two equations. The first one is
\begin{align}
\label{consistency match coeff}
& \alpha \sin(2\pi \Delta_\phi) \\
&= \tfrac qp \, \sin\bigl( \tfrac\pi p - \tfrac{2\pi q}p \Delta_\phi \bigr) \, \frac{ \cosh(2\pi p \cE) - \cos(2\pi \Delta_\phi) }{ \cosh(2\pi q \cE) + \cos\bigl( \frac\pi p - \frac{2\pi q}p \Delta_\phi \bigr) } \,, \nn
\end{align}
which can be used to fix $\Delta_\phi$ in terms of $\cE$. The second one then determines the following combination of the coefficients:
\be
\label{eqn for coeffs susy}
g_\eta^p \, g_\phi^q = \frac{\alpha \, p}{2\pi J} \; \frac{\sin(2\pi\Delta_\phi)}{\cosh(2\pi p\cE)-\cos(2\pi\Delta_\phi)} \;.
\ee
The bounds \eqref{conf limit consistency} on the consistency of the conformal solutions reduce to
\be
\label{delta bound}
0<\Delta_\phi<\tfrac{1}{2q} \;.
\ee
At the lower bound one has $\Delta_\phi = 0$, while at the upper bound one has $\Delta_\eta = 0$. An alternative derivation of the equations in this section is provided in Appendix~\ref{app: superconf sols in superspace}.

\subsection{Superconformal solutions at fixed \tps{\matht{p}}{p} and \tps{\matht{q}}{q}}
\label{sec-existence}

We search for solutions $\Delta_\phi(\cE)$ to (\ref{consistency match coeff}) in the range $\Delta_\phi \in \bigl( 0, \frac1{2q} \bigr)$. To that purpose, we define $R_{p,q,\cE}(\Delta_\phi)$ as the right-hand-side of (\ref{consistency match coeff}) and study the equation $\alpha \sin(2\pi\Delta_\phi) = R_{p,q,\cE}(\Delta_\phi)$. As stressed above, we shall restrict to $\cE \geq 0$.

For $p=q=1$ the model describes free fields with randomly distributed masses. For $\alpha \neq 1$ eqn.~(\ref{consistency match coeff}) has no solutions, while for $\alpha=1$ it is a tautology. We will not consider this model any further. The models with $p=1$ or $p > 1$ have a rather different behavior and we will consider them separately.

\paragraph{Case \tps{$p=1$}{p=1}, \tps{$q\geq 2$}{q>=2}.} In this case the function $R_{1,q,\cE}(\Delta_\phi)$ is zero at the endpoints $\Delta_\phi = 0, \frac1{2q}$ while $\sin(2\pi\Delta_\phi)$ is monotonically increasing from zero, therefore at least one solution exists if the derivative of $R_{1,q,\cE}(\Delta_\phi)$ at $\Delta_\phi=0$ is greater than that of $\alpha\sin(2\pi\Delta_\phi)$. This condition is
\be
\label{p1 solution exist}
\alpha < \frac{q^2 \sinh^2(\pi\cE) }{ \sinh^2(\pi q\cE)} \,.
\ee
It turns out that there is only one solution if \eqref{p1 solution exist} is satisfied, and no solutions otherwise. The function of $\cE$ in \eqref{p1 solution exist} has a maximum of $1$ at $\cE=0$ and it decreases to zero for larger $\cE$. Therefore we must have $\alpha<1$ in order for a solution to exist. At fixed $\alpha < 1$, the solution only exists in the range $\cE \in \bigl[ 0 ,\, \cE_*(\alpha) \bigr)$, where $\cE_*$ saturates (\ref{p1 solution exist}). In this case $\Delta_\phi(\cE)$ has a maximum at $\cE=0$ and it monotonically decreases to zero as $\cE \to \cE_*$. At $\cE = \cE_*$ there is a phase transition signalled by the values $\Delta_\phi = 0$ and $\Delta_\eta = 1/2$ and beyond that point the conformal solution ceases to exist.

This transition is similar to the one found in the two-fermion model of \cite{Heydeman:2022lse}. There it was observed that one of the fermion species saturates its charge while the other one does not. In our model, at the transition point, the charge of the fermion reaches its maximal value $\cQ_\eta=1/2$ while the boson is still above its \textit{minimal} value. This is numerically investigated in Section~\ref{sec: numerics}. Furthermore, in Section~\ref{sec: non conformal} we find gapped solutions that mirror massive free particles. In \cite{Heydeman:2022lse} it was proposed that in the two-fermion model a jump occurs between two distinct sets of solutions in the grand canonical ensemble. However, in Section~\ref{sec: numerics} we illustrate that in our model, for finite $\beta\mu$, the phase after the transition seems to interpolate between the conformal behavior and an exponential behavior of non-conformal solutions. Another possibility, which neither our analytical nor numerical analysis would capture, is that the breakdown of the conformal ansatz signals that a replica-breaking solution is preferred by the path integral. Such a proposal requires further study.

In the special case $p=1$, $q=2$ we can explicitly solve (\ref{consistency match coeff}) and (\ref{p1 solution exist}):
\bea
\label{p1 q2 sol}
\Delta_\phi(\cE) &= \frac1{2\pi} \arccos \biggl( \frac{\alpha}{2-\alpha} \cosh(2\pi \cE) \biggr) \;, \\
\cE_*(\alpha) &= \tfrac1\pi \operatorname{arctanh} \bigl( \sqrt{1-\alpha} \, \bigr) \;.
\eea
The corresponding solution to \eqref{eqn for coeffs susy} is
\be
\label{p1 q2 coeff sol}
g_\eta \, g_\phi^2 = \frac{\alpha}{4\pi J \, (1-\alpha)} \, \frac{ \sqrt{ (2-\alpha)^2 - \alpha^2 \cosh^2(2 \pi \cE) } }{ \cosh(2\pi\cE) } \;.
\ee
For $p=1$, $q=3$ analytic solutions can be found as well, although they are lengthier:
\be
\cos(2\pi \Delta_\phi) = \tfrac{ 2\alpha \cosh(2\pi \cE) + \sqrt{ 36 + 12 \alpha(4-\alpha) \sinh^2(2\pi\cE) } }{ 4(3-\alpha) }
\ee
gives the dimension $\Delta_\phi(\cE)$, while
\be
\cE_*(\alpha) = \tfrac1{2\pi} \operatorname{arccosh} \Bigl( \tfrac{ 3 - \sqrt\alpha}{2\sqrt\alpha} \, \Bigr)
\ee
identifies the phase transition point. The solution for $g_\eta g_\phi^3$ can be easily written.

\paragraph{Case \tps{$p>1$}{p>1}, \tps{$q\geq 2$}{q>=2}.} In this case the function $R_{p,q,\cE}(\Delta_\phi)$ is zero at the endpoints $\Delta_\phi = 0, \frac1{2q}$ if $\cE=0$, while it is positive at $\Delta_\phi=0$ if $\cE> 0$. For $\cE=0$ eqn.~\eqref{consistency match coeff} simplifies to
\be
\label{susy eqn eps zero}
\frac{\alpha\,p}{q} = \tan(\pi\Delta_\phi) \, \tan\biggl( \frac{\pi}{2p}-\frac{\pi q\Delta_\phi}{p} \biggr) \;.
\ee
The right-hand-side is zero at $\Delta_\phi = 0,\frac1{2q}$ and attains a maximum in between. If $\alpha$ is small enough such that $\alpha p/q$ is less than (or equal to) the maximal value then there are two (or one) solutions, otherwise there are no solutions. For $\cE > 0$, since the right-hand-side of \eqref{consistency match coeff} is positive at $\Delta_\phi=0$ and it vanishes at $\Delta_\phi = \frac 1{2q}$, there must be at least one solution regardless of $\alpha$, $\cE$. However, there could be more than one solution. For fixed $p>1$, $q \geq 2$ and small $\alpha$, we empirically find that by varying $\cE$ from above $0$ to $\infty$, we go from having $3$ solutions to $1$ solution through a critical value of $\cE$ where there are $2$ solutions. When $\alpha$ is raised and this scan in $\cE$ is repeated, there is a regime where the number of solutions goes from 1 to 3 to 1. For even larger $\alpha$ there is only one solution regardless of $\cE$. Note that unlike the case of $p=1$ or the model of \cite{Heydeman:2022lse}, there is no region in $\alpha$, $\cE$ where no solutions exist.

\begin{figure}
\includegraphics[width = 0.95\columnwidth]{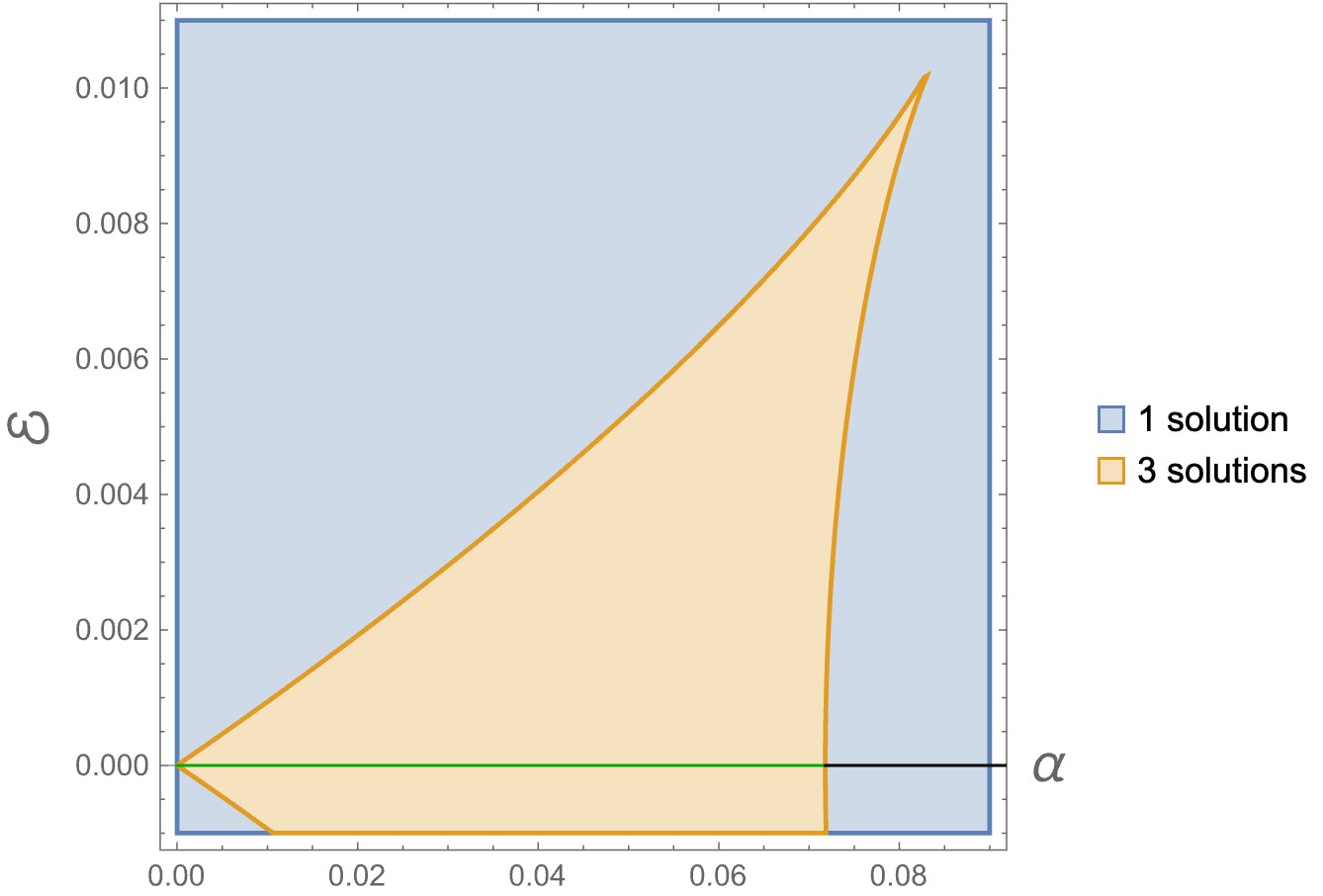} \\
\includegraphics[width = 0.99\columnwidth]{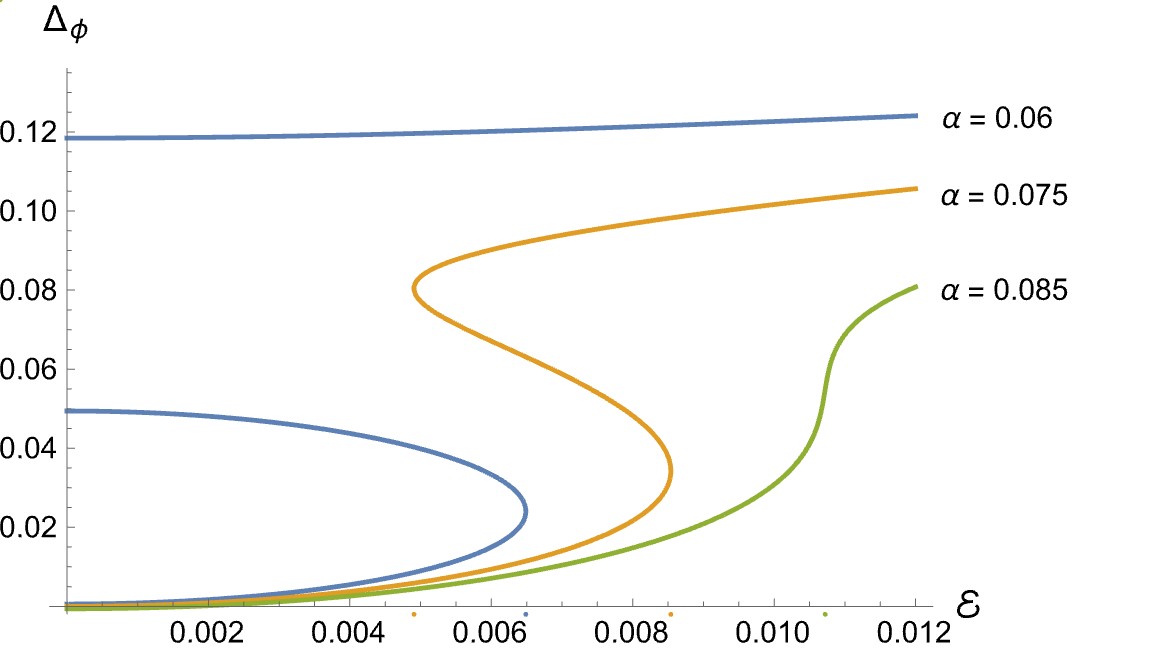}
\caption{\label{fig:sol degen p>1}%
Top: Plot of the number of acceptable solutions to (\ref{consistency match coeff}) for $p=q=3$ as a function of $\alpha$ and $\cE$. For $\cE > 0$ there are 3 solutions in the yellow region, 1 solution in the blue region, and 2 solutions along their boundary. For $\cE=0$ there are two solutions along the green segment, no solutions along the black line, and 1 solution at their boundary. Bottom: Solutions $\Delta_\phi(\cE)$ for $\alpha = 0.06$, $0.075$, $0.085$.}
\end{figure}

The behavior above can be seen explicitly in the case $p=q>1$, where analytic expressions for $\Delta_\phi$ can be obtained by solving the following quartic equation:
\begin{multline}
\label{p eq q quartic}
\cosh(2\pi p \cE) \Bigl( e^{\frac{i\pi}{2p}} (\alpha x^3-x) + e^{-\frac{i\pi}{2p}} (x^3 - \alpha x) \Bigr) \\
+ \tfrac{\alpha-1}2 (x^4-1) + i (\alpha+1) \sin\bigl( \tfrac\pi p \bigr) \, x^2 = 0
\end{multline}
in which $x = \exp \bigl[ i \bigl( 2 \pi \Delta_\phi - \frac{\pi}{2p} \bigr) \bigr]$.
In Fig.~\ref{fig:sol degen p>1} we plot the number of acceptable solutions satisfying $\Delta_\phi\in \bigl( 0, \frac1{2p} \bigr)$ for $p=q=3$ as $\alpha$ and $\cE$ vary. Excluding the horizontal axis $\cE=0$, there are $3$ solutions in the yellow region and $1$ solution in the blue region. On the horizontal axis, there are $2$ solutions along the green segment within the yellow region, and no solution along the black line. On the boundary of the two regions, where the discriminant of the quartic eqn.~\eqref{p eq q quartic} vanishes, there are two solutions for $\cE > 0$ and one solution for $\cE=0$.

There are special values of $\cE$ and $\alpha$ for which \eqref{p eq q quartic} simplifies. For $\alpha=1$, \eqref{p eq q quartic} reduces to a quadratic equation and the unique solution with $\Delta_\phi \in \bigl( 0, \frac1{2p} \bigr)$ is
\be
\label{p eq q Delta sol}
\Delta_\phi = \frac{1}{4p} - \frac{1}{2\pi} \arcsin \biggl[ \frac{\sin \bigl(\frac{\pi}{2p} \bigr) }{ \cosh(2\pi p\cE)} \biggr] \;,
\ee
acceptable whenever $\cE > 0$. The corresponding solution to \eqref{eqn for coeffs susy} is 
\be
\label{p eq q coeff sol}
\rule[-1.8em]{0pt}{0em}
g_\eta^p \, g_\phi^p = \frac{ p \sin \bigl( \frac{\pi}{2p} \bigr) }{ 2\pi J\sqrt{\cosh^2(2\pi p\cE) - \sin^2 \bigl( \frac{\pi}{2p} \bigr) } } \;.
\ee
For $\cE=0$, \eqref{p eq q quartic} can be written as a quadratic equation in $\tan(\pi\Delta_\phi)$ which gives two solutions:
\begin{multline}
\label{p eq q eps zero sol}
\Delta_\phi = \frac{1}{\pi} \arctan \biggl[ \frac{1-\alpha}2 \, \tan \Bigl( \frac{\pi}{2p} \Bigr) \\
\pm \frac12 \sqrt{ (1-\alpha)^2 \tan^2 \Bigl( \frac{\pi}{2p} \Bigr) - 4\alpha} \, \biggr] \,.
\end{multline}
The two solutions are real if and only if $\alpha \leq \tan^2 \bigl( \frac{\pi}{4p} \bigr)$. Indeed $\alpha = \tan^2 \big( \frac{\pi}{4p} \bigr)$ is where the right boundary of the yellow region intersects the horizontal axis in Fig.~\ref{fig:sol degen p>1} and the two solutions merge into one (disappearing for larger values of $\alpha$). It follows from \eqref{p eq q eps zero sol} that both solutions satisfy $\Delta_\phi>0$, that the larger solution becomes $\Delta_\phi = \frac{1}{2p}$ for $\alpha=0$, and that $\Delta_\phi < \frac{1}{2p}$ for the allowed values $\alpha>0$. The corresponding solution to \eqref{eqn for coeffs susy} is
\be
\label{p eq q eps zero coeff sol}
g_\eta^p \, g_\phi^p = \frac{\alpha \, p}{2\pi J \tan(\pi\Delta_\phi) } \;.
\ee

\paragraph{Case \tps{$p>1$}{p>1}, \tps{$q=1$}{q=1}.} The behavior of the function $R_{p,1,\cE}(\Delta_\phi)$ is similar to the one in the previous case, however the function $\sin(2\pi\Delta_\phi)$ vanishes at both endpoints $\Delta_\phi = 0, \frac12$. For very small values of $\alpha$, the number of solutions $\Delta_\phi(\cE)$ to (\ref{consistency match coeff}) as we increase $\cE$ from above $0$ to $\infty$ is 2 (of which only 1 is acceptable at $\cE=0$) and then 0. For larger values of $\alpha$ it is 1 (not acceptable at $\cE=0$), then 2, and then 0. For even larger values of $\alpha$ there is only 1 solution (that becomes $\Delta_\phi=0$ at $\cE=0$ and thus is not acceptable) up to $\cE_*(\alpha)$ and then no solutions for $\cE \geq \cE_*(\alpha)$.

\subsection{Solutions at large \tps{\matht{p}}{p} and fixed \tps{\matht{q}}{q}}
\label{subsec: fixed q large p}

Using the methods in \cite{Maldacena:2016hyu, Fu:2016vas, Davison:2016ngz}, we search for analytic solutions to the Schwinger--Dyson equations at large $p$. We truncate to the bosonic components only, assuming \eqref{zero mixed cpts}, but we do not neglect the kinetic terms.
Let the chemical potentials corresponding to $Q_{\eta,\phi}$ be $\mu_{\eta,\phi}$ respectively. The equations we want to solve are the algebraic ones in \eqref{alg eom zero mixed} and the integro-differential ones in (\ref{integral eom antichiral}) that in components read as in (\ref{integro-differential equations component}):
\begin{widetext}
\bea
\label{int eq with mu}
- \bigl( \partial_{\tau_2} + \mu_\eta \bigr) \, G_{\wb\eta \eta}(\tau_1,\tau_2) + \ts\int\! d\tau_3 \, \Sigma_{\wb\eta \eta}(\tau_2,\tau_3) \, G_{\wb\eta \eta}(\tau_1,\tau_3) &= \delta(\tau_1-\tau_2) \;, \\
G_{\wb{f} f}(\tau_1,\tau_2) + \ts\int\! d\tau_3 \, \Sigma_{\wb{f} f}(\tau_2,\tau_3) \, G_{\wb{f}f}(\tau_1,\tau_3) &= - \delta(\tau_1-\tau_2) \;, \\
-\bigl( \partial_{\tau_2} + \mu_\phi \bigr) \, G_{\wb\phi \phi}(\tau_1,\tau_2) + \ts\int\! d\tau_3 \, \Sigma_{\wb\phi \phi}(\tau_2,\tau_3) \, G_{\wb\phi \phi}(\tau_1,\tau_3) &= -\alpha \, \delta(\tau_1- \tau_2) \;, \\
G_{\wb\psi \psi}(\tau_1,\tau_2) + \ts\int\! d\tau_3 \, \Sigma_{\wb\psi \psi}(\tau_2,\tau_3) \, G_{\wb\psi \psi}(\tau_1,\tau_3) &= \alpha \, \delta(\tau_1-\tau_2) \;.
\eea
\end{widetext}
Assuming that the theory becomes free in the large $p$ limit, we search for solutions in a $1/p$ expansion around the free 2-point functions. Generically, we expand the bilocal fields $G$ (assuming translational invariance) as
\bea
\label{large p expansion}
G(\tau) &= G_0(\tau) + \frac{1}{p^n} G_{1}(\tau) + \cO(p^{-n-1}) \;, \\
G(\omega_k) &= G_0(\omega_k) + \frac{1}{p^n}G_{1}(\omega_k) + \cO(p^{-n-1}) \;.
\eea
Here $G_0$ are the free propagators (we review the correlators of free Fermi and chiral multiplets in Appendix~\ref{app: IR and UV limits}), $\omega_k$ are the appropriate Matsubara frequencies
\footnote{In our conventions $G(\tau) = \protect\sum_{n \,\in\, \bZ + r} G(n) \, e^{\frac{ 2\pi in\tau}{\beta}}$ and $G(n) = \frac{1}{\beta} \protect\int_0^\beta d\tau\, e^{ -\frac{ 2\pi in\tau }{ \beta}} \, G(\tau)$, where $r=0$ for bosons and $r=1/2$ for fermions.},
$n>0$ are suitable integers that we determine below, and we require
\be
\label{large p limit}
\bigl\lvert G_0(\omega_k) \bigr\rvert \,\gg\, \bigl\lvert G_1(\omega_k) \bigr\rvert / p^n \;,\quad \forall \, \omega_k \text{ fixed} \,,\quad p \gg 1 \;.
\ee
More in detail, substituting the free propagators \eqref{free aux psi 2pf}, \eqref{fb eucl 2pt}, \eqref{ff aux 2pt}, \eqref{ff eucl 2pt} into the definition \eqref{def bilocal general} of the bilocal fields, one obtains:
\begin{align}
G_{\bar\eta \eta}^{(0)}(\tau) &= \hspace{0.7em} \frac{ e^{\mu_\eta \tau} }{2 \cosh\bigl( \frac{\mu_\eta \beta}2 \bigr)} \Bigl[ \Theta(\tau) \, e^{- \frac{\mu_\eta \beta}2} - \Theta(-\tau) \, e^{\frac{\mu_\eta \beta}2} \Bigr] \;, \nn\\
G_{\bar{f} f}^{(0)}(\tau) &= -\delta(\tau) \,, \nn\\
G_{\bar\phi \phi}^{(0)}(\tau) &= \alpha \, \frac{ e^{\mu_\phi\tau} }{ 2 \sinh\bigl( \frac{\mu_\phi \beta}2 \bigr) } \Bigl[ \Theta(\tau) \, e^{- \frac{\mu_\phi \beta}2} + \Theta(-\tau) \, e^{ \frac{\mu_\phi \beta}2} \Bigr] \,, \nn\\
G_{\bar\psi \psi}^{(0)}(\tau) &= \alpha \, \delta(\tau) \,. 
\label{biloc UV lim}
\end{align}
Being them free propagators, they satisfy the free equations of motion, obtained from (\ref{int eq with mu}) by setting all $\Sigma$'s to zero. We consider an ansatz for the bilocal fields that is an expansion in $1/p$ around the free result:
\begin{align}
\label{G ansatz fp lq}
G_{\bar\eta \eta}(\tau) &= G_{\bar\eta \eta}^{(0)}(\tau) \Bigl( 1 + p^{-1} \wt G_{\bar\eta \eta}(\tau) + \cO(p^{-2}) \Bigr) \,, \\
G_{\bar{f} f}(\tau) &= - \delta(\tau) + p^{-n_f} \wt G_{\bar{f} f}(\tau) + \cO \bigl( p^{-n_f-1} \bigr) \,, \nn\\
G_{\bar\phi \phi}(\tau) &= G_{\bar\phi \phi}^{(0)}(\tau) \Bigl( 1 + p^{-n_\phi} \wt G_{\bar\phi \phi}(\tau) + \cO \bigl( p^{-n_\phi-1} \bigr) \Bigr) \,, \nn\\
G_{\bar\psi \psi}(\tau) &= \alpha \, \delta(\tau) + p^{-n_\psi} \wt G_{\bar\psi \psi}(\tau) + \cO \bigl( p^{-n_\psi-1} \bigr) \,. \nn
\end{align}
Here we assume that $n_f,n_\phi,n_\psi>0$ and that $\wt G_{\bar\eta \eta}$, $\wt G_{\bar f f}$, $\wt G_{\bar\phi \phi}$, $\wt G_{\bar\psi \psi}$ are $\cO(1)$. We chose the order of the leading correction in $G_{\bar\eta \eta}$ such that various terms in the equations of motion simplify using
\be
\label{Getaeta to power p}
\bigl[ G_{\bar\eta \eta} \bigr]^{p-n} = \bigl[ G_{\bar\eta \eta}^{(0)} \bigr]^{p-n} \, e^{\wt G_{\bar\eta \eta}} \, \bigl( 1 + \cO(p^{-1}) \bigr) \;.
\ee
This in turn leads to the same differential equation as in \cite{Maldacena:2016hyu, Fu:2016vas}, with the benefit of hindsight.
In terms of the chemical potentials $\mu_F$ for $\rU(1)_F$ and $\mu_R$ for any R-charge $\rU(1)_R$, the potentials $\mu_{\eta,\phi}$ are decomposed as
\begin{align}
\mu_\eta &= q \, \mu_F + R[\eta] \, \mu_R \,,\quad& \mu_R &= p \, \mu_\eta + q \, \mu_\phi \,, \nn\\
\mu_\phi &= - p \, \mu_F + R[\phi] \, \mu_R \,.
\label{chemical pots large p}
\end{align}
The last equation follows from the constraint that the J-term has R\nobreakdash-charge $1$, \ie, from $p \, R[\eta] + q \, R[\phi]=1$ . We want to be able to impose the supersymmetry constraint $\mu_R=0$ at large $p$. This implies that the scalings of $\mu_\eta$ and $\mu_\phi$ must be related as $\mu_\phi \sim p \, \mu_\eta$. The simplest scaling is $\mu_\eta\sim\mu_F\sim R[\eta]\sim\cO(p^{-1})$ and $\mu_\phi\sim\mu_R\sim R[\phi]\sim\cO(1)$. In this way both $G_{\bar\eta \eta}^{(0)}$ and $G_{\bar\phi \phi}^{(0)}$ are $\cO(1)$. Besides, the bilocal fields should agree with the free limits at short distances, which implies the following boundary conditions: $\wt G(0) = \wt G(\beta) = 0$ for all components.
Plugging the ansatz (\ref{G ansatz fp lq}) and the boundary conditions into \eqref{int eq with mu} we get:
\begin{align}
\cO(p^{-2}) &= \tfrac1p \, G_{\bar\eta \eta}^{(0)}(\tau_{12}) \, \partial_{\tau_{12}} \wt G_{\bar\eta \eta}(\tau_{12}) \nn\\
&\quad + \ts\int\! d\tau_3 \, \Sigma_{\bar\eta \eta}(\tau_{23}) \, G_{\bar\eta \eta}^{(0)}(\tau_{13})  \;, \nn\\
\cO \bigl( p^{-n_f-1} \bigr) &= \tfrac{1}{p^{n_f}} \wt G_{\bar{f} f}(\tau_{12}) - \Sigma_{\bar{f} f}(\tau_{21})  \;, \nn\\
\cO\bigl( p^{-n_\phi -1} \bigr)  &= \tfrac1{p^{n_\phi}} \, G_{\bar\phi \phi}^{(0)}(\tau_{12}) \, \partial_{\tau_{12}} \wt G_{\bar\phi \phi}(\tau_{12}) \\
&\quad + \ts\int\! d\tau_3 \, \Sigma_{\bar\phi \phi}(\tau_{23}) \, G_{\bar\phi \phi}^{(0)}(\tau_{13}) \;, \nn \\
\cO \bigl( p^{-n_\psi-1} \bigr) &= \tfrac{1}{p^{n_\psi}} \wt G_{\bar\psi \psi}(\tau_{12}) + \alpha \, \Sigma_{\bar\psi \psi}(\tau_{21})   \;, \nn
\end{align}
where $\tau_{ij} = \tau_i - \tau_j$. These equations can be solved to obtain the $\Sigma$'s in terms of the $G$'s at leading order in $\frac1p$.
In particular, one acts with $(\partial_{\tau_1} - \mu_\eta)$ on the first equation and with $(\partial_{\tau_1} - \mu_\phi)$ on the third equation, obtaining
\footnote{We have neglected the terms $ - p^{-1} \, \delta(\tau) \, \partial_\tau \protect\widetilde{G}_{\bar\eta \eta}(\tau)$ in $\Sigma_{\bar\eta \eta}(-\tau)$ and $- p^{-n_\phi} \, \delta(\tau) \, \partial_\tau \protect\widetilde{G}_{\bar\phi \phi}(\tau)$ in $\Sigma_{\bar\phi \phi}(-\tau)$ since we are interested in the equations at $\tau\neq 0$.}:
\begin{align}
\Sigma_{\bar\eta \eta}(-\tau) &= - \tfrac{1}{p} \, G_{\bar\eta \eta}^{(0)}(\tau) \, \partial_\tau^2\wt G_{\bar\eta \eta}(\tau) + \cO(p^{-2}) \,, \nn\\
\label{sigma(g) fp lq}
\Sigma_{\bar{f} f}(-\tau) &= \tfrac{1}{p^{n_f} \rule{0pt}{0.67em}} \, \wt{G}_{\bar{f} f}(\tau) +  \cO ( p^{-n_f-1} ) \,, \\
\Sigma_{\bar\phi \phi}(-\tau) &= \tfrac{1}{\alpha p^{n_\phi} \rule{0pt}{0.67em}} \, G_{\bar\phi \phi}^{(0)}(\tau) \, \partial_\tau^2 \wt G_{\bar\phi \phi}(\tau) + \cO( p^{-n_\phi-1} ) \,, \nn\\
\Sigma_{\bar\psi \psi}(-\tau) &= - \tfrac{1}{\alpha p^{n_\psi} \rule{0pt}{0.67em}} \, \wt G_{\bar\psi \psi}(\tau) +  \cO ( p^{-n_\psi-1} ) \,. \nn
\end{align}

Let us then consider the algebraic equations. Substituting the ansatz \eqref{G ansatz fp lq} into the last two equations of \eqref{alg eom zero mixed} we determine the leading behavior of $\Sigma_{\bar{f} f}$ and $\Sigma_{\bar\psi \psi}$. This can then be used in (\ref{sigma(g) fp lq}) to determine $\wt G_{\bar{f} f}$ and $\wt G_{\bar\psi \psi}$. In order to have a well-defined large $p$ limit, the simplest choice is to scale $J$ as $2^p p^{-n_f}$, so that the combination
\be
\label{def cal J}
\cJ \,\equiv\, \frac{ \alpha^q \, p^{n_f} \, J }{ 2^{p+q-2} \, q \cosh^{p-2} \bigl( \frac{\mu_\eta\beta}{2} \bigr) \sinh^q \bigl( \frac{\mu_\phi\beta}2 \bigr) }
\ee
remains fixed as $p$ becomes large. Recalling that $p$ is odd, we obtain:
\begin{align}
\label{g aux ito g dyn fp}
\wt G_{\bar{f} f}(\tau) &= \frac{ \cJ \,e^{\left(\mu_\eta - \mu_R\right)\tau}}{ 2 \cosh \bigl( \frac{\mu_\eta\beta}2 \bigr)} \, e^{\wt G_{\bar\eta \eta}(-\tau)} 
\times {} \\
&\quad \times \Bigl[ \Theta(\tau) \: e^{- \left(\mu_\eta - \mu_R\right) \frac\beta2} + \Theta(-\tau) \: e^{ \left(\mu_\eta - \mu_R\right) \frac\beta2} \Bigr] \,, \nn\\
\wt G_{\bar\psi \psi}(\tau) &= \frac{ q \, \cJ \,\sinh\bigl( \tfrac{\mu_\phi \beta}{2} \bigr) \, e^{(\mu_\phi - \mu_R)\tau}}{2\cosh^2 \bigl( \frac{\mu_\eta\beta}2 \bigr)}\, e^{\wt{G}_{\bar\eta \eta}(-\tau)} \times {} \nn\\
&\quad \times \Bigl[ \Theta(\tau) \: e^{ - (\mu_\phi - \mu_R) \frac\beta2 } - \Theta(-\tau) \: e^{(\mu_\phi - \mu_R) \frac\beta2} \Bigr] \,, \nn
\end{align}
together with $n_\psi=n_f+1$. Similarly, substituting these expressions and the ansatz \eqref{G ansatz fp lq} into the first two equations of \eqref{alg eom zero mixed}, we determine the leading behavior of $\Sigma_{\bar\eta \eta}$ and $\Sigma_{\bar\phi \phi}$, up to $\cO(p^{-2n_f})$ and $\cO(p^{-2n_f - 1})$ respectively:
\begin{align}
\label{aux in dyn alg}
\Sigma_{\bar\eta \eta}(\tau) &= - \frac{\cJ^2}{ p^{2n_f-1}} \; G^{(0)}_{\bar\eta \eta}(-\tau) \; e^{\wt G_{\bar\eta \eta}(\tau) + \wt G_{\bar\eta \eta}(-\tau)} \,, \\
\Sigma_{\bar\phi \phi}(\tau) &= \frac{ q \, \cJ^2 \,\sinh^2 \bigl( \frac{\mu_\phi \beta}2 \bigr)}{ p^{2n_f} \, \alpha^2\, \cosh^2 \bigl( \frac{\mu_\eta\beta}2 \bigr)} \; G_{\bar\phi \phi}^{(0)}(-\tau) \; e^{\wt G_{\bar\eta \eta}(\tau) + \wt G_{\bar\eta \eta}(-\tau) } . \nn
\end{align}
Equating these expressions with those in  \eqref{sigma(g) fp lq} to leading order in $1/p$ fixes $n_f=1$ and $n_\phi = n_\psi = 2$, and gives the following differential equations:
\bea
\partial_\tau^2 \, \wt G_{\bar\eta \eta}(\tau) &= \cJ^2 \; e^{\wt G_{\bar\eta \eta}(\tau) + \wt G_{\bar\eta \eta}(-\tau)} \,,\\
\partial_\tau^2 \, \wt G_{\bar\phi \phi}(\tau) &= \frac{q \, \cJ^2\, \sinh^2 \bigl( \tfrac{\mu_\phi \beta}2 \bigr) }{ \alpha\, \cosh^2 \bigl( \frac{\mu_\eta\beta}2 \bigr)} \; e^{\wt G_{\bar\eta \eta}(\tau) + \wt G_{\bar\eta \eta}(-\tau)} \,.
\eea
Notice that the free propagators $G_{\bar\eta \eta}^{(0)}$ and $G_{\bar\phi \phi}^{(0)}$ have dropped from the equations.
The only solution to the second equation that satisfies the boundary conditions is
\be
\label{fp lq de Gp}
\wt G_{\bar\phi \phi}(\tau) = \frac{ q \, \sinh^2 \bigl( \tfrac{\mu_\phi \beta}2 \bigr)  }{ \alpha \, \cosh^2 \bigl( \frac{\mu_\eta\beta}2 \bigr)} \; \wt G_{\bar\eta \eta} \;.
\ee
The differential equation for $\wt G_{\bar\eta \eta}$ is the same as the one found in \cite{Maldacena:2016hyu, Davison:2016ngz, Fu:2016vas}. Since the odd part $\wt G_\text{odd}(\tau) = \frac12 \bigl( \wt G_{\bar\eta \eta}(\tau) - \wt G_{\bar\eta \eta}(-\tau) \bigr)$ satisfies $\partial_\tau^2 \, \wt G_\text{odd}(\tau)=0$, the only solution compatible with the boundary conditions is $\wt G_\text{odd}(\tau)=0$ and we can take $\wt G_{\bar\eta \eta}$ to be even. Then the general solution with integration constants $v,b$ is $\exp\bigl[ \wt G_{\bar\eta \eta}(\tau) \bigr] = v / \bigl[ \beta \cJ \sin \bigl( \frac v\beta |\tau| + b \bigr) \bigr]$
where $v>0$, $\beta$ has been introduced for later convenience, and $b$ is defined modulo $2\pi$.
The boundary conditions $\wt G(0) = \wt G(\beta)=0$ imply $\beta \cJ = v / \sin(b)$ and $\sin(b) = \sin(v+b)$. One solution to the second equation is $v = 2\pi m$ for $m \in \bN$, however this does not lead to a function $\exp\bigl[ \wt G_{\bar\eta \eta}(\tau) \bigr]$ that is positive for all values of $\tau$. The other solutions are $b = \bigl( n + \frac12 \bigr) \pi - \frac v2$ for $n\in \bZ$ and without loss of generality we consider $n=0,1$. The other equation reduces to $\beta \cJ = \pm v / \cos\bigl( \frac v2 \bigr)$, where $\pm$ correspond to $n=0,1$, respectively. Only the case $n=0$, $0 < v < \pi$ leads to a positive function $\exp\bigl[ \wt G_{\bar\eta \eta}(\tau) \bigr]$, and the second equation has one and only one solution for all values of $\beta$.
The final solution is thus:
\be
\label{sol large pq const fixed}
e^{\wt G_{\bar\eta \eta}(\tau)} = \frac{v}{ \beta \cJ \cos \bigl( \frac v\beta |\tau| - \frac{v}2 \bigr)} \;,\qquad \beta \cJ = \frac{v}{ \cos \bigl( \frac{v}{2} \bigr)} \;,
\ee
where $0 < v < \pi$.

In the weak coupling limit $\beta\cJ\rightarrow 0$ (keeping all $\mu_I\beta$'s fixed) one has $v\to0$ and $v/\beta \cJ \rightarrow 1$ and therefore $\wt G_{\bar\eta \eta} ,\, \wt G_{\bar\phi \phi} \to 0$. From \eqref{g aux ito g dyn fp} we see that also $\wt G_{\bar{f} f} ,\, \wt G_{\bar\psi \psi} \to 0$ and the solution reduces to the free UV one at leading order in $1/p$. In the strong coupling limit $\beta \cJ \rightarrow\infty$, instead, one has $v\rightarrow\pi$ and, to this order of approximation and away from $\tau=0$, the 2-point functions agree with the conformal ones (\ref{conf 2pt finite beta}) with the following parameters:
\begin{align}
\label{conf param ito uv param}
\Delta_\eta &= \frac1{2p} + \cO(p^{-2}) ,\qquad \Delta_f = \frac12 + \cO(p^{-1}) , \\
\Delta_\phi &= \frac{q \,\sinh^2 \bigl( \tfrac{\mu_\phi \beta}2 \bigr) }{ 2\alpha p^2 \, \cosh^2 \bigl( \frac{\mu_\eta\beta}2 \bigr)} + \cO\bigl( \tfrac1{p^3} \bigr) , \quad  \Delta_\psi = \frac12 + \cO\bigl( \tfrac1p \bigr) , \nn\\
2 \pi \cE_\eta &= -\beta \mu_\eta + \cO(p^{-2}) , \;\, 2 \pi \cE_f = \beta (\mu_R - \mu_\eta) + \cO(p^{-1}) , \nn\\
2 \pi \cE_\phi &= - \beta \mu_\phi + \cO(p^{-3}) , \; 2 \pi \cE_\psi = \beta (\mu_R - \mu_\phi) + \cO(p^{-1}) , \nn
\end{align}
as well as
\begin{align}
\label{g for large p}
g_\eta &= \frac{1 }{ 2 \cosh \bigl( \frac{\mu_\eta\beta}2 \bigr)} \biggl(1 - \frac{1}{p} \log \cJ \biggr) +\cO(p^{-2}) , \\
g_f &= \frac{1}{2 p \cosh \bigl( \frac{\mu_\eta\beta}2 \bigr)} + \cO(p^{-2}) , \nn\\
g_\phi &= \frac{\alpha }{ 2 \sinh \bigl( \frac{\mu_\phi \beta}2 \bigr) } \biggl( 1 - \frac{q \sinh^2 \bigl( \tfrac{\mu_\phi \beta}2 \bigr) }{ \alpha p^2 \cosh^2 \bigl( \frac{\mu_\eta\beta}2 \bigr) } \log \cJ \biggr) + \cO \bigl( \tfrac1{p^3} \bigr) , \nn\\
g_\psi &= \frac{q \sinh \bigl( \tfrac{ \mu_\phi \beta}2 \bigr)}{ 2 p^2 \cosh^2 \bigl( \frac{\mu_\eta\beta}2 \bigr)} + \cO(p^{-3}) . \nn
\end{align}
The conformal dimensions are compatible with the bounds \eqref{conf limit consistency} at leading order in $1/p$, in particular $\Delta_\eta$ saturates $p\Delta_\eta = \frac12 + \cO(p^{-1})$. At higher order in $1/p$ it would be necessary to verify that all bounds are satisfied. We see that the condition $\mu_\phi>0$ (discussed in Section~\ref{sec: model}) is necessary in order to ensure $g_\phi, g_\psi>0$ as required by unitarity. We tentatively see that the solution at large $p$ interpolates between the free UV limit and the IR conformal-like behavior. Notice in particular that the phase transition is not visible in this limit. 

It might be tempting to think that the conformal data in \eqref{conf param ito uv param} solves, at some given order in $1/p$, the self-consistency equations \eqref{conf consistency cond}. This, in general, does not happen, due to incompatibility between the large $p$ and low energy limits. Write \eqref{int eq with mu} in Fourier space:
\be
\label{sd fourier space}
\biggl[ - \frac{s}{G_{0}(\omega_k)} + \beta^2 \, \Sigma(\omega_{-k} ) \biggr] \, G(\omega_k) = -s \,,
\ee
where $\Sigma$ should again be understood as $\alpha^{-1}\Sigma$ for $\phi$ and $\psi$, and we used that $G_0$ is the free propagator. In the large $p$ limit \eqref{large p limit}, using the expansion \eqref{large p expansion}, one has
\begin{align}
\label{sigma large p}
\Sigma(\omega_{-k}) &= \frac{ -s }{ \beta^2 G_0(\omega_k) + \frac{\beta^2}{p^n} G_1(\omega_k) + \ldots 
} + \frac{s}{\beta^2G_{0}(\omega_k)} \nn\\
&= \frac{s}{p^n}\frac{G_{1}(\omega_k)}{\beta^2G_{0}(\omega_k)^2} + \cO \bigl( \tfrac{1}{p^{n+1}} \bigr) \,.
\end{align}
In these terms, as one can see from \eqref{int eq with mu}, the low-energy regime we previously studied for dynamical and auxiliary fields, respectively, corresponds to taking $\bigl\lvert \beta \, \Sigma_{f, \psi}(\omega_{-k}) \bigr\rvert \gg 1$, $\bigl\lvert \mu_{\phi, \eta} - \beta \, \Sigma_{\phi, \eta}(\omega_{-k}) \bigr\rvert \gg |\omega_k|$, and $\omega_k \ll J$.
Combining with \eqref{sigma large p}, for the auxiliary fields we find that
\be
\label{IR limit aux}
p^{-n} \, \bigl\lvert G_1(\omega_k) \bigr\rvert \gg \bigl\lvert G_0(\omega_k) \bigr\rvert \;,
\ee
which directly contradicts \eqref{large p limit} (one has to recall that $\bigl\lvert G_0(\omega_k) \bigr\rvert = 1/\beta$ for auxiliary fields). This means that, as long as we must include auxiliary fields in our considerations, there is no regime of large $p$ and small $\omega_k$ in which the approximation we made in our large $p$ computation and the approximation of dropping the kinetic terms are both reliable.

This result suggests that, at large $p$, the system at low energies first enters into a quasi-conformal regime described by (\ref{conf param ito uv param})--(\ref{g for large p}) in which interactions compete with the kinetic terms, while at extremely low temperatures the system ends up in a different conformal regime that satisfies (\ref{matching coeff in alg})--(\ref{conf consistency cond}) and in which the kinetic terms are negligible.

A particular case, in this respect, is the supersymmetric one for which $\mu_R=0$. In this case the IR conformal solution is supersymmetric since its parameters satisfy the constraints \eqref{susy constr coeff}--\eqref{susy constr dim eps} within the working accuracy. Moreover, \eqref{consistency match coeff} can be independently derived from $\cI$-extremization, as we do in Section~\ref{sec: I ext}, without assuming \eqref{IR limit aux}. At this point, the zero-temperature limit where the index is computed can be taken safely by requiring that $p$ goes to infinity faster than $G_1(\omega)$ as $\omega$ goes to zero, in such a way that \eqref{large p limit} holds. Because of this, \eqref{conf param ito uv param}--(\ref{g for large p}) in the supersymmetric regime do actually solve \eqref{consistency match coeff}.

\subsubsection{Grand potential and entropy at large \tps{$p$}{p}}
\label{logz large p}

It is possible to compute the grand potential $\log Z$ in a large $p$ expansion. To leading order in $1/N$, this is just the averaged action \eqref{bilocal action superspace} (or (\ref{averaged action component}) in components) evaluated on the solution provided by \eqref{sol large pq const fixed}. Instead of a direct evaluation, we follow \cite{Fu:2016vas, Davison:2016ngz} and first compute the derivative of $\log Z$ with respect to $J$. Only the explicit dependence on $J$ in \eqref{bilocal action superspace} matters when taking this derivative, and not the dependence through dimensionful coefficients such as $g_{\eta, \phi}$ in the bilocal fields $G_{\eta, \phi}$ since the fields solve the equations of motion. We get:
\begin{align}
 & J \, \partial_J \, \frac{\log Z}N \\
 &\quad = J \!\int\! d\tau_1 \, d\tau_2 \, \Bigl( \tfrac1q \, G_{\bar{f} f} \, G_{\bar\eta \eta}^{p-1} \, G_{\bar\phi \phi}^q + \tfrac1p \, G_{\bar\eta \eta}^p \, G_{\bar\psi \psi} \, G_{\bar\phi \phi}^{q-1} \Bigr) \nn\\
&\quad = \frac{ \cJ^2 \cos^2 \bigl( \frac{v}2 \bigr) }{ 4 p^2\cosh^2 \bigl( \frac{\mu_\eta \beta}2 \bigr) } \int_{-\beta}^{\beta} \! d\tau \, \frac{\beta-|\tau|}{ \cos^2 \bigl( \frac v\beta |\tau| -\frac{v}2 \bigr) } + \cO(p^{-3}) \nn\\
&\quad = \frac{v \tan \bigl( \frac{v}2 \bigr) }{ 2p^2 \cosh^2 \bigl( \frac{\mu_\eta \beta}2 \bigr)} + \cO(p^{-3}) \,. \nn
\end{align}
The delta functions in $G_{\bar{f}f}$ and $G_{\bar\psi\psi}$ do not contribute to the integral on the second line, and hence only the first term contributes at leading order, since the second term is suppressed by $1/p^2$ with respect to the first one.
Using the relation between $\beta \cJ$ and $v$ in \eqref{sol large pq const fixed} as well as the linearity of $\beta \cJ$ in $J$ in (\ref{def cal J}), the derivative with respect to $J$ can be exchanged for a derivative with respect to $v$ using the formula $J\, \partial_J = v / \bigl[ 1 + \frac v2 \tan\bigl( \frac v2 \bigr) \bigr] \, \partial_v$. We obtain a differential equation in $v$ for $\log Z$:
\be
\label{v de logz}
\frac1N \, \partial_v \log Z = \frac{ \tan\bigl( \frac v2 \bigr) \, \bigl[ 1 + \frac v2 \tan \bigl( \frac v2 \bigr) \bigr] }{ 2 p^2 \cosh^2 \bigl( \frac{\mu_\eta \beta}2 \bigr)} + \cO(p^{-3}) \,.
\ee
The value of $\log Z$ at $v=0$ is known because at that point the theory is free. We can read off $Z \big|_{v=0}$ from \eqref{free b Z} and \eqref{ff z}. Integrating \eqref{v de logz} from $0$ to $v$ at large $p$ then gives
\begin{align}
\frac{\log Z}{N} &= \log \biggl[2\cosh \biggl( \frac{\mu_\eta \beta}2 \biggr)\biggr] - \alpha \log \biggl[2\sinh \biggl( \frac{\mu_\phi \beta}2 \biggr) \biggr]
\nn\\
&\quad + \frac{ v \tan \bigl( \frac v2 \bigr) - \frac{v^2}4 }{ 2 p^2 \cosh^2 \bigl( \frac{\mu_\eta \beta}2 \bigr)} + \cO(p^{-3}) \,.
\label{large p logZ v}
\end{align}
As $p \rightarrow \infty$, $\log Z$ tends to the free value, which is consistent with our starting ansatz \eqref{G ansatz fp lq}.

In order to compute the entropy $S = (1-\beta \partial_\beta) \log Z$, one needs to make the $\beta$ dependence in $v$ explicit. Using $\beta \partial_\beta = \beta \cJ \cos\bigl( \frac v2 \bigr) / \bigl[ 1 + \frac v2 \tan\bigl( \frac v2 \bigr) \bigr] \cdot \partial_v$ we obtain
\begin{align}
\frac{S}N &= \log \bigl[2\cosh \bigl( \tfrac{\mu_\eta \beta}2 \bigr) \bigr] - \tfrac{\mu_\eta \beta}2 \tanh\bigl( \tfrac{\mu_\eta \beta}2 \bigr) \\
&\quad - \alpha \log \bigl[2\sinh \bigl( \tfrac{\mu_\phi \beta}2 \bigr) \bigr] + \tfrac{\alpha \mu_\phi \beta}2 \coth\bigl( \tfrac{\mu_\phi \beta}2 \bigr) \nn\\
&\quad - \frac{v^2}{8p^2\cosh^2 \bigl( \frac{\mu_\eta \beta}2 \bigr)} + \cO(p^{-3}) \,. \nn
\end{align}
On the first two lines is the entropy of the free theory, as in \eqref{free b S entropy} and \eqref{free f S entropy}, while on the third line is the first correction. In order to obtain the low-temperature behavior, we take $\beta\cJ \rightarrow \infty$ while keeping $\beta\mu_{\eta,\phi}$ fixed and expand $v(\beta\cJ)$ in (\ref{sol large pq const fixed}):
\be
v^2 = \pi^2 - \frac{4\pi^2}{\beta\cJ} + \frac{12\pi^2}{(\beta\cJ)^2} + \cO\bigl( (\beta\cJ)^{-3} \bigr) \,.
\ee
The specific entropy $\cS_0 \equiv S_{\beta=\infty}/N$ at zero-temperature can be expressed using the relations $\beta\mu_{\eta,\phi} = -2 \pi \cE_{\eta,\phi}$ between the chemical potentials and the spectral asymmetries of the IR conformal solution, which are valid at zero temperature:
\begin{align}
\cS_0 &= \log \bigl[ 2\cosh(-\pi\cE_\eta) \bigr] - \pi \cE_\eta \tanh(\pi \cE_\eta) \nn\\
&\quad - \alpha \log \bigl[ 2\sinh(-\pi\cE_\phi) \bigr] + \alpha \, \pi \cE_\phi \coth(\pi \cE_\phi) \nn\\
&\quad - \frac{\pi^2 }{ 8p^2 \cosh^2 \bigl( \frac{\mu_\eta \beta}2 \bigr)} + \cO(p^{-3}) \,.
\label{inf p entropy}
\end{align}
We shall see in Section~\ref{sec: I ext} that this quantity, when evaluated on supersymmetric spectral asymmetries satisfying $p\cE_\eta+q\cE_\phi=0$, matches the large $p$ expansion of the entropy extracted from the Witten index.

\subsection{Solutions at large \tps{\matht{p}}{p} and \tps{\matht{q}}{q}}
\label{sec: large p and q}

We can similarly search for analytic solutions to the Schwinger--Dyson equations at large $p$ and $q$, following the same steps as in Section~\ref{subsec: fixed q large p}. We consider an ansatz for the bilocal fields similar to the one in (\ref{G ansatz fp lq}), however we keep the ratio $p/q$ fixed as we send $p \to \infty$:
\begin{align}
G_{\bar\eta \eta}(\tau) &= G_{\bar\eta \eta}^{(0)}(\tau) \Bigl( 1 + p^{-1} \wt G_{\bar\eta \eta}(\tau) + \cO(p^{-2}) \Bigr) \,, \nn\\
\label{G ansatz large pq}
G_{\bar{f} f}(\tau) &= - \delta(\tau) + p^{-1} \wt G_{\bar{f} f}(\tau) + \ldots \,, \\
G_{\bar\phi \phi}(\tau) &= G_{\bar\phi \phi}^{(0)}(\tau) \Bigl( 1 + q^{-1} \wt G_{\bar\phi \phi}(\tau) + \cO \bigl( p^{-2} \bigr) \Bigr) \,, \nn\\
 G_{\bar\psi \psi}(\tau) &= \alpha \, \delta(\tau) + q^{-1} \wt G_{\bar\psi \psi}(\tau) + \ldots \,. \nn
\end{align}
In particular $\bigl[ G_{\bar\phi\phi} \bigr]^{q-n} = \bigl[ G_{\bar\phi\phi}^{(0)} \bigr]^{q-n} \, e^{\wt G_{\bar\phi\phi}} \, \bigl( 1 + \cO(p^{-1}) \bigr)$ besides (\ref{Getaeta to power p}). We insist that no chemical potential is larger than $\cO(1)$, therefore according to (\ref{chemical pots large p}) we shall choose the scaling $\mu_\eta \sim \mu_\phi \sim \mu_R \sim \cO(1)$ while $\mu_F \sim \cO(p^{-1})$, with the possibility of setting $\mu_R = 0$. In particular $\mu_\eta = - (q/p) \, \mu_\phi + \cO(p^{-1})$.

Expanding the Schwinger--Dyson equations determines $\wt G_{\bar{f} f}$ and $\wt G_{\bar\psi \psi}$ algebraically as
\begin{align}
\label{g aux ito g dyn}
\wt G_{\bar{f} f}(\tau) &= \frac{\cJ \, e^{(\mu_\eta - \mu_R)\tau} }{2 \cosh \bigl( \frac{\mu_\eta\beta}2 \bigr)} \, e^{\wt G_{\bar\eta \eta}(-\tau) + \wt G_{\bar\phi \phi}(-\tau)} \times {} \\
&\quad \times \Bigl[ \Theta(\tau) \, e^{-(\mu_\eta - \mu_R)\frac\beta{2}} + \Theta(-\tau) \, e^{(\mu_\eta - \mu_R)\frac\beta{2}} \Bigr] \,, \nn\\
\wt G_{\bar\psi \psi}(\tau) &= \frac{q^2 \, \cJ \sinh \bigl( \frac{\mu_\phi\beta}2 \bigr) \, e^{(\mu_\phi - \mu_R)\tau} }{ 2p^2 \cosh^2 \bigl( \frac{\mu_\eta\beta}2 \bigr)} e^{\wt G_{\bar\eta \eta}(-\tau) + \wt G_{\bar\phi \phi}(-\tau)} \times {} \nn\\
&\quad \times \Bigl[ \Theta(\tau) \, e^{ - (\mu_\phi - \mu_R) \frac\beta2 } - \Theta(-\tau) \, e^{ (\mu_\phi - \mu_R) \frac\beta2 } \Bigr] \,. \nn
\end{align}
We use the same definition of $\cJ$ as in (\ref{def cal J}) (with $n_f=1$), and we keep $\cJ$ fixed as $p \to \infty$.
The algebraic equations then also determine
\begin{align}
\label{aux in dyn alg large pq}
\Sigma_{\bar\eta \eta}(\tau) &= - \frac{\cJ^2}{\gamma\, p} \; G^{(0)}_{\bar\eta \eta}(-\tau) \times {} \\
&\quad \times e^{\wt G_{\bar\eta \eta}(\tau) + \wt G_{\bar\eta \eta}(-\tau) + \wt G_{\bar\phi \phi}(\tau) + \wt G_{\bar\phi \phi}(-\tau)} + \dots \nn\\
\Sigma_{\bar\phi \phi}(\tau) &= \frac{ \cJ^2 \,(1-\gamma) }{ \alpha\,\gamma^2\, q } \; G_{\bar\phi \phi}^{(0)}(-\tau) \times {} \nn\\
&\quad \times e^{\wt G_{\bar\eta \eta}(\tau) + \wt G_{\bar\eta \eta}(-\tau) + \wt G_{\bar\phi \phi}(\tau) + \wt G_{\bar\phi \phi}(-\tau)} + \dots \nn
\end{align}
where we defined
\be
\label{def gamma}
\frac{1}{\gamma} \,\equiv\, 1 + \frac{ q^2 \sinh^2 \bigl( \frac{\mu_\phi\beta}2 \bigr) }{\alpha p^2 \cosh^2 \bigl( \frac{\mu_\eta\beta}2 \bigr)} \;. 
\ee
The dynamical equations give two differential equations for $\wt G_{\bar\eta \eta}$, $\wt G_{\bar\phi \phi}$ that can be recast as:
\begin{align}
\label{large pq de G sum}
\partial_\tau^2 \, \wt G(\tau) &= \cJ^2 \gamma^{-2} \; e^{\wt G(\tau)+\wt G(-\tau)} \,, \\
\wt G_{\bar\eta \eta}(\tau) &= \gamma \, \wt G(\tau) \,,\qquad
\wt{G}_{\bar\phi \phi}(\tau) = (1-\gamma) \, \wt{G}(\tau) \,. \nn
\end{align}
Note that $\gamma\in(0,1)$. The solution for $\wt G$ is again \eqref{sol large pq const fixed}, with the substitution $\cJ \to \cJ \gamma^{-1}$. In the weak coupling limit $\beta\cJ \rightarrow 0$, the solution reduces to the free UV solution to first order in $1/p$. In the strong coupling limit, instead, we reproduce the conformal 2-point functions with the following parameters, at this order of approximation:
\begin{align}
\Delta_\eta &= \frac{\gamma}{2p} + \cO(p^{-2}), & \Delta_f &= \frac12 + \cO(p^{-1}), \\
\Delta_\phi &= \frac{1-\gamma}{2q} + \cO(p^{-2}), \!\! & \Delta_\psi &= \frac12 + \cO(p^{-1}), \nn\\
2\pi \cE_\eta &= -\beta \mu_\eta + \cO(p^{-2}), \!\! & 2\pi \cE_f &= \beta( \mu_R - \mu_\eta) + \cO(p^{-1}), \nn\\
2\pi\cE_\phi &= - \beta\mu_\phi+\cO(p^{-2}), \!\! & 2\pi \cE_\psi &= \beta ( \mu_R - \mu_\phi) + \cO(p^{-1}), \nn
\end{align}
as well as
\begin{align}
g_\eta &= \frac{ 1 }{ 2 \cosh \bigl( \frac{\mu_\eta \beta}2 \bigr) } \Bigl(1 + \tfrac{\gamma}{p}\log\gamma \Bigr) + \cO(p^{-2}) \,, \\
g_f &= \frac{\gamma}{ 2p \cosh \bigl( \frac{\mu_\eta \beta}2 \bigr) } + \cO(p^{-2}) \,, \nn \\
g_\phi &= \frac{\alpha }{ 2 \sinh \bigl( \frac{\mu_\phi \beta}2 \bigr)} \Bigl(1 + \tfrac{1-\gamma}{q} \log\gamma \Bigr) + \cO(p^{-2}) \,, \nn\\
g_\psi &= \frac{ \alpha \, (1-\gamma) }{ 2q \sinh \bigl( \frac{\mu_\phi \beta}2 \bigr)} + \cO(p^{-2}) \,. \nn
\end{align}
These values satisfy the consistency bounds \eqref{conf limit consistency}. 

As in the large $p$ fixed $q$ case, as long as we must include auxiliary fields in our considerations, there is no regime of large $p,q$ and small $\omega_k$ for which the large $p,q$ solution and the conformal solution are both reliable. Again, $\cI$-extremization gives us an independent derivation of \eqref{consistency match coeff} in the supersymmetric case.

\subsubsection{Grand potential and entropy at large \tps{$p$}{p} and \tps{$q$}{q}}
\label{logz large p and q}

We now compute the grand potential $\log Z$ to leading order in $1/N$ in a large $p$ expansion, at fixed $p/q$, by evaluating the averaged action on the solution of the large $p$ and $q$ Schwinger--Dyson equations. By differentiating $\log Z$ with respect to $J$ we get:
\begin{align}
& J \, \partial_J \, \frac{\log Z}N \\
&\;\; = J \!\int\! d\tau_1 \, d\tau_2 \, \Bigl( \tfrac1q \, G_{\bar{f} f} \, G_{\bar\eta \eta}^{p-1} \, G_{\bar\phi \phi}^q + \tfrac1p \, G_{\bar\eta \eta}^p \, G_{\bar\psi \psi} \, G_{\bar\phi \phi}^{q-1} \Bigr) \nn\\
&\;\; = \frac{ \cJ^2 \cos^2 \bigl( \frac{v}2 \bigr) }{ 4 p^2\gamma\cosh^2 \bigl( \frac{\mu_\eta \beta}2 \bigr) } \int_{-\beta}^{\beta} \! d\tau \, \frac{\beta-|\tau|}{ \cos^2 \bigl( \frac v\beta |\tau| -\frac{v}2 \bigr) } + \cO(p^{-3}) \nn\\
&\;\; = \frac{\gamma v \tan \bigl( \frac{v}2 \bigr) }{ 2 p^2 \cosh^2 \bigl( \frac{\mu_\eta \beta}2 \bigr) } + \cO(p^{-3}) \,, \nn
\end{align}
where $\gamma$ is defined in \eqref{def gamma}. Notice that, this time, both terms in the action contribute at leading order.
We follow the same steps as in Section~\ref{logz large p}, we exchange the derivative with respect to $J$ for a derivative with respect to $v$, we integrate from $0$ to $v$, and obtain
\begin{align}
\frac{\log Z}{N} &= \log \bigl[ 2 \cosh \bigl( \tfrac{\mu_\eta \beta}2 \bigr) \bigr] - \alpha \log \bigl[ 2 \sinh \bigl( \tfrac{\mu_\phi \beta}2 \bigr) \bigr] \nn\\
&\quad + \frac{ \gamma v \tan \bigl( \frac v2 \bigr) - \gamma\frac{v^2}4 }{ 2 p^2 \cosh^2 \bigl( \frac{\mu_\eta \beta}2 \bigr) } + \cO(p^{-3}) \,.
\label{large p and q logZ v}
\end{align}
In order to compute the entropy $S = (1-\beta \partial_\beta) \log Z$ we rewrite the $\beta$ dependence through $v$. Using the relation $\beta \partial_\beta = \bigl(1-\beta\gamma^{-1}\partial_\beta\gamma\bigl)\,v\, /\, \bigl[ 1 + \frac v2 \tan\bigl( \frac v2 \bigr) \bigr] \cdot \partial_v$, we get
\begin{align}
\frac{S}N &= \log \bigl[ 2\cosh \bigl( \tfrac{\mu_\eta \beta}2 \bigr) \bigr] - \tfrac{\mu_\eta \beta}2 \tanh\bigl( \tfrac{\mu_\eta \beta}2 \bigr) \\
& - \alpha \log \bigl[ 2 \sinh\bigl( \tfrac{\mu_\phi \beta}2 \bigr) \bigr] + \tfrac{\alpha \mu_\phi \beta}2 \coth\bigl( \tfrac{\mu_\phi \beta}2 \bigr) \nn\\
& - \frac{\gamma\,v^2}{8p^2 \cosh^2 \bigl( \frac{\mu_\eta \beta}2 \bigr)} \Bigl[ 1 + (1-\gamma) \, \mu_\phi \beta \coth \bigl( \tfrac{\mu_\phi \beta}2 \bigr) \nn\\
& + \gamma \, \mu_\eta \beta \tanh\bigl( \tfrac{\mu_\eta \beta}2 \bigr) \Bigr] + \frac{\gamma \, v \, \mu_\eta \beta \, \tan \bigl( \frac{v}2 \bigr) \tanh\bigl( \frac{\mu_\eta \beta}2 \bigr) }{ 2p^2 \cosh^2 \bigl( \frac{\mu_\eta \beta}2 \bigr)} \nn
\end{align}
up to $\cO(p^{-3})$. We now take $\beta\cJ \rightarrow \infty$ while keeping $\beta\mu_{I}$ fixed, in order to get the zero-temperature specific entropy. Expressing everything in terms of the spectral asymmetries of the IR conformal solution, and keeping only finite terms, we get
\begin{align}
\label{inf p and q entropy}
\cS_0 &= \log \bigl[ 2 \cosh(\pi \cE_\eta) \bigr] - \pi \cE_\eta \tanh(\pi \cE_\eta) \\
&\quad - \alpha \log \bigl[ 2 \sinh( - \pi \cE_\phi) \bigr]  + \alpha \, \pi \cE_\phi \coth(\pi \cE_\phi) \nn\\
&\quad - \frac{\gamma \, \pi^2 }{ 8p^2 \cosh^2 \bigl(\pi \cE_\eta \bigr)} \Bigl[1 + 2 (1-\gamma) \pi \cE_\phi \coth( \pi \cE_\phi ) \nn\\
&\quad + 2 \gamma \pi \cE_\eta \tanh( \pi \cE_\eta) \Bigr] + \cO(p^{-3}) \,. \nn
\end{align}
This result matches the large $p,q$ expansion of the entropy extracted from the Witten index in Section~\ref{sec: I ext} for supersymmetric chemical potentials.

\subsection{Luttinger--Ward relation}
\label{sec-LW}

The ``Luttinger--Ward'' (LW) relation \cite{Luttinger:1960ua} gives the total charge $\langle Q \rangle$ of a conformal solution in terms of a sum of contributions indexed by $A$, each associated with a field of abundance $N_A$, charge $Q_A$, statistics $s_A = \pm1$, dimension $\Delta_A$, and spectral asymmetry $\cE_A$:
\begin{align}
\label{LW master}
& \langle Q \rangle = \sum\nolimits_A \, N_A \, Q_A \, \fq_{s_A}(\Delta_A,\cE_A) + C \,,\quad C = \text{const.} \nn\\
& \fq_s(\Delta,\cE) = s \Biggl[ \bigl( \Delta - \tfrac12 \bigr) \frac{ \sinh(2\pi\cE) }{ \cosh(2\pi\cE) - \cos \bigl[ 2\pi \bigl( \Delta + \frac{1-s}4 \bigr) \bigr] } \nn\\
& \qquad\qquad\quad\;\;
+ \frac{1}{2\pi i}\log \frac{ \sin \bigl[ \pi \bigl( \Delta + \frac{1-s}4 + i\cE \bigr) \bigr] }{ \sin \bigl[ \pi \bigr( \Delta + \frac{1-s}4 - i\cE \bigr) \bigr] } \Biggr] .
\end{align}
The charge is only determined up to a constant $C$, because of ambiguities in matching the UV limit (see also Appendix~A of \cite{Heydeman:2022lse}). 

For our model, the charges of fields are in Table~\ref{tab: charges}. For the superconformal solutions, using (\ref{susy constr coeff})--(\ref{susy constr dim eps}), the charges $\langle Q_\eta \rangle$, $\langle Q_\phi \rangle$, $\langle Q_F \rangle$ are:
\begin{widetext}
\begin{align}
\!\! \frac{\langle Q_\eta \rangle}N &= \frac{q \, \Delta_\phi \sinh(2\pi\cE_\eta) }{ \cosh(2\pi\cE_\eta) {+} \cos(2\pi\Delta_\eta) } {+} \frac{\alpha \, p \, \Delta_\phi \, \sinh(2\pi \cE_\phi) }{ \cosh(2\pi\cE_\phi) {-} \cos(2\pi \Delta_\phi) } {-} \frac{p}{2\pi i} \log \frac{ \cos[ \pi(\Delta_\eta {+} i\cE_\eta)] }{ \cos[ \pi ( \Delta_\eta {-} i\cE_\eta)] } {+} \frac{\alpha \, p}{2\pi i} \log \frac{ \sin[\pi(\Delta_\phi {+} i\cE_\phi)] }{ \sin[\pi (\Delta_\phi {-} i\cE_\phi)] } {+} C_\eta \nn \\
\!\! \frac{\langle Q_\phi \rangle}N &= - \frac{ q \, \Delta_\eta \sinh(2\pi\cE_\eta) }{ \cosh(2\pi \cE_\eta) {+} \cos(2\pi \Delta_\eta) } {-} \frac{ \alpha \, p \, \Delta_\eta \sinh(2\pi \cE_\phi) }{ \cosh(2\pi \cE_\phi) {-} \cos(2\pi \Delta_\phi) } {-} \frac{q}{2\pi i} \log \frac{ \cos[ \pi ( \Delta_\eta {+} i\cE_\eta)] }{ \cos[ \pi( \Delta_\eta {-} i\cE_\eta)] } {+} \frac{\alpha \, q}{2\pi i} \log \frac{ \sin[ \pi( \Delta_\phi {+} i\cE_\phi)] }{ \sin( \pi( \Delta_\phi {-} i\cE_\phi)] } {+} C_\phi \nn \\
\!\! \frac{\langle Q_F \rangle}N &= \frac{q}{2} \, \frac{ \sinh(2\pi \cE_\eta) }{ \cosh(2\pi \cE_\eta) + \cos(2\pi \Delta_\eta) } + \frac{\alpha p}{2} \, \frac{ \sinh(2\pi \cE_\phi) }{ \cosh(2\pi \cE_\phi) - \cos(2\pi \Delta_\phi) } + q \, C_\eta - p \, C_\phi \;.
\label{non-R LW}
\end{align}
\end{widetext}
The constants $C_{\eta,\phi}$ will be determined by analyzing the Witten index in Section \ref{sec: I ext}.

\subsection{Non-conformal solutions}
\label{sec: non conformal}

In addition to the conformal solutions we discussed so far, which can describe the system at low energies, we also find non-conformal solutions. This is not unusual: they have previously been identified in SYK-like models, \eg{} in \cite{Davison:2016ngz, Heydeman:2022lse}. These solutions can be concurrent with conformal solutions at a given fixed value of $\mu_{\eta,\phi}$, although they differ in their charge. However, unlike the conformal solutions, they are exact solutions to the full Schwinger--Dyson equations at $\beta = \infty$.

We identify two families of solutions that are compatible with the requirement $\mu_\phi > 0$
\footnote{We further require $G_{\bar\phi \phi}>0$, since the sign of the Euclidean $G_{\bar\phi \phi}$ is the sign of the Lorentzian spectral density which must be non-negative, as we discuss in Appendix~\ref{app: IR and UV limits}. This rules out other exponential solutions of the Schwinger--Dyson equations.}.
Furthermore, we require that these solutions do not diverge for $\tau \to \pm\infty$.
Such solutions only exist when the chemical potentials $\mu_{\eta, \phi}$ are in certain domains.
When both chemical potentials are above some critical value, \ie, when $\mu_\eta > \mu_\eta^\text{c}$ and $\mu_\phi > 0$, we find:
\begin{align}
\label{eq:non-conformal-positive-eta}
G_{\bar\eta \eta} &= - \Theta (-\tau) \, e^{ ( \mu_\eta - \mu^\text{c}_\eta ) \tau} , \\
G_{\bar{f}f} &= - \delta(\tau) + a_f \, \Theta(\tau) \, e^{ - \left[ (p-1) (\mu_\eta - \mu^\text{c}_\eta ) + q ( \mu_\phi - \mu^\text{c}_\phi ) + a_f \right] \tau } \nn\\
G_{\bar\phi \phi} &= \alpha \, \Theta (-\tau) \, e^{ ( \mu_\phi - \mu^\text{c}_\phi ) \tau } , \nn\\
G_{\bar\psi \psi} &= \alpha \, \delta (\tau) {+} \alpha \, a_\psi \Theta(\tau) \, e^{ - \left[ p (\mu_\eta - \mu^\text{c}_\eta ) + (q-1) ( \mu_\phi - \mu^\text{c}_\phi ) - a_\psi \right] \tau } \nn
\end{align}
where the constants take the following values
\footnote{We used the rule $\Theta(\tau)^m \, \Theta(-\tau)^n \, \delta(\tau) = 2^{-m-n} \, \delta(\tau)$, namely $\Theta(0) = \frac12$, which follows from using the function $\text{P} \frac1{i\omega} + \pi\, \delta(\omega)$ as the Fourier transform of $\Theta(\tau)$, and will be consistent with our numerical results.}:
\bea
\mu_\eta^\text{c} &= \frac{(p+q-1) \, \alpha^q J}{ 2^{p+q-2} \, q} \,,\quad&  a_f &= \frac{J \alpha^q}q \,, \\
\mu_\phi^\text{c} &= - \frac{ (p+q-1) \, \alpha^{q-1} J }{ 2^{p+q-2} \, p} \,,\quad& a_\psi &= \frac{J \alpha^{q-1}}p \,.
\eea
Note that the chemical potentials must also satisfy $p \, (\mu_\eta - \mu^\text{c}_\eta ) + (q-1) ( \mu_\phi - \mu^\text{c}_\phi ) > a_\psi$.
On the other hand, when $\mu_\eta$ is below a critical value but $\mu_\phi$ is above, we find:
\begin{align}
\label{eq:non-conformal-negative-eta}
G_{\bar\eta \eta} &= \Theta(\tau) \, e^{ -(\mu^\text{c}_\eta - \mu_\eta)\tau} \,,\\
G_{\bar{f}f} &= -\delta(\tau) + \delta_{p,1} \, \Theta(\tau ) \, a_f \, e^{-\left[q (\mu_\phi - \mu^\text{c}_\phi) + a_f \right] \tau} \nn\\
G_{\bar\phi \phi} &= \Theta(-\tau) \, \alpha \, e^{( \mu_\phi - \mu^\text{c}_\phi) \tau } \,,\qquad
G_{\bar\psi \psi} = \alpha \, \delta(\tau) \,, \nn
\end{align}
with
\begin{align}
\mu_\eta^\text{c} &=  \frac{(q-p+1) \, \alpha^q J}{2^{p+q-2}q} \,,\qquad a_f = \frac{\alpha^q J}{q} \,, \nn\\
\mu_\phi^\text{c} &= \frac{(q-p-1) \, \alpha^{q-1} J }{2^{p+q-2}p} \,.
\end{align}
In particular, for $p=1$ and $q=2$ the critical chemical potentials are $\mu_\eta^\text{c} = \frac12 J \alpha ^2$ and $\mu_\phi^\text{c} = 0$. In this solution, the boson behaves like a free boson while the fermion seems to acquire an effective mass.
This makes this solution similar to the non-conformal solutions found in the two-fermion model of \cite{Heydeman:2022lse}. 
In the particular case where the chemical potentials do not break supersymmetry, \ie{} when $\mu_\eta = -2 \mu_\phi$, the solution \eqref{eq:non-conformal-positive-eta} cannot be realized. However, the solution \eqref{eq:non-conformal-negative-eta} is consistent for any $\mu_\phi > 0$.

We can also explicitly calculate the entropy of these solutions. We take the off-shell action at finite $\beta$ and apply $\bigl( 1 - \beta J \partial_{\beta J} - \beta \mu \partial_{\beta \mu} \bigr)$. After simplifying with the equations of motion, we find
\begin{align}
\label{eq:entropy-explicit}
& \frac SN =  \log \Bigl[ 2 \cosh \bigl( \tfrac{\beta\mu_\eta }2 \bigr) \Bigr] - \alpha  \log \Bigl[ 2 \sinh \bigl( \tfrac{\beta\mu_\phi }2 \bigr) \Bigr] \\
&\;\; + \sum_k \Biggl[ \log \Biggl( \frac{1 + \frac{ \Sigma_{\bar\eta \eta}(\omega_k) }{ i\omega_k - \mu_\eta} }{ 1 + \Sigma_{\bar{f} f}(\omega_k)} \Biggr) - \alpha \log \Biggl( \frac{1 + \frac{\Sigma_{\bar\phi \phi}(\omega_k) }{ i\omega_k - \mu_\phi} }{ 1 + \Sigma_{\bar\psi \psi}(\omega_k)} \Biggr) \Biggr] \nn \\
&\;\; + \beta \mu_\eta \, Q_\eta + \beta \mu_\phi \, Q_\phi - \beta \int_0^\beta \! d\tau \Bigl[\Sigma_{\bar\eta \eta}(\tau) \, G_{\bar\eta \eta}(\tau) \nn\\
&\;\; + \Sigma_{\bar{f} f}(\tau) \, G_{\bar{f} f}(\tau) + \Sigma_{\bar\phi \phi}(\tau) \, G_{\bar\phi \phi}(\tau) + \Sigma_{\bar\psi \psi}(\tau) \, G_{\bar\psi \psi}(\tau) \Bigr] \nn
\end{align}
where $\omega_k$ are either the fermionic or bosonic Matsubara frequencies, depending on the field.
The first two terms are the free action, which appears from regulating the logarithmic term.
Promoting the sum to an integral, we can plug \eqref{eq:non-conformal-positive-eta} or \eqref{eq:non-conformal-negative-eta} in, so that the sum cancels with the integral. This leaves the entropy to be solely the ``free part'':
\begin{align}
\frac SN &= \lim_{\beta \,\rightarrow\, \infty} \Bigl\{ \log \Bigl[ 2 \cosh \bigl( \tfrac{\beta \mu_\eta}2 \bigr) \Bigr] - \alpha \log \Bigl[ 2 \sinh \bigl( \tfrac{\beta \mu_\phi}2 \bigr) \Bigr] \nn\\
&\quad + \tfrac12 \bigl( \beta \mu_\eta + \alpha \, \beta \mu_\phi \bigr) \Bigr\} = 0 \,,
\end{align}
where we assumed that $\mu$ does not scale with $\beta$. We obtain this result for both families, irrespective of $p,q$.

\subsection{Witten index and \tps{\matht{\cI}}{I}-extremization}
\label{sec: I ext}

In supersymmetric quantum mechanics, a protected quantity is the Witten index $\cI^\text{W} = \Tr_\cH \, (-1)^F e^{-\beta H}$, which does not depend on $\beta$ \cite{Witten:1982df}. In the presence of a flavor symmetry $\rU(1)_F$, the index can be refined by inserting a complex fugacity $x$ for the flavor charge $Q_F$. Besides, with $\cN=2$ supersymmetry and in the presence of a $\rU(1)$ R-symmetry $R$, one can construct an alternative index in which $e^{i\pi R}$ is used in place of $(-1)^F$:
\be
\label{R charge index}
\cI(x) = \Tr_\cH \, e^{i\pi R} \, e^{-\beta H} \, x^{Q_F} \,.
\ee
Let us explain the relation between the two indices, for the models considered in this paper.

When $\gcd(p,q)=1$ and there is no discrete $\bZ_{\gcd(p,q)}$ flavor symmetry, there exists an assignment of R-charges such that $(-1)^F = e^{i\pi R}$ and hence \eqref{R charge index} is identical to the ordinary Witten index. Such an assignment is such that
\be
\label{F number R-charge assignment}
R[\eta] = 2m+1 \,,\;\; R[\phi] = 2n \,,\;\; p\, R[\eta] + q \, R[\phi] = 1
\ee
for some $m,n \in \bZ$ (and recall that $p$ is odd). These equations have a solution if and only if $\gcd(p,q) = 1$. Any other R-charge assignment is then related to the one in (\ref{F number R-charge assignment}) by mixing with the flavor symmetry, namely, by rotating the phase of $x$.

When $d \equiv \gcd(p,q) > 1$, although the ordinary Witten index is not equal to \eqref{R charge index}, one can consider Witten indices refined by an additional twist $g$ for the $\bZ_d$ flavor symmetry, where $g$ was defined in (\ref{def g Z_gcd}) and $g^d$ is equivalent to a $\rU(1)_F$ rotation:
\be
\cI^\text{W}_r(x) = \Tr_\cH \, (-1)^F \, g^r \, e^{-\beta H} \, x^{Q_F} \quad\text{for}\quad r \in \bZ_d \,.
\ee
The R-charge assignment such that $(-1)^F g^r = e^{i\pi R}$ and thus $\cI(x) = \cI^\text{W}_r(x)$ should satisfy $R[\eta] = 1 + \frac{2r}p + 2m$, $R[\phi] = 2n$ and $p \, R[\eta] + q\, R[\phi] = 1$. There is always one and only one solution for $r$ in the range $0 \leq r < d$ which is $r = \frac{1+d}2 \text{ mod } d$, with $r \neq 0$ whenever $d>1$. Thus \eqref{R charge index} is equivalent to one of the refined Witten indices.

Writing the $\rU(1)_F$ fugacity as $x=e^{2\pi iu}=e^{2\pi i y+2\pi\cE}$ with $y,\cE \in \bR$, the insertions in \eqref{R charge index} can be recast in the following way:
\be
\label{index grouped phases}
\cI=\Tr_\cH \, e^{i\pi(R \,+\, 2yQ_F)} \, e^{-\beta (H \,+\, \mu Q_F)}
\ee
where $2\pi \cE = - \mu \beta$. Here we used the symbol $\mu$ in place of $\mu_F$ in order to avoid cluttering, and later on we will identify $\cE$ with the same quantity introduced in \eqref{cE solution}.
The $\cI$-extremization principle introduced in \cite{Benini:2015eyy, Benini:2016rke} states that the values $\hat{y}, \hat\mu$ that extremize $\cI$ under Laplace transform select the infrared Hamiltonian $H_\text{IR} = H + \hat\mu \, Q_F$ and the infrared superconformal R-charge
\be
\label{Rsc ito R}
R_\text{sc} = R + 2 \hat{y} \, Q_F \,.
\ee
Using the relation $\bigl\{ \cG_{\frac12}, \wb\cG_{-\frac12} \bigr\} = L_0 - R_\text{sc}/2$ in the superconformal algebra  --- see \eqref{superconf algebra} --- we observe that chiral primaries annihilated by $\cG_{\frac12} = Q$ and $\wb\cG_{-\frac12} = Q^\dag$ must have $\Delta = \frac12 R_\text{sc}$. Applying this to $\phi$ and $\eta$ we get
\be
\label{y dim rel}
\Delta_\phi = - p \, \hat{y} + \tfrac12 \, R[\phi] \,,\qquad \Delta_\eta = q \, \hat{y} + \tfrac12 \, R[\eta] \,,
\ee
which allow us to trade $\hat{y}$ for the conformal dimensions $\Delta_{\phi,\eta}$. The fact that $R$ is an R-charge guarantees that the dimensions satisfy the supersymmetry constraint \eqref{susy constr dim eps}. Using \eqref{y dim rel} and Table~\ref{tab: charges}, $R_\text{sc}$ can also be written as
\be
\label{Rsc ito Qeta Qphi}
R_\text{sc} = 2\Delta_\eta Q_\eta + 2\Delta_\phi Q_\phi \,.
\ee

The index has an ambiguity given by overall multiplication by a power of $x$. In the Hamiltonian formalism this corresponds to the ambiguity in the assignment of charges to the Fock vacuum (\ie, the normal ordering ambiguity in the definition of charge operators). In the path-integral formalism it corresponds to the ambiguity in the regularization of 1\nobreakdash-loop determinants
\footnote{This is similar to the parity anomaly in 3d theories. For a complex fermion of charge 1, the fermionic Fock space has two states, and if we insist on assigning integer charges then there is no canonical choice and one is forced to break charge conjugation. On the other hand, one could assign charges $\pm\frac12$ to the two states in a charge-conjugation invariant fashion, but then the charges are not integer.}.
We fix the ambiguity by demanding that the charge operators be written as (anti-)commutators in the Hamiltonian formalism. Each chiral multiplet contributes $\bigl( e^{-\frac{i\pi R[\phi]}2} x^{\frac{p}{2}} - e^{ \frac{i\pi R[\phi]}2} x^{-\frac{p}{2}} \bigr){}^{-1}$ while each Fermi multiplet $e^{- \frac{i\pi R[\eta]}2} x^{-\frac{q}{2}} + e^{\frac{i\pi R[\eta]}2} x^{\frac{q}{2}}$. The result is      
\be
\label{index for p,q model}
\cI(x) = \Biggl[ \frac{ e^{ - \frac{i\pi R[\eta]}2} x^{-\frac q2} + e^{ \frac{i\pi R[\eta]}2} x^{ \frac q2} }{ \bigl( e^{-\frac{i\pi R[\phi]}2} x^{\frac p2} - e^{ \frac{ i \pi R[\phi]}2} x^{- \frac p2} \bigr)^\alpha } \Biggr]^N \,.
\ee
In order to extract the degeneracy of BPS states $d(Q_F,R)$ at fixed charges, that is captured by the index written as
\be
\cI(x) = \sum_{Q_F, R} d(Q_F, R) \; e^{i\pi R} \, x^{Q_F} \,,
\ee
it is useful to rewrite it as a constrained partition function for BPS states. Let $x_\eta=e^{2\pi iu_\eta}$ and $x_\phi=e^{2\pi iu_\phi}$ be the two fugacities for $Q_\eta$ and $Q_\phi$, respectively, and consider
\begin{align}
\Tr_\text{BPS} x_\eta^{Q_\eta} \, x_\phi^{Q_\phi} &= \Tr_\text{BPS} \exp\Bigl\{ 2\pi i \Bigl[ \bigl( p\, u_\eta + q\, u_\phi \bigr) R \nn\\
&\quad + \bigl( R[\phi] \, u_\eta - R[\eta] \, u_\phi \bigr) Q_F \Bigr] \Bigr\}
\end{align}
where we used \eqref{Qeta Qphi ito QF R}. When the chemical potentials are constrained to $p \, u_\eta + q \, u_\phi = \frac12$, and besides we identify $u \equiv R[\phi]\, u_\eta - R[\eta]\, u_\phi$ according to \eqref{def muF muR}, that quantity reduces to the index $\cI(x)$. Therefore, the Laplace transform of the index computes:
\begin{align}
\label{laplace transf}
& \oint\! \frac{dx}{2\pi i x} \: \cI(x) \,x_\eta^{-Q_\eta} \, x_\phi^{-Q_\phi} \Big|_{pu_\eta+qu_\phi=\frac12} \\
&\quad = \int_0^1 \! du \, \exp\Bigl[ \log\cI(x) - i \pi R - 2\pi i u Q_F \Bigr] \nn \\
&\quad \equiv \int_0^1 \! du \; e^{N\, \cS(u; \, Q_F, R)} = \sum\nolimits_{\tilde R} d(Q_F, \tilde R) \, e^{i\pi (\tilde R - R)} \,. \nn
\end{align}
Here the function $\cS(u; Q_F,R)$, sometimes called the \emph{entropy function}, is $1/N$ times the quantity in brackets on the second line. At large $N$ and $Q_F$, one can compute the integral in the saddle-point approximation extremizing $\cS$ with respect to $u$. On the right-hand-side of \eqref{laplace transf} appears a weighted sum over R-charges of the degeneracy of states within a sector of fixed $Q_F$. Assuming that this sum is dominated by one value, and comparing with the saddle point computation, one observes that the R-charge $R$ of each saddle is fixed by requiring that $\cS$ is real. Then the value of $N\cS$ is the zero-temperature entropy, $\log d(Q_F,R)$.

The saddle point in terms of $u$ is given by
\be
\alpha p \cot \bigl( p \pi u - \tfrac\pi2 R[\phi] \bigr) + q \tan \bigl( q \pi u + \tfrac\pi2 R[\eta] \bigr) + 2i \cQ_F = 0
\ee
where we introduced $\cQ_F = Q_F/N = \cO(1)$. Substituting \eqref{y dim rel} and \eqref{susy constr dim eps}, and using $y=u-i\cE$, the real and imaginary parts of this equation can be written in terms of $\Delta_\phi$ and $\cE$. The real part turns out to be precisely the relation (\ref{consistency match coeff}) between the conformal dimension $\Delta_\phi$ and the spectral asymmetry $\cE$ in superconformal solutions. The imaginary part reads:
\begin{multline}
\label{LW relation}
2 \cQ_F + \frac{ \alpha \, p \sinh(2\pi p\cE) }{ \cosh(2\pi p\cE) - \cos(2\pi \Delta_\phi) } \\
- \frac{ q \sinh(2\pi q\cE) }{ \cosh(2\pi q\cE) + \cos\bigl( \frac{\pi}{p} 
- \frac{2\pi q }p \Delta_\phi \bigr)}  = 0 \,.
\end{multline}
This coincides with the LW relation that determines $Q_F$ in \eqref{non-R LW}, if the constants satisfy $q \, C_\eta - p \, C_\phi=0$. Setting $\im\cS=0$ determines the R-charge to be
\begin{align}
& \frac{R}{N} = \frac{1}{2\pi i} \log \frac{\cos(\pi\Delta_\eta - i \pi \cE_\eta) }{ \cos(\pi\Delta_\eta + i \pi \cE_\eta)} \\
& - \frac{\alpha}{2\pi i} \log \frac{ \sin(\pi\Delta_\phi - i \pi \cE_\phi) }{ \sin(\pi\Delta_\phi + i \pi \cE_\phi)} 
+ \frac{\alpha}2 - 2 \biggl(\! \frac{R[\phi]}{2p} - \frac{\Delta_\phi}{p} \!\biggr)\cQ_F \nn
\end{align}
where the principal value is taken in the logarithms. Using \eqref{Rsc ito R} and \eqref{y dim rel}, we can determine the superconformal R-charge of the BPS states:
\begin{align}
\label{Rsc conf param}
\frac{R_\text{sc}}{N} &= -\frac{1}{2\pi i} \log \biggl[ \frac{ \cos(\pi\Delta_\eta + i \pi \cE_\eta) }{ \cos(\pi\Delta_\eta - i \pi \cE_\eta)} \biggr] \\
&\quad + \frac{\alpha}{2\pi i} \log \biggl[ \frac{ \sin(\pi\Delta_\phi + i\pi \cE_\phi) }{ \sin(\pi\Delta_\phi - i \pi \cE_\phi)} \biggr] + \frac{\alpha}{2} \,. \nn
\end{align}
On the other hand, substituting the LW relations for $Q_\eta$ and $Q_\phi$ from \eqref{non-R LW} into \eqref{Rsc ito Qeta Qphi} gives another determination of $R_\text{sc}$, which agrees with \eqref{Rsc conf param} if $ 2\Delta_\eta C_\eta+2\Delta_\phi C_\phi=\frac{\alpha}{2}$. Solving this and the previous constraint determines the constants in the Luttinger--Ward relations to be
\be
\label{constants-index}
C_\eta = \frac{\alpha \, p}{2} \;,\qquad\qquad C_\phi = \frac{\alpha \, q}{2} \;.
\ee
We shall see that this is consistent with the numerical result in Fig.~\ref{fig-luttinger-ward}. In particular, we find out that $R_\text{sc}$ takes a common non-zero value among the supersymmetric ground states
\footnote{The fact that the ground states have a non-vanishing but well-defined R-charge (meaning that they are still eigenvectors of the R-charge operator) means that the R-symmetry is unbroken in those states.}.
Finally, the real part of the entropy function that gives the entropy is
\begin{align}
\label{re and im S}
\re \cS &= \frac12 \log \Biggl[ 2^{1-\alpha} \, \frac{ \cosh(2\pi q\cE) + \cos \bigl( \frac{\pi}p - \frac{2\pi q}p \Delta_\phi \bigr) }{ \bigl[ \cosh(2\pi p\cE)-\cos(2\pi\Delta_\phi) \bigr]^\alpha } \Biggr] \nn\\
&\quad - 2\pi\cE\cQ_F \,.
\end{align}

\subsubsection{Entropy of various solutions}

The index $\cI(x)$ in (\ref{index for p,q model}) is the grand canonical partition function of the theory, while $d(Q_F, R)$ in (\ref{laplace transf}) is the microcanonical degeneracy of states. It follows that after substituting $\cE(\cQ_F)$, the quantity $\re\cS$ in (\ref{re and im S}) becomes the zero-temperature microcanonical entropy $S_0(\cQ_F)$. We compute it in various cases, within the regime of validity of the IR conformal ansatz.

\begin{figure}
\includegraphics[width = 0.95\columnwidth]{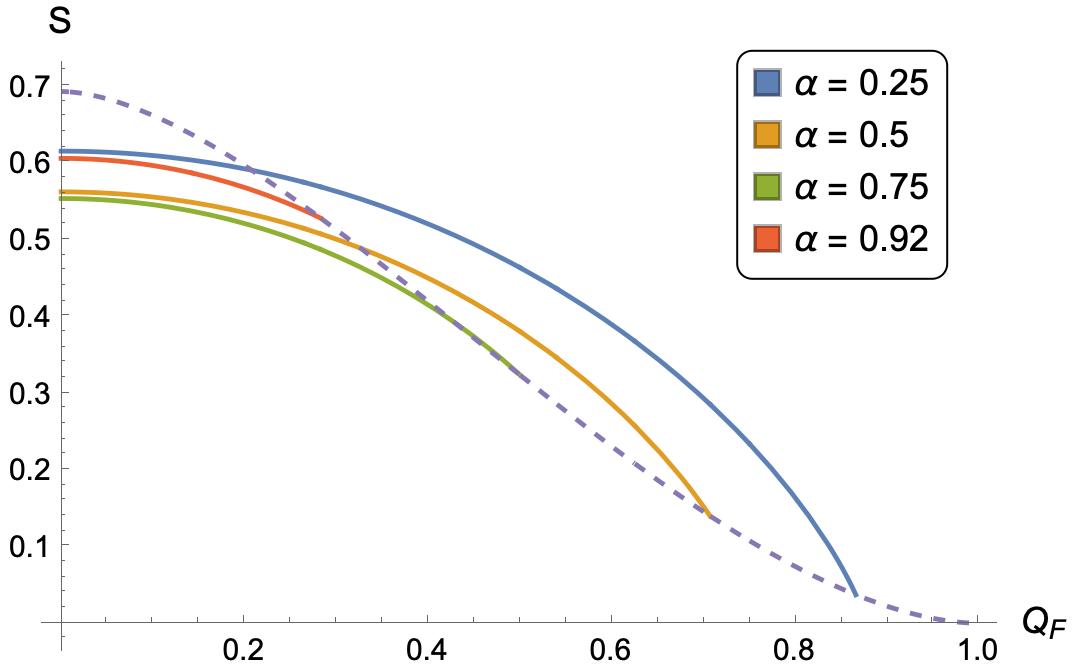}
\caption{\label{fig: S(QF) p1 q2}%
Plot of the entropy $S_0(\cQ_F)$ in the range $0 \leq \cQ_F < \sqrt{1-\alpha}$ for different values of $\alpha$, in the case $p=1$, $q=2$. The dashed line is the value of the entropy at the upper critical value of $\cQ_F$ as $\alpha$ is varied.}
\end{figure}

\paragraph{Solutions for \tps{$p=1$}{p=1}, \tps{$q=2$}{q=2}.}
Plugging the solution \eqref{p1 q2 sol} for $\Delta_\phi(\cE)$ into the LW relation \eqref{LW relation} we determine the spectral asymmetry $\cE$ and $\Delta_\phi$ as functions of $\cQ_F$:
\bea
\label{p1 q2 eps sol}
\cE &= \frac{1}{2\pi} \operatorname{arctanh} \biggl( \frac{2 \cQ_F}{2-\alpha} \biggr) \,,\\
\Delta_\phi &= \frac1{2\pi} \arccos \biggl( \frac{\alpha}{ \sqrt{(2-\alpha)^2 - 4 \cQ_F^2} } \biggr) \,.
\eea
Due to the bound $0 \leq \tanh\bigl( \pi \cE \bigr) < \sqrt{1-\alpha}$ in \eqref{p1 q2 sol} for conformal solutions, the charge is bound by $0 \leq \cQ_F < \sqrt{1-\alpha}$ at fixed $\alpha<1$.
Evaluating the entropy function \eqref{re and im S} on the solution, one obtains the entropy $S_0 = \re \cS$:
\begin{align}
\label{re S on p1 q2 sol}
& S_0 = \\
& \frac12 \log \Biggl[ \frac{ 4^{2-\alpha} \, (1-\alpha)^{1-\alpha} }{ \rule{0pt}{1em} \bigl( 2-\alpha + 2\cQ_F \bigr){}^{1-\frac\alpha2 + \cQ_F} \bigl( 2-\alpha - 2 \cQ_F \bigr){}^{1-\frac\alpha2 - \cQ_F} } \Biggr] . \nn
\end{align}
The quantity $e^{NS_0(\cQ_F)}$ is the degeneracy of ground states at fixed charge $\cQ_F$. In Fig.~\ref{fig: S(QF) p1 q2} we plot the entropy $S_0(\cQ_F)$ for various values of $\alpha$ and we observe that it is always positive, with a maximum at $\cQ_F=0$. The entropy does not vanish as $\cQ_F$ reaches the critical value $\sqrt{1-\alpha}$, and the dashed line is the envelop of the entropy at those values as $\alpha$ is varied.

\begin{figure}
\includegraphics[width=0.90\columnwidth]{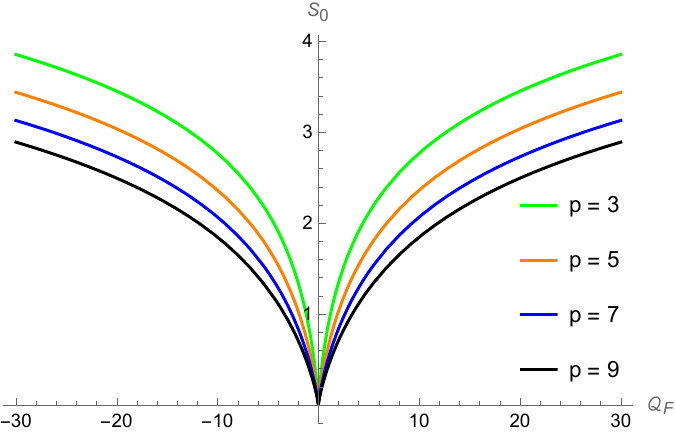}
\caption{\label{peq reS plot}%
Plot of $S_0(\cQ_F)$ against $\cQ_F$ for the solution in the cases $p=q=3,5,7,9$ and $\cE\neq 0$.}
\end{figure}

\paragraph{Solutions for \tps{$p=q>1$}{p=q>1}, \tps{$\alpha=1$}{a=1} and \tps{$\cE \neq 0$}{E>=0}.}
Similarly, plugging the explicit solution (\ref{p eq q Delta sol}) for $\Delta_\phi(\cE)$ into the LW relation \eqref{LW relation} we get
\bea
\label{peq eps sol}
\cE &= -\frac{1}{2\pi p} \operatorname{arcsinh} \biggl( \frac{ B }{ \sqrt2\, \cQ_F} \biggr) \,, \\
\Delta_\phi &= \frac1{4p} - \frac1{2\pi} \arcsin \biggl[ \frac{\sqrt{2}\, |\cQ_F|}C \sin\Bigl( \frac{\pi}{2p} \Bigr) \biggr] \,,
\eea
where we defined the following functions of $\cQ_F$:
\begin{align}
A &= \cos\bigl( \tfrac\pi{2p} \bigr) \sqrt{ \bigl( p^2 - \cQ_F^2 \bigr)\rule{0pt}{0.7em}^2 \cos^2 \bigl( \tfrac{\pi}{2p} \bigr) + 4p^2\cQ_F^2} \;, \nn \\
B &= \sqrt{ \bigl( p^2 - \cQ_F^2 \bigr) \cos^2 \bigl( \tfrac{\pi}{2p} \bigr) + A} \;, \\
C &= \sqrt{ p^2 + \cQ_F^2 - \bigl( p^2 - \cQ_F^2 \bigr) \sin^2 \bigl( \tfrac{\pi}{2p} \bigr) + A} \;. \nn
\end{align}
Note that $C^2 - B^2 = 2\cQ_F^2$. Evaluating the entropy function \eqref{re and im S} on the solution we get
\be
\label{re S on peq sol}
S_0 = \log \Biggl[ \frac{ \cos\bigl( \frac{\pi}{2p} \bigr) \, \bigl( p C + \lvert \cQ_F \rvert B \bigr) \, \bigl( C + \sgn(\cQ_F) B)^\frac{\cQ_F}{p} }{ 2^{\frac{\cQ_F}{2p}-\frac{1}{2}} \, |\cQ_F|^\frac{\cQ_F}{p} \, B^2} \Biggr] .
\ee
Fig.~\ref{peq reS plot} shows $S_0(\cQ_F)$ against $\cQ_F$ for various values of $p=q$, and it is always positive.

\begin{figure}
\includegraphics[width=0.90\columnwidth]{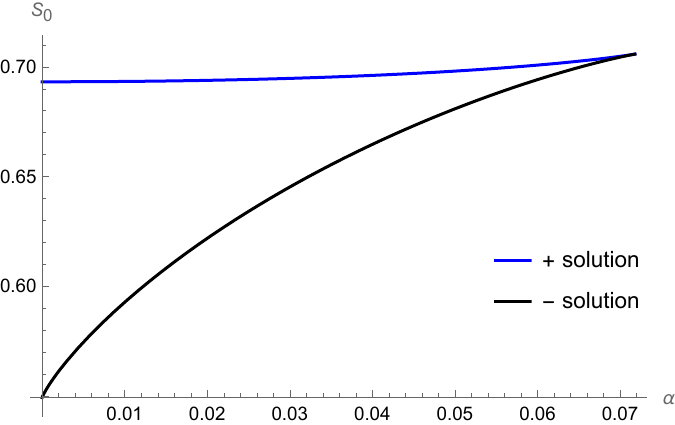}
\caption{\label{peq e0 plot}%
Plot of $S_0$ against $\alpha\in \bigl( 0, \tan^2\frac{\pi}{4p} \bigr)$ for the solutions in the case $p=q=3$, $\cE=0$. The blue/black curves correspond to the solutions with a plus/minus sign in \eqref{p eq q eps zero sol}.}
\end{figure}

\paragraph{Solutions for \tps{$p=q>1$}{p=q>1} and \tps{$\cE=0$}{E=0}.}
In this case, whenever $\alpha <\tan^2 \bigl( \frac\pi{4p} \bigr)$ there are two acceptable conformal solutions \eqref{p eq q eps zero sol} which coincide at $\alpha =  \tan^2 \bigl( \frac\pi{4p} \bigr)$. The only solution to the LW relation (\ref{LW relation}) is $\cQ_F=0$. Evaluating \eqref{re and im S} on \eqref{p eq q eps zero sol} and with $\cQ_F=0$ gives
\begin{widetext}
\be
\label{reS on e0 sol}
\! S_0 = \log \left[ \frac{ 2 \, \Bigl( (1-\alpha) \tan \bigl( \frac{\pi}{2p} \bigr) \pm \sqrt{ (1-\alpha)^2 \tan^2 \bigl( \frac{\pi}{2p} \bigr) - 4\alpha} \, \Bigr)^\frac{1-\alpha}2 }{ (1-\alpha)^\frac{1-\alpha}2 \cos^\alpha \bigl( \frac{\pi}{2p} \bigr) \, \Bigl( (1+\alpha) \tan \bigl( \frac{\pi}{2p} \bigr) \pm \sqrt{ (1-\alpha)^2 \tan^2 \bigl( \frac{\pi}{2p} \bigr) -4\alpha} \, \Bigr)^\frac{1+\alpha}2 } \right]\! . \!
\ee
\end{widetext}
The signs above are correlated with the sign in \eqref{p eq q eps zero sol}. Fig.~\ref{peq e0 plot} shows $S_0$ against $\alpha$ for $p=3$, and it is always positive. In addition, we see that the solution with the plus sign always has a higher entropy, and is expected to be the dominant saddle at large $N$. 

\paragraph{Solution at large \tps{$p$}{p}.}
One can check that the conformal parameters given in \eqref{conf param ito uv param} solve the Schwinger--Dyson equation \eqref{consistency match coeff} up to $\cO(p^{-3})$. Substituting the large $p$ values of $\Delta_{\eta}$ and $\Delta_{\phi}$ into \eqref{re and im S}, using \eqref{non-R LW} and 
\begin{align}
\!\!\!\! \cosh(2\pi\cE_\eta) + \cos(2\pi\Delta_\eta) &= 2\cosh^2(\pi q\cE) - \tfrac{\pi^2}{2p^2}+\cO \bigl( \tfrac{1}{p^4} \bigr) , \nn\\
\!\!\!\! \cosh(2\pi\cE_\phi) - \cos(2\pi\Delta_\phi) &= 2\sinh^2(\pi p\cE) + \cO \bigl( \tfrac{1}{p^4} \bigr) , \!\!\!
\end{align}
reproduces the expansion in \eqref{inf p entropy}. This is a consistency check for the large $p$ computation.

\section{Numerical results}
\label{sec: numerics}

In this section we present some results we obtained by numerically solving the Schwinger--Dyson equations. Our first goal  is to check that the analytic approximations we carried out are sensible. Indeed we verify that the numerical solutions exhibit the conformal behavior of equations \eqref{conf 2pt finite beta} in the appropriate parameter region, and the non-conformal behavior of \eqref{eq:non-conformal-negative-eta} outside that region. Second, we extract additional information which is analytically unreachable: we check that the zero-temperature limit of the entropy matches the prediction from the supersymmetric index and we estimate the Schwarzian coupling.

\subsection{Summary of the numerical method}

The SD equations can be solved numerically by adapting the method used in \cite{Maldacena:2016hyu}. We discretize the interval $[0,\beta]$ into $M$ points, with $M$ a power of $2$
\footnote{One can sample the interval at either $\tau_i = i/M$ or $\tau_i = \bigl( i + \frac12 \bigr)/M$, for $i=0,\dots,M-1$. For sufficiently high number of points, they give identical solutions within the numerical precision, however we found the latter choice to be numerically more stable. Note that since the model depends on $\beta$ only through $\beta J$ and $\beta \mu$, one can always take $\beta=1$ for numerical purposes.}.
In frequency space we pick the $M$ fermionic and bosonic Matsubara frequencies with lowest absolute value, namely
\be
\omega^{\rm f}_k = \tfrac{2\pi}{\beta} \bigl( k - \tfrac{M}{2} + \tfrac{1}{2} \bigr) \;,\quad
 \omega^{\rm b}_k = \tfrac{2\pi}{\beta} \bigl( k - \tfrac{M}{2} + 1 \bigr) \;, 
\ee
where $k=0 , \dots, M-1$ and the discretization is manifest because we take a finite number of frequencies
\footnote{Since $M$ is even, the truncation is not symmetric around zero for the bosonic Matsubara frequencies:  $\omega^{\rm b}_k = - \frac{\pi (M-2)}{\beta},\dots, 0, \dots, \frac{\pi M}{\beta}$. This creates an apparent difficulty when calculating $\Sigma(-\omega_k)$ for $k=M-1$. We fix this by assuming that the bilocal fields are real, so that $\Sigma(-\omega_k) = \overline{\Sigma}(\omega_k)$.}.
We transform between the two descriptions with Fast Fourier Transforms. We update each iteration with a weighing factor $x$, as in \cite{Maldacena:2016hyu}, for which we found $x=2/3$ to be a good general choice. Schematically,
\begin{align}
\label{eq:num-sketch}
& G(\omega_k)_i \rightarrow \Sigma(\tau_k)_i = \Sigma_\text{EOM} \bigl[ G(\tau_k)_i \bigr] \rightarrow {} \\
&\quad {} \rightarrow G(\omega_k)_{i+1} = (1-x) \, G(\omega_k)_i + x  \, G_\text{EOM} \bigl[ \Sigma(-\omega_k)_i \bigr] \,. \nn
\end{align}
We iterate until $\sum_k \bigl\lvert G(\omega_k)_i -G(\omega_k)_{i+1} \bigr\rvert{}^2$ is smaller than some precision goal. To begin the iteration, we take the free solution.

For small $\mu_\phi$ and finite $J$, the equation of motion for $G_{\bar\phi\phi}(\omega=0)$ is highly sensitive to numerical errors in $\Sigma_{\bar \phi \phi}(\omega=0)$. One possible workaround, proposed for example in \cite{Christos:2022lma}, is to change the equation of motion for $G_{\bar\phi\phi}$ by replacing the parameter $\mu_\phi$ with $\rho$ so that $G_{\bar\phi\phi}(\omega=0)=\alpha\rho$. We replace the integro-differential equation for $G_{\bar\phi\phi}$ with
\be
\label{eq:num-sachdev-trick}
G_{\bar\phi\phi} \bigl[ \Sigma( -\omega^\text{b}_k)_i \bigr] = \frac{ - \alpha \rho}{\rho \bigl[ -i \omega^\text{b}_k + \Sigma_{\bar\phi \phi}(-\omega^\text{b}_k)_i - \Sigma_{\bar \phi \phi}(0)_i \bigr] - 1} \,.
\ee
One can then use the original equation of motion at the end to verify which value of $\mu_\phi$ is realized. In principle, one could take $\mu_\phi = \Sigma_{\bar \phi \phi}(0)+1/\rho$, however it is best to average the EOM in a window of frequency space. If one desires a specific value of $\mu_\phi$, this approach incurs a high additional numerical cost, since the solver must itself be iterated many times to carry out a bisection search or equivalent technique. Nevertheless, we found this to be a useful approach for those cases that are more numerically unstable, such as $(p,q)\neq (1,2)$.

\subsection{Conformal behavior for \tps{\matht{(p,q)=(1,2)}}{(p,q)=(1,2)}}

In this section we focus on the case $(p,q)=(1,2)$. Numerically, this turns out to be the most accessible case. It is also of particular theoretical interest, as it is the closest choice to the quantum mechanical model in \cite{Benini:2022bwa}.

\begin{figure}
\includegraphics[width=\columnwidth]{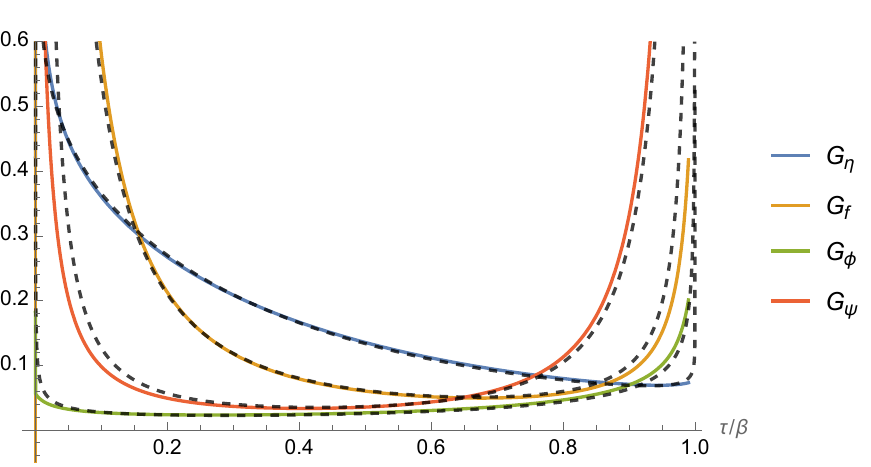} \\[.2em]
(a) $\bigl( \beta\mu_\eta ,\, \beta\mu_\phi \bigr) = (-2 ,\, 1) $
\\
\includegraphics[width=\columnwidth]{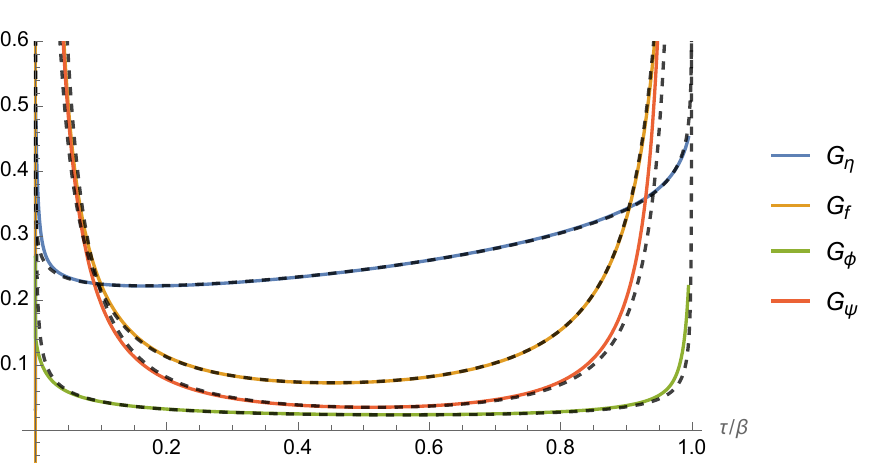} \\[.2em]
(b) $\bigl( \beta\mu_\eta ,\, \beta\mu_\phi \bigr) = (1 ,\, 2) $
\caption{\label{fig-conformal-12}%
Two solutions for $\beta J =1000$, $\alpha=1/4$, $(p,q)=(1,2)$. The colored full lines are the numerical solution of the Schwinger--Dyson equations with $M = 2^{18}$ points. The black dashed lines are a conformal solution for which the parameters $\cE_{\eta,\mu}$ and $g_{\eta,\mu}$ have been fitted numerically. (a): Supersymmetric chemical potentials. (b): Non-supersymmetric.}
\end{figure}

We first test that the conformal ansatz is realized at low energies. We take a solution with large $\beta J$ and scan the numerical solutions to \eqref{matching coeff in alg} for the best quadratic fit. These solutions are parametrized by $\cE_\eta$, $\cE_\phi$, $g_\eta$, and $g_\phi$. In principle, one could just fit $G_{\bar\eta\eta}$ and $G_{\bar\phi\phi}$, which can be beneficial since the auxiliary fields tend to converge slower to the conformal solution. However, we found that this results in less precise and less accurate estimates of the $\cE$'s, so we fit the four bilocal fields together.
Clearly, we must exclude from the fit the regions with $\tau$ close to $0$ and $\beta$, where the approximation is invalid and singular. Typically we exclude $\bigl\lvert \frac{1}{2} - \frac{\tau}{\beta} \bigr\rvert > 3/8$.

\begin{figure}
\includegraphics[width=0.95\columnwidth]{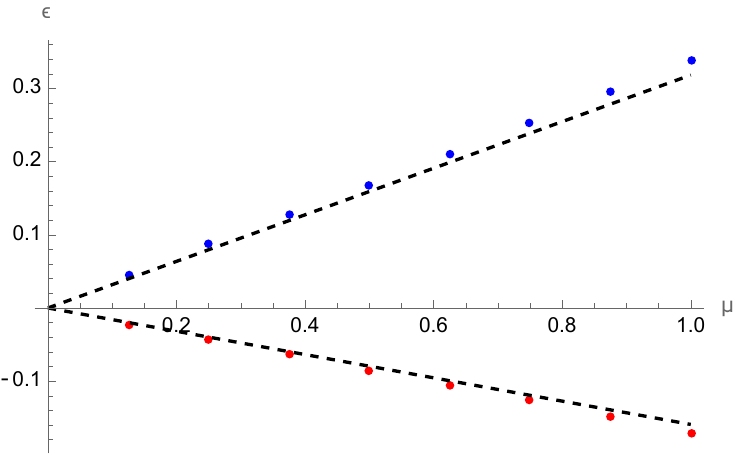} \\[.2em]
(a) $\bigl( \beta\mu_\eta ,\, \beta\mu_\phi \bigr) = (-2\mu ,\, \mu)$
\\
\includegraphics[width=0.95\columnwidth]{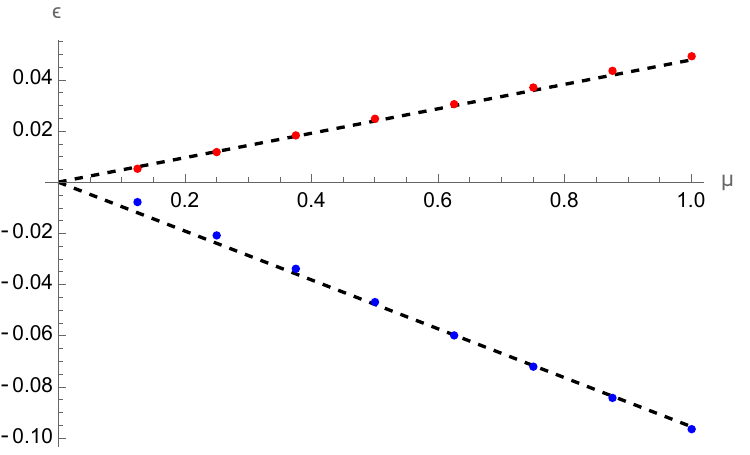} \\[.2em]
(b) $\bigl( \beta\mu_\eta ,\, \beta\mu_\phi \bigr) = (\mu ,\, 2\mu)$
\caption{\label{fig-epsilon-mu}%
Plot of $\cE_\eta$ (blue points) and $\cE_\phi$ (red points) as a function of $\mu$. The data was generated with $2^{20}$ points, $\beta J = 600$, and $\alpha=1/5$. The dashed lines in each plot correspond to \eqref{eq:susy-epsilon} (a: SUSY chemical potentials) and \eqref{eq:nonsusy-epsilon} (b: non-SUSY), respectively.}
\end{figure}

We present two illustrative examples in Fig.~\ref{fig-conformal-12}. Clearly, the conformal solution is realized and it matches our proposed ansatz. For $(p,q)=(1,2)$ and supersymmetric chemical potentials, we verified numerically that, at sufficiently large $\beta J$, we have
\be
\label{eq:susy-epsilon}
2\pi \cE_\eta \,\simeq\, - \beta \mu_\eta = 2 \beta \mu_\phi \;,\qquad 2\pi \cE_\phi \,\simeq\, -\beta \mu_\phi \;,
\ee
which matches the expectation from the supersymmetric index. This suggests that, even in the deep IR, SUSY is at most lightly broken at small temperatures. For other choices of chemical potentials, we also find a linear relation at low temperatures. The ansatz
\bea
\label{eq:nonsusy-epsilon}
2\pi \cE_\eta \,&\simeq\, - \bigl( 1 - \tfrac{2\alpha}{5} \bigr) \beta \mu_\eta + \tfrac{4\alpha}{5} \beta \mu_\phi \;,\\
2\pi\cE_\phi \,&\simeq\, \tfrac12 \bigl( 1 - \tfrac{2\alpha}{5} \bigr) \beta \mu_\eta - \tfrac{2\alpha}{5} \beta \mu_\phi
\eea
was numerically successful for large $\beta J$ and finite chemical potentials well below the critical value, testing with different values of $\alpha$. We plot a supersymmetric and a non-supersymmetric example in Fig.~\ref{fig-epsilon-mu}.
This ansatz has the curious property of being supersymmetric irrespectively of the chemical potentials, suggesting an emergent IR supersymmetry. When fitting the low energy behavior, $\Delta_{\phi,\eta}$ and $g_{\phi,\eta}$ also satisfy the respective supersymmetry constraints within $1\%-5\%$ at $\beta J$ between $200$ and $600$.
This ansatz was further confirmed when comparing the numerical values of the charge and entropy to the Luttinger--Ward relation and the zero-temperature entropy for non-supersymmetric cases in Section~\ref{sec:num-Q-and-S}. However, we note that one expects, for physical fields at fixed $\mathcal{Q}_{\eta,\phi}$, to have
\be
\mu_I(\mathcal{Q}_\eta,\mathcal{Q}_\phi,T) = \mu_{0,I}(\mathcal{Q}_\eta,\mathcal{Q}_\phi) + 2\pi \cE_I(\mathcal{Q}_\eta,\mathcal{Q}_\phi) T + \ldots
\ee
see for instance \cite{Parcollet:1997ysb, Sachdev:2015efa, Davison:2016ngz}. Here the index $I=\eta,\phi$ labels the $\rU(1)$ symmetries. Since we work at fixed $\mu$ and the function $\mathcal{Q}_{I}(\mu)$ is a priori complicated, eqn.~\eqref{eq:nonsusy-epsilon} might be the linearized behavior of a more intricate function $\mu_{0,I} \bigl( \mathcal{Q}_\eta( \mu_\eta, \mu_\phi),\, \mathcal{Q}_\phi( \mu_\eta, \mu_\phi) \bigr)$.

\subsection{Charge and entropy for \tps{\matht{(p,q)=(1,2)}}{(p,q)=(1,2)}}
\label{sec:num-Q-and-S}

\begin{figure}
\includegraphics[width=\columnwidth]{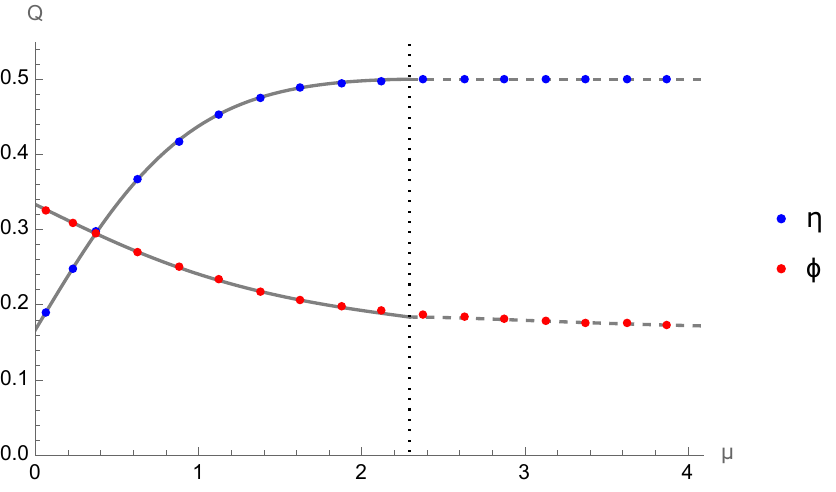} \\[.2em]
(a) $\alpha = 1/3$
\\
\includegraphics[width=\columnwidth]{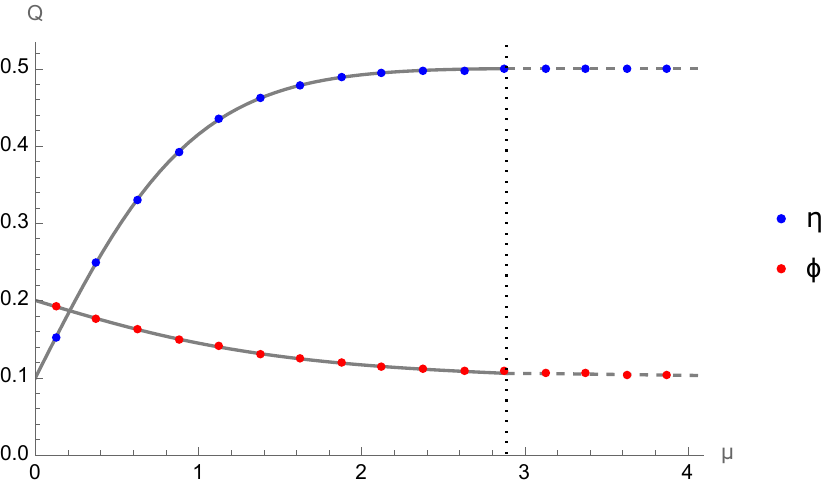} \\[.2em]
(b) $\alpha = 1/5$
\caption{\label{fig-luttinger-ward}%
We plot $\mathcal{Q}_\eta$ (blue points) and $\mathcal{Q}_\phi$ (red points) as a function of $\mu_F$. The data was generated with $2^{18}$ points, $\beta J = 200$ and $(\beta\mu_\eta,\beta\mu_\phi)=(-2\mu,\mu)$. The gray full line is the Luttinger--Ward formula \eqref{LW-corrected}, assuming \eqref{eq:susy-epsilon}. The gray dashed lines are the continuations in \eqref{eq:LW-beyond-conformal}. The vertical dotted line is the value $\mu_F$ for which $\cE_*$ is reached.}
\end{figure}

An important check of the conformal ansatz is the Luttinger--Ward relation between $\cE$ and $\mathcal{Q}$, which we presented in Section~\ref{sec-LW}. As previously noted, the Luttinger--Ward calculation determines $\mathcal{Q}(\cE)$ up to undetermined constants, which we fixed exploiting the index in \eqref{constants-index}. Numerically, we verify these constants to be:
\bea
\label{LW-corrected}
\mathcal{Q}_\eta &= \mathfrak{q}_-(\cE _\eta, \Delta_\eta) + (1-p) \, \mathfrak{q}_+( \cE_f, \Delta_f ) \\
&\quad - \alpha p \, \mathfrak{q}_- (\cE_\psi, \Delta_\psi ) + \tfrac{\alpha p}{2} \;, \\
\mathcal{Q}_\phi &= \alpha \, \mathfrak{q}_+( \cE _\phi, \Delta_\phi ) + \alpha (1-q) \, \mathfrak{q}_-( \cE_\psi, \Delta_\psi ) \\
&\quad - q \, \mathfrak{q}_+( \cE_f, \Delta_f ) + \tfrac{\alpha q}{2} \;,
\eea
where the function $\mathfrak{q}$ is given in \eqref{LW master}. This matches the index prediction \eqref{constants-index}.
We note that the charges do not vanish when $\mu_{\eta,\phi} \rightarrow 0$, which is supported by the numerics, as can be seen in Fig.~\ref{fig-luttinger-ward}.
The charges also converge very quickly to their low-temperature limit, since they depend on $\beta J$ through $\cE$.
Thus we can take $\beta J$ to be large and assume \eqref{eq:susy-epsilon}.

In the superconformal solution, as we take $\cE$ close to $\cE_*(\alpha)$ defined in \eqref{p1 q2 sol}, we approach the phase transition discussed in Section~\ref{sec-existence}. In order to avoid working with $\cE$ which is numerically more indirect, we label as $\beta\mu_{\raisebox{0pt}[0pt][0pt]{\scriptsize$\phi$}}^* = 2 \arccosh \bigl( \alpha^{-\frac12}\bigr)$ the critical chemical potential, which follows from \eqref{eq:susy-epsilon}.

\begin{figure}
\includegraphics[width=\columnwidth]{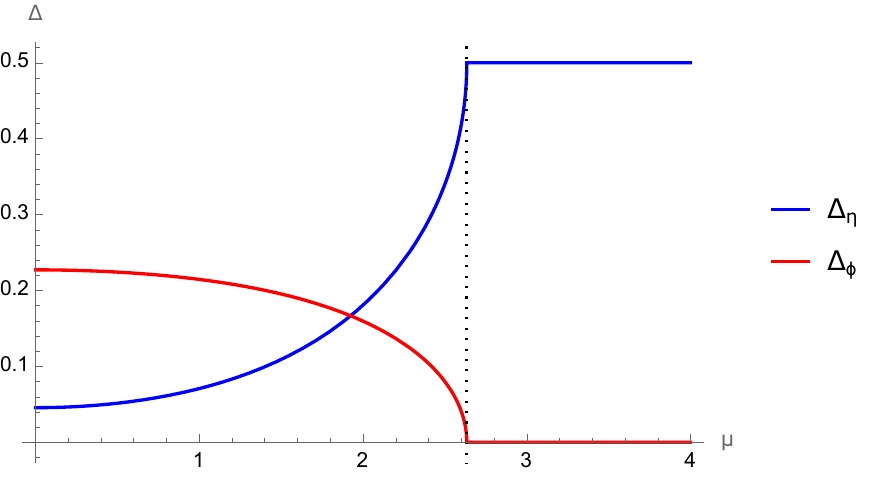}
\caption{\label{fig-deltas}%
We plot $\Delta_\eta$  and $\Delta_\phi$ for $\alpha=1/4$ and $(\beta\mu_\eta,\beta\mu_\phi) = (-2\mu,\mu)$, assuming \eqref{eq:susy-epsilon}. The data was generated by solving \eqref{consistency match coeff} numerically.
The vertical dotted line is the value of $\mu$ for which $\cE_*$ is reached.}
\end{figure}

Recall that, as $\cE \rightarrow \cE_*$, the conformal dimensions converge to $\Delta_\eta=1/2$ and $\Delta_\phi=0$, see Fig.~\ref{fig-deltas}. This is a solution --- albeit a trivial one --- to the consistency equations \eqref{conf consistency cond}, or equivalently \eqref{consistency match coeff}, for any value of $\cE$. Focusing only on the consistency equations, the absence of other nearby solutions for the conformal dimensions when $\cE > \cE_*$  suggests that the would-be conformal solution should have these marginal values. 
While, technically, this fact renders the conformal approximation no longer valid, since the kinetic and interacting terms are of the same order in $\omega$, we can conjecturally extend the Luttinger--Ward relation by plugging the fixed dimensions in. We conjecture
\be
\label{eq:LW-beyond-conformal}
\mathcal{Q}_\eta = \tfrac{1}{2} \,, \quad 
\mathcal{Q}_\phi = 1 - \coth( \beta \mu_F) + \tfrac{\alpha}{2} \coth \bigl( \tfrac{\beta \mu_F}{2} \bigr) \,.
\ee
As we can see in the dashed line in Fig.~\ref{fig-luttinger-ward}, the numerical results do not deviate significantly from such an analytic continuation.

We can further combine this result with \eqref{re S on p1 q2 sol} in order to obtain a zero-temperature analytic prediction for the entropy as a function of the flavor chemical potential, within the conformal phase and, conjecturally, beyond.
We use \eqref{eq:entropy-explicit} to calculate the entropy numerically at each finite temperature, and then we take $\beta J \rightarrow \infty$ with $\beta\mu$ fixed.

\begin{figure}
\includegraphics[width=\columnwidth]{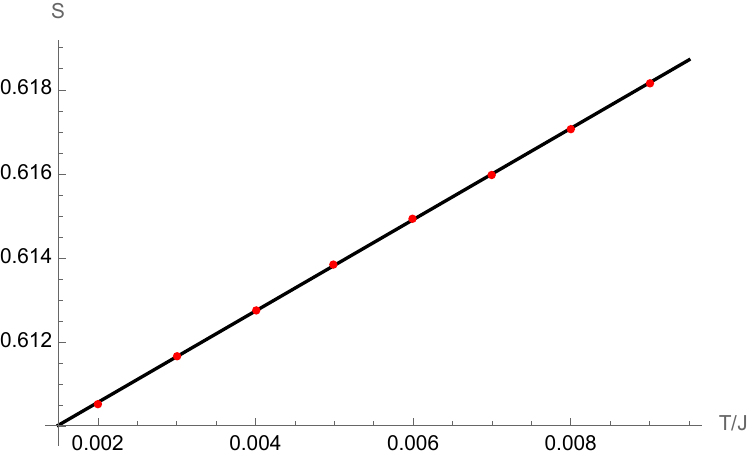} \\[.2em]
(a) $\alpha = \frac14$, $\mu = 0.125$, $r=0$, $M=2^{22}$
\\
\includegraphics[width=\columnwidth]{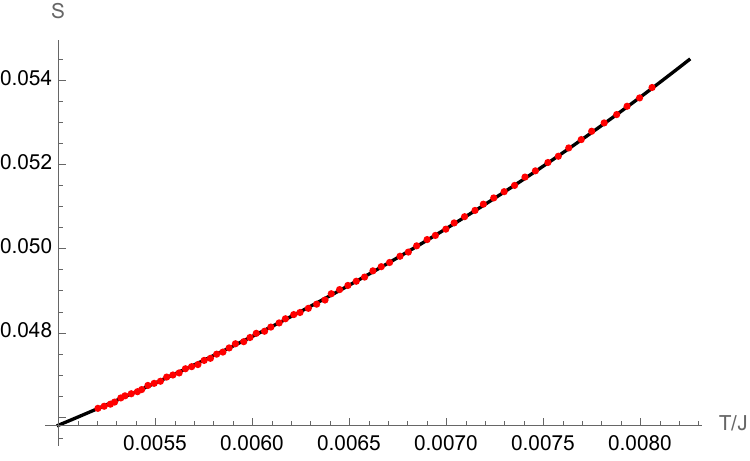} \\[.2em]
(b) $\alpha = \frac14$, $\mu = 2.58$, $r=3$, $M=2^{24}$
\caption{\label{fig-fits}%
We plot two examples of the extrapolation of the zero-temperature entropy for $(\beta \mu_\eta,\beta\mu_\phi)=(-2\mu,\mu)$. In plot (a) we take a small value of $\mu$ for which it suffices to fit with a line, allowing us also to estimate the Schwarzian coupling. Meanwhile, in plot (b) we take a value of $\mu$ close to the phase transition. Instead of fitting the points directly, we fit the third Richardson transform with the ansatz \eqref{eq:rich-fit}. Here $M$ is the number of points in the discretized Schwinger--Dyson equation.}
\end{figure}

\begin{figure}
\includegraphics[width=\columnwidth]{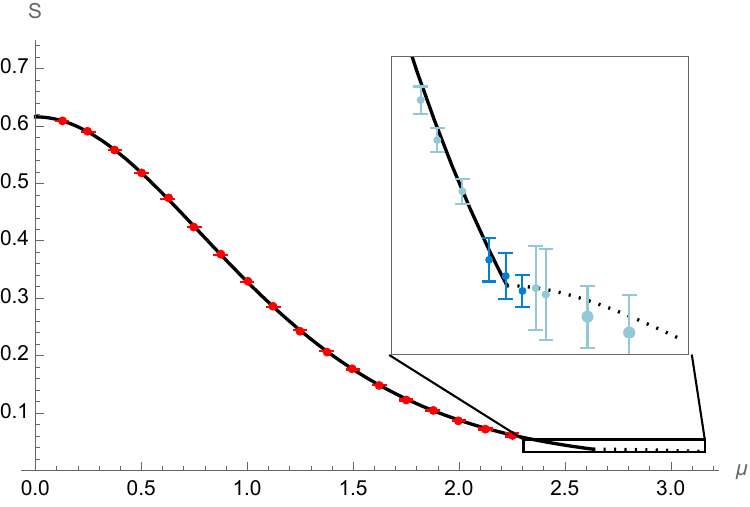}
\caption{\label{fig-entropy-index}%
Plot of the zero-temperature entropy $S_0$ as a function of the chemical potentials $(\beta\mu_\eta,\beta\mu_\phi)=(-2\mu,\mu)$ at $\alpha=1/4$. The red dots were generated with $2^{22}$ points, $\beta J$ spanning from $200$ to $1000$, and the entropy was obtained with simple linear extrapolation. The light blue and dark blue dots were generated with $2^{23}$ and $2^{24}$ points, respectively, with $\beta J$ spanning from $100$ to $400$, and extrapolated using the Richardson-transform fit described in eqn.~\eqref{eq:rich-fit}. The black line is the prediction from combining the index and the Luttinger--Ward formula \eqref{LW-corrected}, while the dashed line uses the index and the continuation \eqref{eq:LW-beyond-conformal}.}
\end{figure}

For the extrapolation to zero temperature, the precision strongly depends on how close to the phase transition we are. For values of the chemical potential sufficiently below $\beta\mu_\phi^*$,
it suffices to probe values of $\beta J$ between $50$ and $500$ and use a polynomial fit to extract the constant term, as can been seen in Fig.~\ref{fig-fits}~(a). The error is estimated by dropping points at both ends of the sample. The resulting extrapolation of the zero-temperature entropy are the red dots in Fig.~\ref{fig-entropy-index}.
However, as we move closer to the critical point, this extrapolation becomes harder. The main limiting factor is that we cannot probe arbitrarily low temperatures due finite-size effects and numerical instabilities. Furthermore, if we assume that the entropy behaves as
\be
\label{entropy-ansatz}
S(T) \,\approx\, S_0 + g_\text{Sch} (\beta J)^{-1} + g_\gamma (\beta J)^{-\gamma} + \ldots \;,
\ee
what we observe numerically (as well as in the analysis of the spectrum in Section~\ref{sec: spectrum}) is that $\gamma$ becomes close to $1$.
Thus, we generate points at integer values of $\beta J$ between $100$ and $400$ and then take one or more Richardson transforms, see \cite{bender-book, Marino:2007te}. After the $r$-th Richardson transform, we expect to converge to $S_0$ as
\be
\label{eq:rich-fit}
\bigl[ S(T) \bigr]_r \,\approx\, S_0  + g_{\gamma,r} (\beta J)^{-\gamma} + g_r (\beta J)^{-(1+r)} + \ldots \;.
\ee
Fitting the form \eqref{eq:rich-fit} for all the constants is more reliable at low temperature than \eqref{entropy-ansatz} since the exponents are more distinct, and we estimate the numerical error by varying the points included in the fit and the order $r$ of the Richardson transform. We used this strategy for the light-blue points in Fig.~\ref{fig-entropy-index}, with a step of $10$ for those on the left and a step of $4$ for those on the right (a smaller step increases the convergence of the Richardson acceleration).
Lastly, for points very close to the critical point the convergence to $S_0$ is particularly slow, so we considered only points with $\beta J$ between $200$ and $400$ with a step of $2$. These are the dark-blue points in Fig.~\ref{fig-entropy-index}. The error is estimated by varying the degree of the Richardson transform and by comparing with the previous method, taking the largest estimate. An example of such a fit is given in Fig.~\ref{fig-fits}~(b). The exponent $\gamma$ in \eqref{entropy-ansatz} seems to be close to $1$ at the transition point, but unfortunately the precision attainable with our points is not enough to extract meaningful estimates, including whether it is smaller or larger than $1$. 

Unfortunately, $\beta J\sim 500$ is still slowly converging to the low-temperature behavior and the above techniques have limited success in isolating $S_0$. Furthermore, the results are sensitive to small changes in the extrapolation method, which we attempt to capture with the error estimates in Fig.~\ref{fig-entropy-index}, but which are likely underestimated. With these significant caveats, what we can at best observe is that, in
Fig.~\ref{fig-entropy-index}, the kink that follows from extrapolating $\mathcal{Q}_F(\mu)$ with \eqref{eq:LW-beyond-conformal} seems to be close to the numerical results. This could suggest some form of second-order phase transition. 
However, note that this would-be kink only appears when working at fixed chemical potential, while at fixed charge we can just apply \eqref{re S on p1 q2 sol} directly, which is smooth.

\begin{table}
\arraycolsep=2.5mm
$\ds
\begin{array}{c | cc | cc}
\toprule
 & \multicolumn{2}{c|}{\beta\mu_\phi = 1/8} & \multicolumn{2}{c}{\beta\mu_\phi = 3/2} \\
\alpha & \mathcal{Q}_F & g_\text{Sch} & \mathcal{Q}_F & g_\text{Sch} \\
\midrule
1/4 & 0.1087 & 1.09 \pm 0.03 & 0.7920 & 1.09 \pm 0.01 \\
1/5 & 0.1119 & 0.97 \pm 0.03 & 0.8147 & 0.96 \pm 0.01 \\
1/6 & 0.1140 & 0.90 \pm 0.03 & 0.8298 & 0.90 \pm 0.02 \\
1/7 & 0.1155 & 0.87 \pm 0.04 & 0.8405 & 0.86 \pm 0.02 \\
\bottomrule
\end{array}
$
\caption{\label{tab-schwarzian}%
Estimates of the Schwarzian coupling, for $(\beta\mu_\eta, \beta\mu_\phi)=(-2\mu,\mu)$.}
\end{table}

\begin{figure}
\vspace{-1.2em}
\includegraphics[width=\columnwidth]{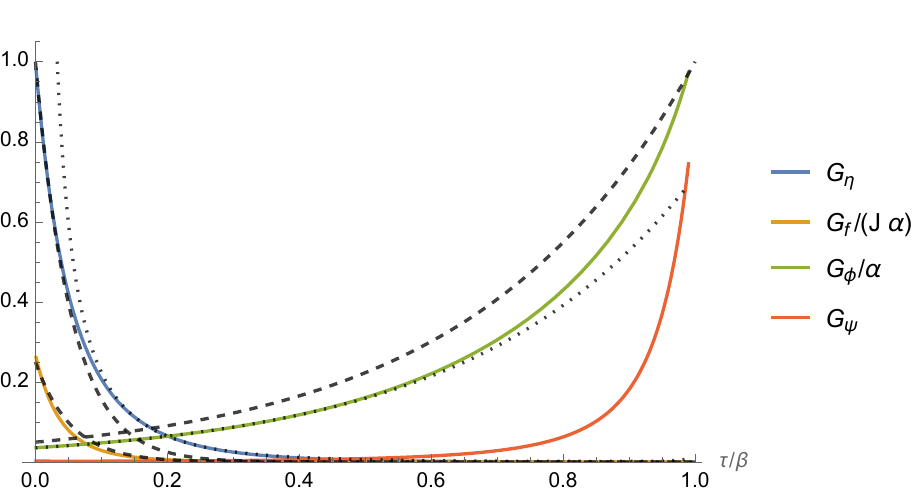}
\caption{\label{fig-nonconformal-phase}%
Solutions for $\beta J = 400$, $\alpha=1/4$, $(p,q)=(1,2)$, $(\beta\mu_\eta,\beta\mu_\phi) = (-6,3) $. The colored full lines are the numerical solution to the Schwinger--Dyson equations with $2^{22}$ points. The black dashed lines are the zero-temperature prediction \eqref{eq:non-conformal-negative-eta}. The dotted lines are the prediction from the conformal ansatz with $\Delta_\eta=\frac{1}{2}$, $\Delta_\phi=0$, $\cE_{\eta,\phi}$ as in \eqref{eq:susy-epsilon}, and $g_{\eta,\phi}$ fitted numerically. In the conformal solution $G_{\bar f f}=G_{\bar\psi\psi}=0$.}
\end{figure}

For chemical potentials sufficiently lower than the transition point, we can also extract the coefficient of the linear term in $T$ from the entropy, which should be proportional to the Schwarzian coupling. However, as we move closer to the phase transition, the linear behavior is less clear as it seems to require even lower temperatures.  We show some of the estimates of $g_\text{Sch}$ for different values of $\cQ_F$ in Table~\ref{tab-schwarzian}.

\begin{figure}
\includegraphics[width=\columnwidth]{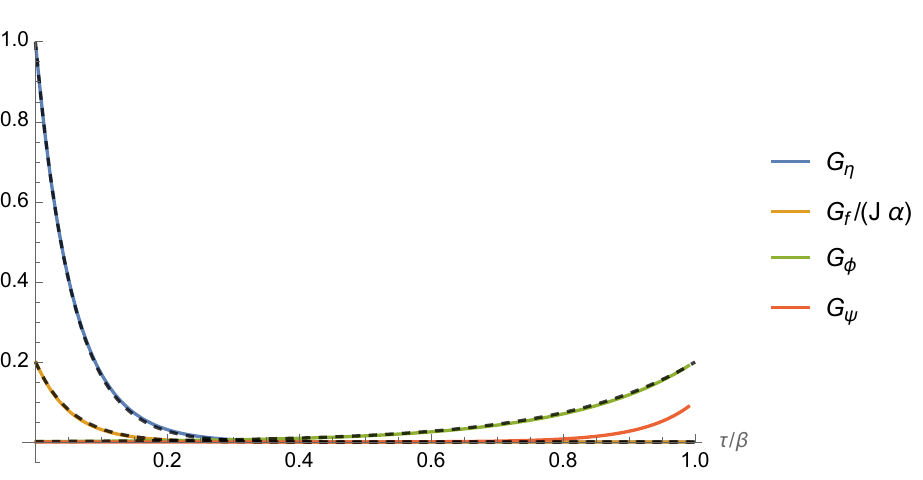}
\caption{\label{fig-nonconformal-high}%
Solutions for $\beta J = 400$, $\alpha=1/4$, $(p,q)=(1,2)$, $(\beta\mu_\eta,\beta\mu_\phi) = (-10,5) $. The colored full lines are the numerical solution to the Schwinger--Dyson equations with $2^{18}$ points. The black dashed lines are the zero-temperature prediction \eqref{eq:non-conformal-negative-eta}.}
\end{figure}

As for the solutions themselves, at $\beta\mu_\phi>\beta\mu_\phi^*$ they seem to interpolate between the conformal solution and the exponential behavior of the non-conformal solution \eqref{eq:non-conformal-positive-eta}, as we illustrate in Fig.~\ref{fig-nonconformal-phase}. For $\tau$ in the middle of the interval (the IR regime), we see that a conformal ansatz with $\Delta_\eta=1/2$ and $\Delta_\phi=0$ is a good approximation of the physical fields $\eta, \phi$ (for the auxiliary fields $f, \psi$, the ansatz gives $0$), while at $\tau$ closer to $0$ and $\beta$ the non-conformal solution is a better match.
As we take $\beta\mu_F\rightarrow\infty$, the solution seems to converge to \eqref{eq:non-conformal-positive-eta}, which analytically corresponds to the case $\beta\rightarrow\infty$ and $\mu$ finite. As shown in Fig.~\ref{fig-nonconformal-high}, we can match $G_{\bar \eta \eta}$, $G_{\bar f f}$, and $G_{\bar \phi \phi}$ with their zero-temperature limit. $G_{\bar\psi\psi}$ vanishes at zero temperature, so we observe that it is dominated by the subleading contribution, which we did not establish analytically.

\subsubsection{Entropy for non-supersymmetric solutions}

\begin{figure}
\includegraphics[width=0.95\columnwidth]{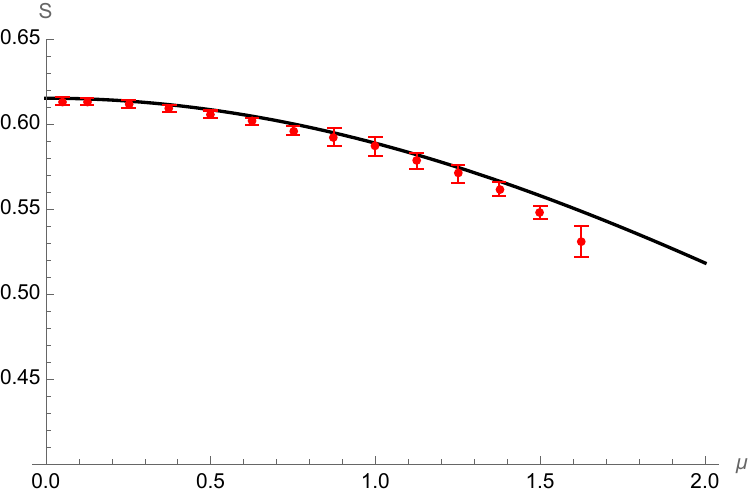}
\caption{\label{fig-entropy-nonsusy}%
Plot of the zero-temperature entropy $S_0$ as a function of the chemical potentials $(\beta\mu_\eta,\beta\mu_\phi)=(\mu,2\mu)$ at $\alpha = 1/4$. The red dots were generated with $2^{23}$ points, $\beta J$ spanning from $10$ to $200$, and linear extrapolation of the entropy. The black line is the prediction obtained from the index, the Luttinger--Ward formula \eqref{LW-corrected}, and the ansatz \eqref{eq:nonsusy-epsilon}.}
\end{figure}

One can also solve the Schwinger--Dyson equations for non-supersymmetric chemical potentials. Using a similar extrapolation as before, one can then obtain the zero-temperature entropy. Since we empirically found a supersymmetric low-energy behavior in \eqref{eq:nonsusy-epsilon}, we can conjecturally compare the non-supersymmetric zero-temperature entropy with the entropy from $\cI$-extremization using the values of $\cE$ in \eqref{eq:nonsusy-epsilon}. In Fig.~\ref{fig-entropy-nonsusy} we plot such an example. The theoretical prediction seems compatible with the numerical result within the extrapolation error for smaller values of $\mu_{\eta}$ and $\mu_\phi$. Unfortunately, the numerical iteration is noticeably more unstable for non-supersymmetric potentials, and thus we could neither probe a wider range of chemical potentials nor work with greater precision.

Even within the small band of values of $\mu$ probed numerically, these results are surprising. They reinforce the SUSY IR behavior found in \eqref{eq:nonsusy-epsilon}. Unlike the check in Fig.~\ref{fig-epsilon-mu}, this test does not rely on fitting $\cE_{\phi,\eta}$ at late $\tau$, which is a delicate task. So, this figure is independent evidence of the emergent SUSY. Due to numerical limitations, however, we cannot ascertain whether this behavior is valid only for small values of $\mu$ or whether the deviation from the line is due to numerical errors.

\subsection{Results for other values of \tps{\matht{(p,q)}}{(p,q)}}

\begin{figure}
\includegraphics[width=\columnwidth]{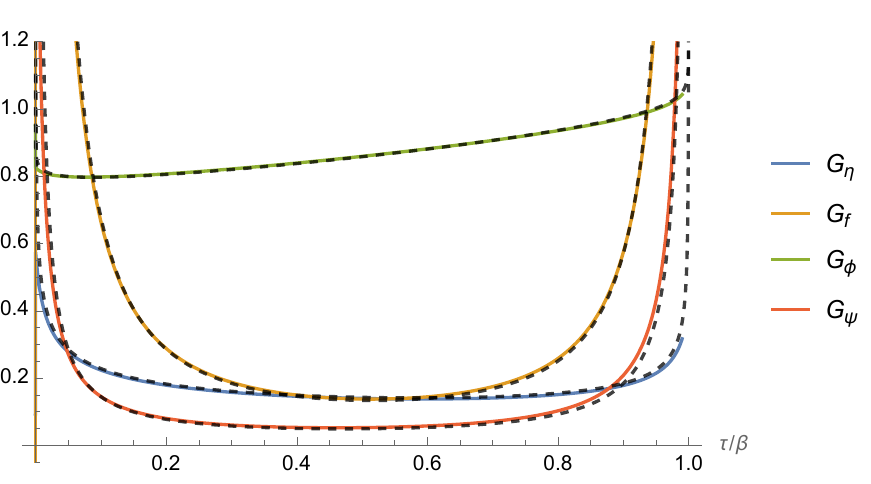} \\[.2em]
Conformal-like solution
\\
\includegraphics[width=\columnwidth]{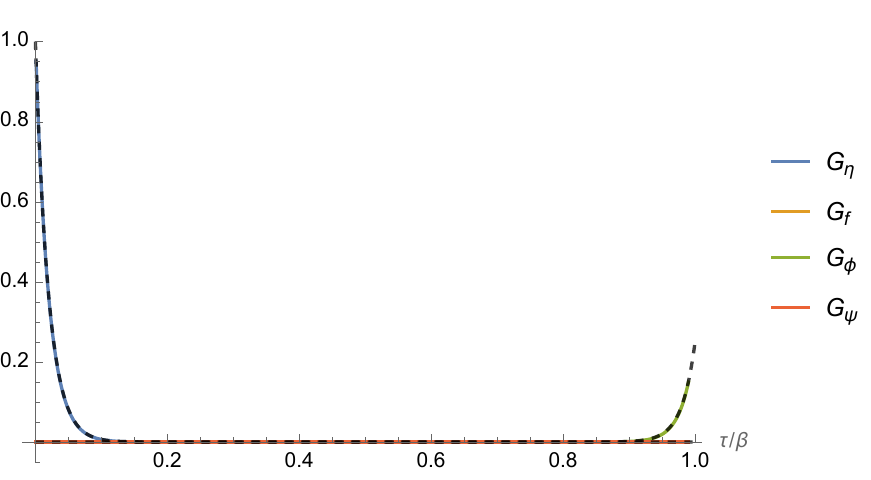} \\[.2em]
Exponential-like solution
\caption{\label{fig-pq-33}%
Two solutions for $\beta J = 450$, $\alpha = 1/4$, $(p,q)=(3,3)$, and $(\beta\mu_\eta,\beta\mu_\phi)=(-30,30)$. The colored full lines are the numerical solution to the Schwinger--Dyson equations with $2^{20}$ points, using \eqref{eq:num-sachdev-trick} on top and the standard approach at the bottom. 
(a): The black dashed lines are a fit of the conformal ansatz. 
(b): The black dashed lines represent the zero-temperature non-conformal solution \eqref{eq:non-conformal-negative-eta}.}
\end{figure}

Solving the Schwinger--Dyson equation for values of $(p,q)$ other than $(1,2)$ turns out to be numerically more unstable. The standard prescription, when it stabilises, often lands on the exponential-like non-conformal solutions.
We can use the prescription \eqref{eq:num-sachdev-trick} to find conformal-like solutions, but this makes the study of supersymmetric solutions much more costly since it requires iterating for many values of $\rho$, since for each value of $\mu_\eta$ it is \textit{a priori} unknown which value of $\rho$ corresponds to the supersymmetric value of $\mu_\phi$. In Fig.~\ref{fig-pq-33} we show some examples of the conformal solution being realized for $p=q=3$. In this case we can even compare a conformal and non-conformal solution for the same values of $(\beta\mu_\eta, \beta\mu_\phi)$, obtained with \eqref{eq:num-sketch} instead.

\section{Spectrum of low-lying operators}
\label{sec: spectrum}

In order to understand the IR dynamics of the model, we find the spectrum of physical excitations around conformal solutions. For simplicity, we present the full derivation only for superconformal solutions, while we describe the non-supersymmetric case at the end.
The goal in this section is first to ensure that the Schwarzian action is present in the spectrum of near-conformal corrections, and then to investigate when it is dominant at low energies, as expected in gravitational models.
Another feature that we expect and verify in these models is the presence of ``soft modes'' associated to the $U(1)$ flavor and R symmetries. The flavor current is part of the ``$\mathcal{N}=2$ current multiplet'', while the R-symmetry is part of the Schwarzian multiplet. Although the current modes are technically lower-energy than the Schwarzian modes, they are compatible with a gravitational interpretation.

\subsection{Expanding the action}
\label{subsubsec: 4pt from saddle point}

In order to derive the spectrum, we expand the fields around solutions of the equations of motion, and then diagonalize the quadratic fluctuations. In superspace, one expands the bilocal fields around the superconformal solutions as
\begin{align}
\label{G perturbation}
\cG_\cY(T_1,T_2) &= G^*_{\bar\eta \eta}(T_{12}) + N^{-1/2} \delta \cG_{\cY}(T_1,T_2) \,, \\
\cG_\Phi(T_1,T_2) &= G^*_{\bar\phi \phi}(T_{12}) + N^{-1/2} \delta \cG_{\Phi}(T_1,T_2) \,, \nn\\
\label{Sigma perturbation}
\Sigma_\cY(T_1,T_2) &= \Sigma^*_{\bar{f} f}(T_{12}) + N^{-1/2} \delta \Sigma_{\cY}(T_1,T_2) \,, \\
\Sigma_\Phi(T_1,T_2) &= \Sigma^*_{\bar\psi \psi}(T_{12}) + N^{-1/2} \delta \Sigma_{\Phi}(T_1,T_2) \,. \nn
\end{align}
Here $G^*$ and $\Sigma^*$ indicate the superconformal solutions. Due to supersymmetry, the bilocal superfields are determined by their lowest components as in \eqref{superconformal solutions}. When expanding the action \eqref{bilocal action superspace}, the terms containing $\delta\Sigma$'s at quadratic order are:
\begin{widetext}
\begin{multline}
\frac{1}{2} \int\! d^2T_2 \, d^2\overline{T}_1 \, d^2T_4 \, d^2\overline{T}_3 \; \delta\Sigma(T_1,T_2)^\sT \begin{pmatrix}
- G^*_{\bar\eta \eta} G^*_{\bar\eta \eta} & 0 \\
0 & \frac{1}{\alpha} G^*_{\bar\phi \phi} G^*_{\bar\phi \phi}
\end{pmatrix} \! (T_{14},T_{32}) \; \delta\Sigma(T_3,T_4) \\
+ \frac{1}{2} \int\! d^2T_2 \, d^2\overline{T}_1 \; \delta \cG(T_1,T_2)^\sT \, \delta\Sigma(T_1,T_2)
+ \frac{1}{2} \int\! d^2T_2 \, d^2\overline{T}_1 \; \delta\Sigma(T_1,T_2)^\sT \, \delta\cG(T_1,T_2) \;,
\end{multline}
where we defined $\delta\Sigma(T_1,T_2) = \smat{ \delta \Sigma_\cY \\ \delta \Sigma_\Phi }(T_1,T_2)$ and $\delta\cG(T_1,T_2) = \smat{ \delta \cG_\cY \\ \delta \cG_\Phi } (T_1,T_2)$.
In the matrix on the first line, $T_{14}$ and $T_{32}$ are the arguments of the first and second $G^*$, respectively. Since we are interested in the action for $\delta\cG$ to leading order in $N$, it is sufficient to integrate out $\delta\Sigma$ classically. Combining the result with terms in the action involving $\delta\cG$ only, we get the quadratic terms
\be
\label{superspace action quad exp}
- \tfrac{1}{2} \ts\int\! d^2T_2 \, d^2\overline{T}_1 \, d^2T_4 \, d^2\overline{T}_3 \; \delta \cG(T_1,T_2)^\sT \, \cF_0^{-1}\cdot\bigl[\unit-\cK\bigr](T_1,T_2;T_3,T_4) \; \delta \cG(T_3,T_4) \,.
\ee
The operator $\cK$, which we call the superspace kernel, is given by
\be
\label{kernel definition 1}
\cK(T_1,T_2;T_3,T_4) \equiv \begin{pmatrix}
\cK_{\cY\cY}(T_1,T_2;T_3,T_4) & \cK_{\cY\Phi}(T_1,T_2;T_3,T_4)\\
\cK_{\Phi\cY}(T_1,T_2;T_3,T_4) & \cK_{\Phi\Phi}(T_1,T_2;T_3,T_4)
\end{pmatrix}
\ee
where
\bea
\label{kernel definition 2}
\cK_{\cY\cY}(T_1,T_2;T_3,T_4) &= \tfrac{(p-1)J}{q} \, G^*_{\bar\eta \eta}(T_{34})^{p-2} \, G^*_{\bar\phi \phi}(T_{34})^q \, G^*_{\bar\eta \eta}(T_{14}) \, G^*_{\bar\eta \eta}(T_{32}) \;, \\
\cK_{\cY\Phi}(T_1,T_2;T_3,T_4) &= J \, G^*_{\bar\eta \eta}(T_{34})^{p-1} \, G^*_{\bar\phi \phi}(T_{34})^{q-1} \, G^*_{\bar\eta \eta}(T_{14}) \, G^*_{\bar\eta \eta}(T_{32}) \;, \\
\cK_{\Phi\cY}(T_1,T_2;T_3,T_4) &= - \tfrac{J}{\alpha} \, G^*_{\bar\phi \phi}(T_{34})^{q-1} \, G^*_{\bar\eta \eta}(T_{34})^{p-1} \, G^*_{\bar\phi \phi}(T_{14}) \, G^*_{\bar\phi \phi}(T_{32}) \;, \\
\cK_{\Phi\Phi}(T_1,T_2;T_3,T_4) &= - \tfrac{(q-1)J}{\alpha p} \, G^*_{\bar\phi \phi}(T_{34})^{q-2} \, G^*_{\bar\eta \eta}(T_{34})^p \, G^*_{\bar\phi \phi}(T_{14}) \, G^*_{\bar\phi \phi}(T_{32}) \;.
\eea
After plugging the supersymmetric conformal ansatz \eqref{G conformal ansatz} in, we obtain
\bea
\cK_{\cY\cY} &= \tfrac{(p-1)}q \, g_\eta^{p} g_\phi^{q} J \, G_{-}(T_{14};\Delta_\eta,\cE_\eta) \, G_{-}(T_{32};\Delta_\eta,\cE_\eta) \, G_{-} \bigl(T_{34}; \tfrac12 - 2 \Delta_\eta, - 2 \cE_\eta \bigr) \,, \\
\cK_{\cY\Phi} &= g_\eta^{p+1} g_\phi^{q-1}J \, G_{-}(T_{14};\Delta_\eta,\cE_\eta) \, G_{-}(T_{32};\Delta_\eta,\cE_\eta) \, G_{+} \bigl( T_{34}; \tfrac12 - \Delta_\eta-\Delta_\phi, -\cE_\eta - \cE_\phi \bigr) \,, \\
\cK_{\Phi\cY} &= - \tfrac1\alpha \, g_\eta^{p-1} g_\phi^{q+1} J  \, G_{+}(T_{14};\Delta_\phi,\cE_\phi) \, G_{+}(T_{32};\Delta_\phi,\cE_\phi) \, G_{+} \bigl( T_{34}; \tfrac12 - \Delta_\eta - \Delta_\phi, -\cE_\eta -\cE_\phi \bigr) \,, \\
\cK_{\Phi\Phi} &= - \tfrac{(q-1)}{\alpha p} \, g_\eta^{p} g_\phi^{q} J \, G_{+}(T_{14};\Delta_\phi,\cE_\phi) \, G_{+}(T_{32};\Delta_\phi,\cE_\phi) \, G_{-} \bigl( T_{34}; \tfrac12 - 2\Delta_\phi, -2\cE_\phi \bigr) \,.
\eea
We also defined the matrix $\cF_0$ as
\be
\cF_0(T_1,T_2;T_3,T_4) \equiv \begin{pmatrix}
- G^*_{\bar\eta \eta} G^*_{\bar\eta \eta}& 0 \\
0 & \frac{1}{\alpha} G^*_{\bar\phi \phi} G^*_{\bar\phi \phi} \end{pmatrix}(T_{14},T_{32}) \,.
\ee
From \eqref{superspace action quad exp} we see that the spectrum is determined by the zeros of the kinetic operator $\unit-\cK$, which is determined by the eigenvalue equation 
\be
\label{spectral problem}
\delta \cG(T_1,T_2) = \int\! d^2\overline{T}_3 \, d^2T_4 \; \cK(T_1,T_2;T_3,T_4) \; \delta \cG(T_3,T_4) \,.
\ee

\subsubsection{Deriving the spectral problem from the Schwinger--Dyson equations}

A shortcut to obtain the same eigenvalue equation is to expand the Schwinger--Dyson equations around superconformal solutions. The eqns.~\eqref{superspace algebraic eq} and \eqref{integral eom antichiral} can be recast as
\bea
\label{chiral/anti-chiral conv superspace eoms IR}
J \int\! d^2T_4 \; \cG_{\cY}(T_{34})^{p-1} \, \cG_{\Phi}(T_{34})^{q-1} \begin{pmatrix}
\cG_\cY(T_{14}) \, \cG_{\Phi}(T_{34}) \\
\cG_\Phi(T_{14}) \, \cG_{\cY}(T_{34})
\end{pmatrix} &= \begin{pmatrix}
q \, \overline{\delta}{}^2(T_1,T_3) \\
-p \, \alpha \, \overline{\delta}{}^2(T_1,T_3)
\end{pmatrix} \,, \\
J \int\! d^2\overline{T}_3 \; \cG_{\cY}(T_{34})^{p-1} \, \cG_{\Phi}(T_{34})^{q-1} \begin{pmatrix}
\cG_\cY(T_{32}) \, \cG_{\Phi}(T_{34}) \\
\cG_\Phi(T_{32}) \, \cG_{\cY}(T_{34})
\end{pmatrix} &= \begin{pmatrix}
q \, \delta^2(T_4,T_2) \\
-p \, \alpha \, \delta^2(T_4,T_2)
\end{pmatrix} \,.
\eea
The kinetic terms in \eqref{integral eom antichiral} have been justifiably neglected in the IR. The upper component of the first equation in \eqref{chiral/anti-chiral conv superspace eoms IR} can be expanded to linear order in $\delta\cG$ as:
\begin{align}
& \tfrac{J}{q} \ts\int\! d^2T_4 \bigl[ (p-1) \, G^*_{\bar\eta \eta}(T_{34})^{p-2} \, G^*_{\bar\phi \phi}(T_{34})^{q} \, G^*_{\bar\eta \eta}(T_{14}) \, \delta \cG_{\cY}(T_3,T_4) + {} \\
& \quad + q \, G^*_{\bar\eta \eta}(T_{34})^{p-1} \, G^*_{\bar\phi \phi}(T_{34})^{q-1} \, G^*_{\bar\eta \eta}(T_{14}) \, \delta \cG_{\Phi}(T_3,T_4)
+ G^*_{\bar\eta \eta}(T_{34})^{p-1} \, G^*_{\bar\phi \phi}(T_{34})^{q} \, \delta \cG_\cY(T_1,T_4) \bigr] = 0 \,. \nn
\end{align}
Multiplying by $G^*_{\bar\eta \eta}(T_{32})$, integrating over $T_3$, and using the second equation in \eqref{chiral/anti-chiral conv superspace eoms IR}, we obtain the upper component of (\ref{spectral problem}). An analogous equation for $\cG_\Phi$ can be obtained from the bottom component of the first equation in \eqref{chiral/anti-chiral conv superspace eoms IR}, so reproducing the full (\ref{spectral problem}).

\subsubsection{4-point functions in the large \tps{$N$}{N} limit}

With the quadratic action \eqref{superspace action quad exp} we can also derive 4-point functions to order $\cO(1/N)$. After the disorder average, the 4-point functions we can compute are those expressed in terms of bilocal fields, which we group into a matrix:
\begin{multline}
\label{4-pt saddle pt}
\frac{1}{N^2} \biggl\langle \, \begin{matrix}
\overline{\cY}_a \cY_a \overline{\cY}_b \cY_b & \overline{\cY}_a \cY_a \overline{\Phi}_b \Phi_b\\
\overline{\Phi}_a \Phi_a \overline{\cY}_b \cY_b & \overline{\Phi}_a \Phi_a \overline{\Phi}_b \Phi_b
\end{matrix} \, \biggr\rangle (T_1,T_2;T_3,T_4) = \biggl\langle \, \begin{matrix}\cG_\cY\cG_\cY & \cG_\cY\cG_\Phi \\ \cG_\Phi\cG_\cY & \cG_\Phi\cG_\Phi \end{matrix} \, \biggr\rangle (T_1,T_2;T_3,T_4) = {} \\
=\,\begin{pmatrix}G_{\overline{\eta}\eta}^*G_{\overline{\eta}\eta}^* & G_{\overline{\eta}\eta}^*G_{\overline{\phi}\phi}^* \\ G_{\overline{\phi}\phi}^*G_{\overline{\eta}\eta}^* & G_{\overline{\phi}\phi}^*G_{\overline{\phi}\phi}^*\end{pmatrix}(T_{12},T_{34})+\frac{1}{N}\left\langle\begin{matrix}\delta\cG_\cY\delta\cG_\cY & \delta\cG_\cY\delta\cG_\Phi \\ \delta\cG_\Phi\delta\cG_\cY & \delta\cG_\Phi\delta\cG_\Phi\end{matrix}\right\rangle(T_1,T_2;T_3,T_4)\,.
\end{multline}
The notation here is that $T_1$ is the coordinate of the first field from the left, $T_2$ of the second one, and so on. In the last equality we inserted the expansion \eqref{G perturbation}. The 4-point functions of fundamental fields are therefore computed by the 2-point functions of $\delta\cG$ to leading order in $N$. By adding sources to \eqref{superspace action quad exp}, completing the square, and taking functional derivatives with respect to the sources, one obtains
\be
\label{4 pt fn SD}
\left\langle \begin{matrix} \delta\cG_\cY \delta\cG_\cY & \delta\cG_\cY \delta\cG_\Phi \\ \delta\cG_\Phi \delta\cG_\cY & \delta\cG_\Phi \delta\cG_\Phi \end{matrix}\right\rangle \! (T_1,T_2;T_3,T_4) 
= [\unit-\cK]^{-1} \cdot \cF_0(T_1,T_2;T_3,T_4) + \cO \left(\tfrac{1}{N} \right) \;.
\ee
The product $\cdot$ includes both matrix multiplication as well as integration over chiral and anti-chiral superspace coordinates, like in \eqref{spectral problem}.
\end{widetext}

\subsection{Computing the spectrum}

We search for eigenvectors of $\cK$ with unit eigenvalue, as in (\ref{spectral problem}). We denote the components of $\delta\cG$ as $\delta G_{\bar{A}B}$ where $A$, $B$ range over the fields $\eta$, $f$, $\phi$, and $\psi$. Eqn.~\eqref{spectral problem} separates into three independent equations: one for bosonic fluctuations $\delta G^\rB$, and two for fermionic fluctuations $\delta G^\rF$ and $\delta \overline{G}{}^\rF$. These are defined as
\begin{align}
\delta G^\rB &= \bigl( \delta G_{\bar\eta \eta}, \delta G_{\bar{f} f}, \delta G_{\bar\phi \phi}, \delta G_{\bar\psi \psi} \bigr){}^\sT  , \\
\delta G^\rF &= \bigl( \delta G_{\bar\eta f}, \delta G_{\bar\phi \psi} \bigr){}^\sT , \qquad
\delta\overline{G}{}^\rF = \bigl( \delta G_{\bar{f} \eta}, \delta G_{\bar\psi \phi} \bigr) \,. \nn
\end{align}
The vectors $\delta G^\rB(\tau_1, \tau_2)$, $\delta G^\rF(\tau_1, \tau_2)$, and $\wb{G}{}^\rF(\tau_1, \tau_2)$ satisfy three equations as in (\ref{spectral problem}), but in terms of $K^\rB(\tau_1, \tau_2; \tau_3, \tau_4)$, $K^\rF(\tau_1, \tau_2; \tau_3, \tau_4)$ and $\wb{K}{}^\rF(\tau_1, \tau_2; \tau_3, \tau_4)$, respectively. 
The bosonic and fermionic fluctuations do not mix in these equations because the superconformal solutions for fermionic 2-point functions are zero.

We follow the same steps as in \cite{Maldacena:2016hyu, Polchinski:2016xgd, Fu:2016vas, Heydeman:2022lse} to diagonalize the kernel operators $K^\rB$, $K^\rF$, $\overline{K}{}^\rF$, so we shall be brief. Due to the conformal symmetry of the solutions $G^*$, the kernels commute with all conformal generators and with the 2-particle conformal Casimir with eigenvalue $h(h-1)$. Since Casimir and kernels can be simultaneously diagonalized, one can focus on a subspace of fixed $h$. For each $\delta G_{\bar{A}B}$, the corresponding space is spanned by two eigenfunctions:
\be
\label{casimir eigenfn}
G_s \biggl( \tau_{12}; \, \frac{\Delta_A+\Delta_B-h}{2}, 0 \biggr) = \frac{ \Theta(\tau_{12}) + s \, \Theta(-\tau_{12})}{|\tau_{12}|^{\Delta_A+\Delta_B-h}}\;.
\ee
Here $G_s$ is the 2-point function defined in (\ref{G conformal ansatz}). We therefore expand each perturbation $\delta G_{\bar AB}$ as
\begin{align}
\label{spectrum basis ansatz}
& \delta G_{\bar{A}B}(\tau_1,\tau_2) \\
&\;\; = \sum\nolimits_{s = \pm1} g_{L(A, B)} \, G_s\bigl( \tau_{12}; \, \tfrac{\Delta_A + \Delta_B - h}{2}, 0 \bigr) \, \delta g_{\bar{A}B;s} \,. \nn
\end{align}
The factors $g_{L(A,B)}$ denote the coefficient $g$ in the conformal ansatz \eqref{G conformal ansatz} for the lowest multiplet component between $A$ and $B$. For example, if $A=\eta$ and $B=f$, then $L(A,B)=\eta$. One can think of these factors as rescalings of the expansion coefficients $\delta g_{\bar{A}B;s}$. They are included so that the matrix elements of the kernels only depend on the combinations $g_\eta^{p-1}g_\phi^q g_f$ and $g_\eta^{p}g_\phi^{q-1}g_\psi$ determined by the equations of motion \eqref{matching coeff in alg} (even in the non-supersymmetric case), and not on the individual coefficients. As expected, the kernels act within the subspace of fixed $h$, and can be represented by ordinary matrices acting on the coefficients $\delta g_{\bar{A}B;s}$.
However, the size of each matrix is doubled since each subspace is two-dimensional and spanned by \eqref{casimir eigenfn}, thus $K^\rB$ is represented by a $8\times 8$ matrix while $K^\rF$ and $\overline{K}{}^\rF$ are represented by $4\times 4$ matrices.

We are left with the ordinary problem of determining the values of $h$ such that $K^\rB$, $K^\rF$ or $\overline{K}{}^\rF$ have eigenvalue $1$, which can be solved numerically. For each eigenvalue, we plot $k(h)-1$ as a function of $h$ and look for zeros, where the graphs intersect the horizontal axis. The number of coincident intersections tells us the number of modes at a given value of $h$. Since $h$ is interpreted as the conformal dimension of operators $O_h$ around the given conformal solution \cite{Polchinski:2016xgd, Heydeman:2022lse}, we are effectively finding the spectrum of operator dimensions. The modes in $K^\rB$ correspond to bosonic operators, while those in $K^\rF$ or $\overline{K}{}^\rF$ to fermionic operators. In addition, it turns out that $\overline{K}{}^\rF$ is identical to $K^\rF$ as a matrix, so it is sufficient to consider $K^\rF$ only, keeping in mind that the fermionic spectrum is doubled. Since the solution we expand around is supersymmetric, the spectrum is organized into multiplets, as we will verify. In the following figures, the graphs of $k(h)-1$ for $K^\rB$ are displayed in blue while those for $K^\rF$ in black. We now present our results for a few representative cases.

\begin{figure}
\includegraphics[width=\columnwidth]{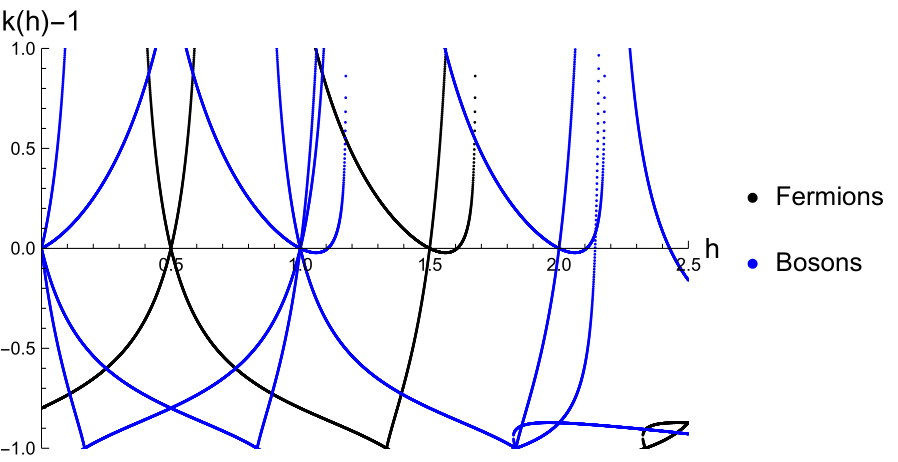}
\caption{\label{fig: 331 spectrum}%
Case $p=q=3$, $\alpha=1$, $\cE=1/3$. Shifted eigenvalues $k(h)-1$ of $K^\rB$ (blue) and $K^\rF$ (black) as the operator dimension $h$ varies. There is an operator $O_h$ in the spectrum whenever there is a horizontal intercept and $k(h)=1$. The total number of fermionic operators is doubled, due to the presence of $\overline{K}{}^\rF$ as well.}
\end{figure}

\paragraph{Case \tps{$p=q=3$}{p=q=3}.} For simplicity, we consider supersymmetric solutions with $\alpha=1$, so that there is a unique solution given in \eqref{p eq q Delta sol}. We also fix $\cE = 1/3$ for definiteness. We read off the spectrum from Fig.~\ref{fig: 331 spectrum}. Naively one would infer the presence of two $\cN=2$ Schwarzian multiplets at $h = \bigl( 1, 2\times\frac{3}{2}, 2 \bigr)$, however one of the two multiplets is spurious, due to an emergent IR reparametrization symmetry which is however incompatible with the UV boundary conditions \cite{Fu:2016vas, Heydeman:2022lse}. There appear also two $\cN=2$ current multiplets at $h = \bigl( 2 \times \frac12, 1 \bigr)$, but one is again spurious for the same reason. Lastly, we notice the presence of a multiplet at $h \simeq \bigl( 1.11, 2 \times 1.61, 2.11 \bigr)$. As pointed out in \cite{Maldacena:2016upp} and shown in \cite{Milekhin:2021cou, Milekhin:2021sqd}, the contribution of a bosonic operator $O_h$ with $1< h<\frac32$ to the free energy is proportional to $\beta^{2-2h}$, which dominates over the $\beta^{-1}$ contribution from the Schwarzian at low temperatures. This implies that the IR physics in this solution should be non-universal and not described by the Schwarzian theory at leading order. We shall refer to the range $1<h<\frac{3}{2}$ as the ``dangerous region". Fermionic operators are excluded from this discussion since only bosonic operators can enter the Lagrangian as deformations.

\begin{figure}
\includegraphics[width=\columnwidth]{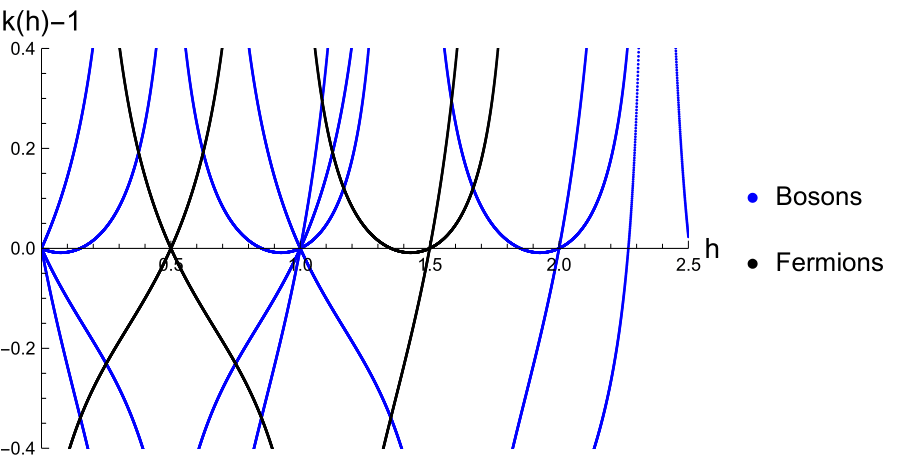}
\\
\vspace{0.6em}
\includegraphics[width=\columnwidth]{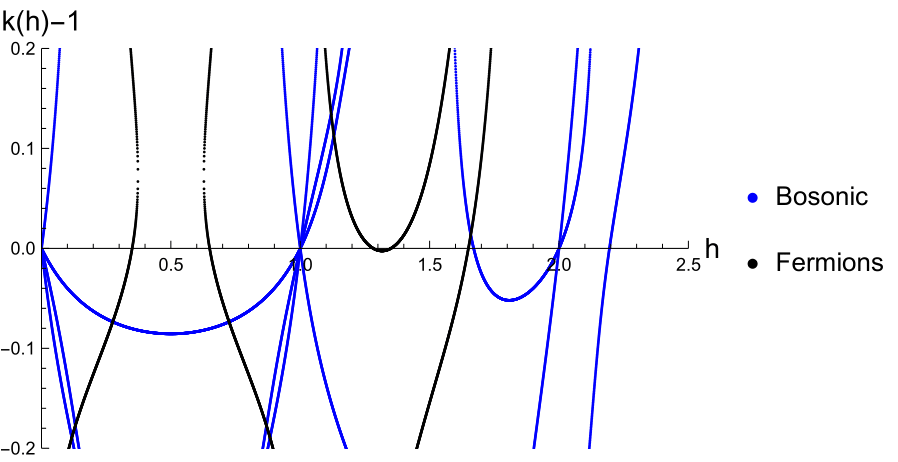}
\caption{\label{fig: 12 spectrum}%
Case $p=1$, $q=2$, $\alpha = 2/3$. Shifted eigenvalues $k(h)-1$ of $K^\rB$ (blue) and $K^\rF$ (black).
Top: supersymmetric case with $\cE_\eta=2\cE$, $\cE_\phi=-\cE$, and $\cE=0.05 < \cE_* \simeq 0.21$. Bottom: non-supersymmetric case with $\cE_\eta=0.10$ and $\cE_\phi=0.05$.}
\end{figure}

\paragraph{Case \tps{$p=1$}{p=1}, \tps{$q=2$}{q=2}.} In the supersymmetric case the unique solution is given in \eqref{p1 q2 sol} and \eqref{p1 q2 coeff sol}, which is valid for $0<\cE<\cE_*$. For concreteness, we fix $\alpha = 2/3$ so that $\cE_* \simeq 0.21$, and $\cE=0.05$.
Looking at the top of Fig.~\ref{fig: 12 spectrum}, there appear two $\cN=2$ Schwarzian multiplets at $h = \bigl(1, 2 \times \frac32, 2 \bigr)$ and two $\cN=2$ current multiplets at $h = \bigl( 2 \times \frac12, 1 \bigr)$. Of these, only one Schwarzian multiplet and one current multiplet is physical while the other copy is spurious, as explained above. In addition, we notice a multiplet with $h \simeq \bigl( 0.84, 2 \times 1.34, 1.84 \bigr)$. The presence of a bosonic operator $O_h$ with $h\simeq0.84<1$ is not a cause for concern because this is a relevant deformation that can be tuned to zero by appropriately choosing the UV parameters. Therefore, we expect the IR physics of this solution to be dominated by the Schwarzian to leading order in $1/\beta$, and this is confirmed by the numerical results in Section~\ref{sec:num-Q-and-S}. In the non-supersymmetric case we take $\cE_\eta=0.1$ and $\cE_\phi=0.05$ (they do not satisfy \eqref{susy constr dim eps}) and the solution is obtained by solving (\ref{matching coeff in alg}) numerically. Looking at the bottom of Fig.~\ref{fig: 12 spectrum}, we do not find a multiplet structure anymore. The presence of two Schwarzian modes ($h=2$) and four current modes ($h=1$) is still required by the symmetries of the theory --- and half of these modes are spurious. We also discern two bosonic modes with $h \simeq 1.67, 2.20$ outside the dangerous region. As expected, the fermionic modes split and move away from $h = \frac12, \frac32$; there are in fact modes with $h \simeq 0.35, 0.65, 1.28, 1.35, 1.65$, all with multiplicity two.

\begin{figure}
\includegraphics[width=\columnwidth]{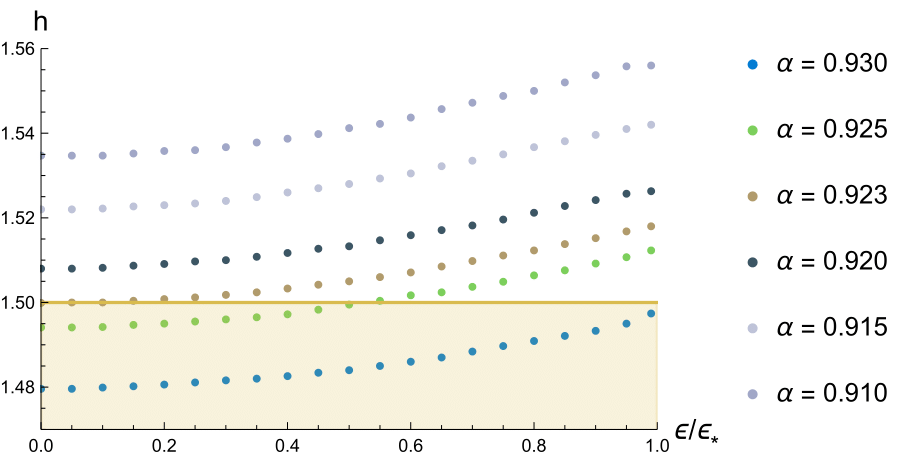} \\[.3em]
\includegraphics[width=\columnwidth]{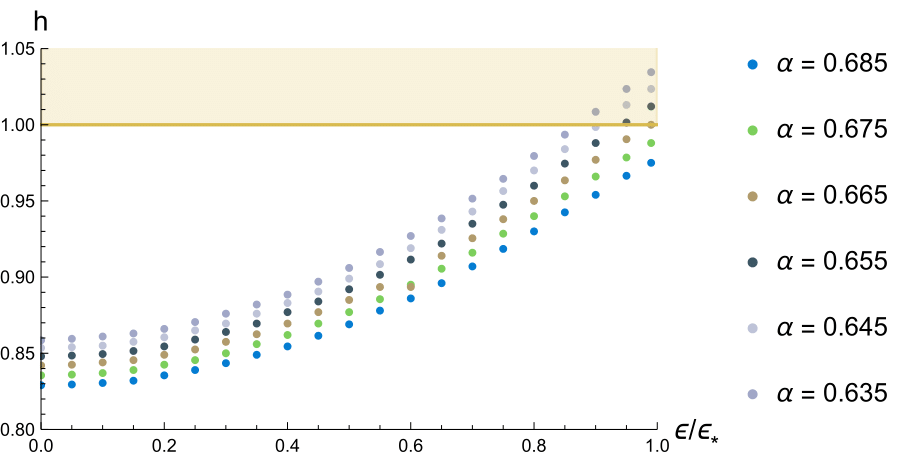}
\caption{\label{fig: 12 dangerous modes}%
Case $p=1$, $q=2$. Bosonic operator dimensions closest to the dangerous region $1 < h < \frac32$, as $\alpha$ and $\cE$ are varied. Different values of $\alpha$ are color coded. For $\alpha$ in the range $0.665 \leq \alpha \leq 0.923$, no dangerous modes are present in the spectrum.}
\end{figure}

Since the Schwarzian mode can be the dominant one in the IR or not depending on the parameters $p,q,\alpha, \cE$, we study in which ranges the two behaviors take place. As we will see, the spectrum is free of dangerous modes for $p=1$ and $0.665 \leq \alpha \leq0.923$, but not otherwise. It is reasonable to expect a qualitative difference between the models with $p=1$ and $p>1$, since the former have a scalar potential while the latter always contain fermions in their interaction terms.

In Fig.~\ref{fig: 12 dangerous modes} we consider the case $p=1$, $q=2$.
We plot the dimensions of the bosonic operators which lie closest to the dangerous region $1<h<\frac32$, as a function of $\cE$. This is repeated for various values of $\alpha$. We observe that when the abundance parameter $\alpha$ is bigger than a critical value $\alpha_\text{max} \simeq 0.923$, dangerous modes appear at small values of $\cE$. In this regime, we expect the IR physics not to be captured by the Schwarzian action, although we have not been able to observe this phenomenon numerically. Similarly, when $\alpha$ is smaller than a critical value $\alpha_\text{min} \simeq 0.665$, dangerous modes appear for large values of $\cE$.

\begin{figure}
\includegraphics[width=\columnwidth]{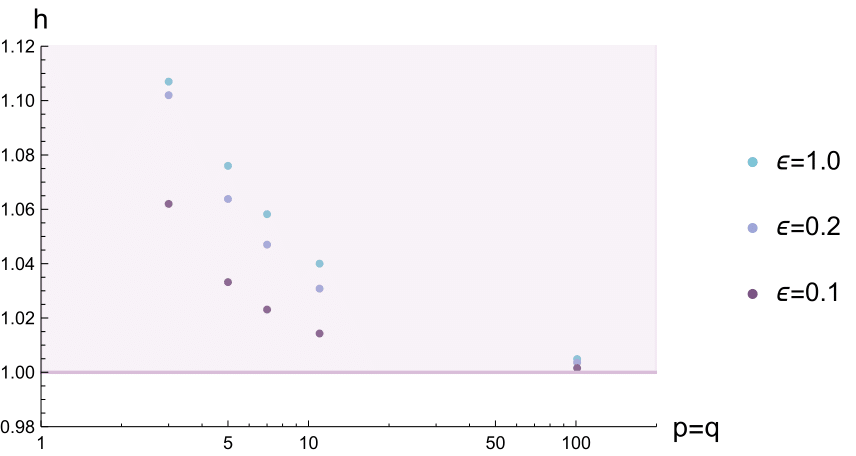}
\caption{\label{fig: pp dangerous modes}%
Cases with $p=q>1$ and $\alpha=1$. Bosonic operator dimensions in the dangerous region $1 < h < \frac32$ as $p=q$ are varied. Differently colored points correspond to different values of $\cE$.}
\end{figure}

In Fig.~\ref{fig: pp dangerous modes} we study the bosonic operators close to the dangerous region in the case $p=q>1$ and with $\alpha=1$. As $p=q$ is increased, one observes that the operator dimensions tend to $1$ from above, but are always within the dangerous region. This remains qualitatively unchanged for various values of $\cE$, therefore we expect the IR physics not to be dominated by the Schwarzian mode for $p>1$.

\begin{figure}
\includegraphics[width=\columnwidth]{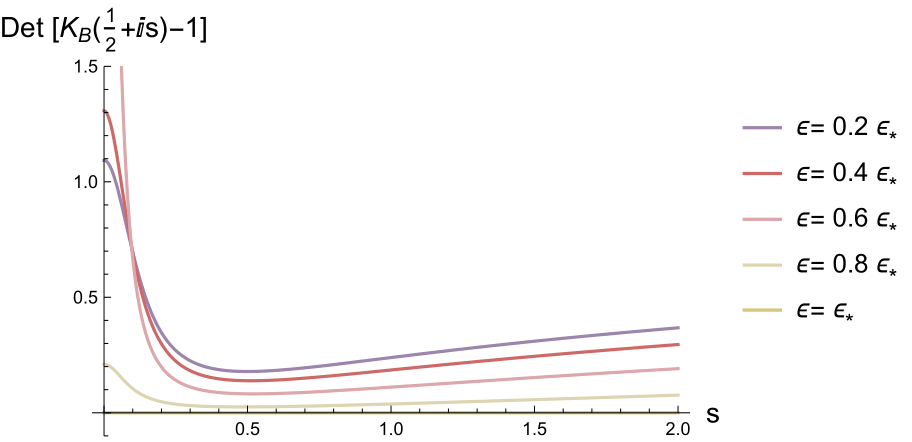}
\caption{\label{fig: 12 complex modes}%
Plot of $\det(K^\rB - \unit)$ evaluated at $h = \frac12 + is$. 
We fix $p=1$, $q=2$, $\alpha = 2/3$ and vary $\cE$ up to $\cE_*$.}
\end{figure}

Since reality of the conformal Casimir $h(h-1) $ does not exclude complex values $h = \frac12 + is$ with $s \in \bR$, the reality of the operator dimensions is a nontrivial consistency check of the spectrum. In Fig.~\ref{fig: 12 complex modes} we plot $\det(K^\rB - \unit)$ evaluated at $h = \frac12 + is$ against $s$ for $p=1$, $q=2$, $\alpha=\frac{2}{3}$ and various values of $\cE$. An operator with complex dimension can exist only if $\det(K^\rB - \unit) = 0$. For $\cE < \cE_*$ it is clear that $\det( K^\rB - \unit)$ is strictly positive and there are no such operators. As $\cE\rightarrow\cE_*$, the graph gets increasingly flatter and closer to the horizontal axis.
This phenomenon was also present at a phase transition of the $\cN=2$ model of \cite{Fu:2016vas}, as discussed in \cite{Heydeman:2022lse}.

\section{Chaos exponents}
\label{sec: chaos exp}

An interesting feature to study, both in quantum many-body systems and in toy models of quantum gravity, is the quantum chaotic behavior. From the gravity side, it has long been expected that black holes are maximally chaotic, see \eg{} \cite{Sekino:2008he, Shenker:2013pqa, Hayden:2007cs, Maldacena:2015waa}, a feature that is captured by the standard SYK model \cite{Maldacena:2016hyu}. In this section we compute the chaos exponents of various out-of-time-order correlators (OTOCs) in our models, using the retarded kernel approach of \cite{Murugan:2017eto}. Working in components, we first introduce the various component 4-point functions and the integral equations that they satisfy, which follow from the superspace integral equation \eqref{4 pt fn SD}. We then define the OTOCs as analytic continuations of the 4-point functions and find the continued versions of the integral equations. Finally, analyzing the equations at late times, we determine the chaos exponents. It turns out that one of these exponents saturates the maximal chaos bound $2\pi/\beta$ of \cite{Maldacena:2015waa}.

\subsection{Integral equations for 4-point functions}

In components, eqn.~\eqref{4 pt fn SD} decomposes into 3 equalities: one for the $4\times4$ matrix of bosonic 4-point functions:
\be
\label{LO bos 4pt fn}
F^\rB(\tau_1,\tau_2;\tau_3,\tau_4) \equiv \lim_{N\to\infty} \Bigl\langle \delta G_{\bar{A}A} \, \delta G_{\bar{B}B} \Bigr\rangle_{4\times4} (\tau_1,\tau_2;\tau_3,\tau_4)
\ee
with $A, B \in \{\eta, f, \phi, \psi\}$ and computed from the 2-point functions of bosonic bilocal fluctuations $\delta G_{\bar{A}A}$, and two for the $2\times 2$ matrices of fermionic 4-point functions:
\bea
\label{LO fer 4pt fn}
F^\rF &\equiv \lim_{N\to\infty} \left\langle \begin{matrix}
\delta G_{\bar{\eta}f} \delta G_{\bar{f}\eta} & \delta G_{\bar{\eta}f} \delta G_{\bar{\psi}\phi} \\
\delta G_{\bar{\phi}\psi} \delta G_{\bar{f}\eta} & \delta G_{\bar{\phi}\psi} \delta G_{\bar{\psi}\phi}
\end{matrix} \right\rangle \,, \\
\overline{F}{}^\rF &\equiv \lim_{N\to\infty} \left\langle \begin{matrix}
\delta G_{\bar{f}\eta} \delta G_{\bar{\eta}f} & \delta G_{\bar{f}\eta} \delta G_{\bar{\phi}\psi}\\
\delta G_{\bar{\psi}\phi} \delta G_{\bar{\eta}f} & \delta G_{\bar{\psi}\phi} \delta G_{\bar{\phi}\psi}
\end{matrix} \right\rangle \,,
\eea
computed from the 2-point functions of fermionic bilocal fluctuations $\delta G_{\bar{A}B}$. As in the previous section, this decomposition comes about because the kernel $\cK$ does not mix bosonic and fermionic components. Specifically, \eqref{4 pt fn SD} implies that $F^\rB$, $F^\rF$, $\overline{F}{}^\rF$ satisfy the equations:
\begin{align}
\label{4 point via kernel}
F^\rB &= \frac{1}{1-K^\rB} \cdot F^\rB_0 = \sum\nolimits_{n=0}^\infty \bigl( K^\rB \bigr)^n \cdot F_0^\rB \,, \\
F^\rF &= \frac{1}{1-K^\rF} \cdot F_0^\rF = \sum\nolimits_{n=0}^\infty \bigl( K^\rF \bigr)^n \cdot F_0^\rF \,, \nn \\
\overline{F}{}^\rF &= \frac{1}{1 - \overline{K}{}^\rF} \cdot \bigl( F_0^\rF \bigr)^\sT = \sum\nolimits_{n=0}^\infty \bigl( \overline{K}{}^\rF \bigr)^n \cdot \bigl( F_0^\rF \bigr)^\sT \,, \nn
\end{align}
where $\cdot$ indicates the convolution $f\cdot g \, (\tau_1,\tau_2;\tau_3,\tau_4) = \int\! d\tau_5 d\tau_6 \, f(\tau_1,\tau_2; \tau_5,\tau_6) \, g(\tau_5,\tau_6; \tau_3,\tau_4)$ together with the standard matrix product, while $\sT$ denotes both a matrix transpose and the swapping of arguments from $(\tau_1,\tau_2; \tau_3,\tau_4)$ to $(\tau_3,\tau_4; \tau_1,\tau_2)$. Lastly, $F_0^\rB$ and $F_0^\rF$ are the diagonal matrices
\begin{widetext}
\bea
F_0^\rB(\tau_1,\tau_2; \tau_3,\tau_4) &= \diag \Bigl\{ - G_{\bar\eta \eta}^* (\tau_1,\tau_4) \, G_{\bar\eta \eta}^*(\tau_3,\tau_2) \,,\, G_{\bar{f}f}^* (\tau_1,\tau_4) \, G_{\bar{f}f}^* (\tau_3,\tau_2) \,, \\
& \hspace{1.40cm}
\tfrac{1}{\alpha} \, G_{\bar\phi \phi}^* (\tau_1,\tau_4) \, G_{\bar\phi \phi}^* (\tau_3,\tau_2) \,,\, - \tfrac{1}{\alpha} \, G_{\bar\psi \psi}^* (\tau_1,\tau_4) \, G_{\bar\psi \psi}^* (\tau_3,\tau_2) \Bigr\} \,, \\
F_0^\rF(\tau_1,\tau_2;\tau_3,\tau_4) &= \diag \Bigl\{ G^*_{\bar{f}f} (\tau_3,\tau_2) \, G^*_{\bar\eta \eta} (\tau_1,\tau_4) \,,\, - \tfrac{1}{\alpha} \, G_{\bar\psi \psi}^* (\tau_3,\tau_2) \, G_{\bar\phi \phi}^* (\tau_1,\tau_4) \Bigr\} \,.
\eea
Equivalently, \eqref{4 point via kernel} can be written as the integral equations 
\be
\label{integral eqns F^BF}
F(\tau_1,\tau_2; \tau_3,\tau_4) - \int\! d\tau_5 \, d\tau_6 \, K(\tau_1,\tau_2; \tau_5,\tau_6) \, F(\tau_5,\tau_6; \tau_3,\tau_4) = F_0(\tau_1,\tau_2; \tau_3,\tau_4) \;,
\ee
\end{widetext}
written in terms of $\{ F^\rB, K^\rB, F_0^\rB \}$, $\{ F^\rF, K^\rF, F_0^\rF \}$ and $\{ \wb{F}{}^\rF, \wb{K}{}^\rF, (F_0^\rF)^\sT \}$, respectively.
Each component of these matrix equations is represented by a Feynman diagram of the form in Fig.~\ref{schematic Feynman integral eqn}.

\begin{figure*}[t]
\begin{tikzpicture}
	\draw [shorten < = 0.3cm, shorten > = 0.3cm] (-1, 0.7) node {$\bar{A}$} to (1, -0.7) node {$C$};
	\draw [shorten < = 0.3cm, shorten > = 0.3cm] (-1, -0.7) node {$B$} to (1, 0.7) node {$\bar{D}$};
	\fill [white] (0,0) circle [radius = 0.45];
	\filldraw [pattern = north west lines] (0,0) circle [radius = 0.45] node [fill = white, inner sep = 1pt] {$F$};
\node at (2, 0) {$-$};
\begin{scope}[shift = {(4,0)}]
	\draw [shorten < = 0.3cm] (-1, 0.7) node {$\bar{A}$} to (0, 0.7);
	\draw [shorten < = 0.3cm] (-1, -0.7) node {$B$} to (0, -0.7);
	\draw [shorten > = 0.3cm] (0, 0.7) arc [y radius = 1.4, x radius = 1, start angle = 90, end angle = 0] node {$P$};
	\draw [shorten > = 0.3cm] (0, -0.7) arc [y radius = 1.4, x radius = 1, start angle = -90, end angle = 0] node {$\bar{Q}$};
	\filldraw [pattern = crosshatch dots] (0, 0.7) to[out = -140, in = 140] (0, -0.7) to[out = 40, in = -40] (0,0.7) to cycle;
	\node [fill = white, inner sep = 1pt] at (0,0) {$K$};
\end{scope}
\node at (5.5, 0) {$\cdot$};
\begin{scope}[shift = {(7,0)}]
	\draw [shorten < = 0.3cm, shorten > = 0.3cm] (-1, 0.7) node {$\bar{Q}$} to (1, -0.7) node {$C$};
	\draw [shorten < = 0.3cm, shorten > = 0.3cm] (-1, -0.7) node {$P$} to (1, 0.7) node {$\bar{D}$};
	\fill [white] (0,0) circle [radius = 0.45];
	\filldraw [pattern = north west lines] (0,0) circle [radius = 0.45] node [fill = white, inner sep = 1pt] {$F$};
\end{scope}
\node at (9, 0) {$=$};
\begin{scope}[shift = {(11,0)}]
	\draw [shorten < = 0.3cm, shorten > = 0.3cm] (-1, 0.7) node {$\bar{A}$} to (1, 0.7) node {$\bar{D}$};
	\draw [shorten < = 0.3cm, shorten > = 0.3cm] (-1, -0.7) node {$B$} to (1, -0.7) node {$C$};
	\node at (0,0) {$F_0$};
\end{scope}
\end{tikzpicture}
\caption{\label{schematic Feynman integral eqn}%
Feynman diagram representation of the integral equations (\ref{integral eqns F^BF}).}
\end{figure*}

\subsection{OTOCs and chaos exponents}

The OTOCs we want to compute are double commutators such as 
\begin{widetext}
\begin{align}
\label{double comm def 1}
& \tfrac{1}{N^2} \Tr \bigl( \,e^{-\beta H}\, \bigl[ \, \wb{C}_a \bigl( \tfrac{\beta}{2} \bigr) , B_b \bigl( \tfrac{\beta}{2}+it_2 \bigr) \bigr] \,\bigl[ \, \wb{A}_b (it_1), D_a(0) \,\bigr] \, \bigr) = {} \\
&\quad = \tfrac{1}{N^2} \lim_{\epsilon\rightarrow 0^+} \bigl\langle \bigl( \overline{C}_a \bigl( \tfrac{\beta}{2} + \epsilon \bigr) - \overline{C}_a \bigl( \tfrac{\beta}{2} - \epsilon \bigr) \bigr) \, B_b \bigl( \tfrac{\beta}{2} + it_2 \bigr) \,
\overline{A}_b(it_1) \, \bigl( D_a(-\epsilon) - D_a(\epsilon) \bigr) \bigr\rangle_\beta \nn \\
&\quad = (-1)^{\xi_1[\bar A,B,\bar C]} 
\lim_{\epsilon\rightarrow 0^+} \bigl\langle G_{\bar{A}B} G_{\bar{C}D} \bigl( it_1, \tfrac{\beta}{2} + it_2; \tfrac{\beta}{2} + \epsilon, -\epsilon \bigr) 
- G_{\bar{A}B} G_{\bar{C}D} \bigl( it_1, \tfrac{\beta}{2} + it_2; \tfrac{\beta}{2} + \epsilon, \epsilon \bigr) \nn \\
&\hspace{10.4em} + G_{\bar{A}B} G_{\bar{C}D} \bigl( it_1, \tfrac{\beta}{2} + it_2; \tfrac{\beta}{2}
 - \epsilon, \epsilon \bigr) 
 - G_{\bar{A}B} G_{\bar{C}D} \bigl( it_1, \tfrac{\beta}{2} + it_2; \tfrac{\beta}{2} - \epsilon, -\epsilon \bigr) \bigr\rangle \,. \nn
\end{align}
\end{widetext}
Here $A$ is from the same multiplet as $B$, and $C$ is from the same multiplet as $D$, while the indices $a,b$ are summed over. If both $A$ and $D$ (or $B$ and $C$) are fermionic, the bracket $[ \; , \, ]$ between them should be understood as an anti-commutator. In the second line, the shifts of Euclidean time by $\pm\epsilon$ implement the operator ordering. We also defined the parity operator $\xi_1[\bar A,B,\bar C] \equiv F[\bar A] \, F[B] + F[\bar C] \, F[\bar A B]$, where $F[\cdot]$ is the fermion number of a field, set to 0 for bosons and 1 for fermions. The OTOC is constructed so that it is computed by expectation values of bilocal fields in the third line, which are the only accessible observables after the disorder average. In addition, it is especially convenient to consider double commutators rather than a more general OTOC since we will see that the locations of operator insertions in the contributing diagrams are heavily constrained \cite{Murugan:2017eto}. As seen from \eqref{4-pt saddle pt}, such 4-point functions are computed  at order $\cO(1/N)$ by the expectation values $\langle \delta G_{\bar{A}B} \, \delta G_{\bar{C}D} \rangle$ of fluctuations around the conformal solutions, which are grouped into the matrices $F^\rB$ and $F^\rF$ in \eqref{LO bos 4pt fn} and \eqref{LO fer 4pt fn}. The double commutators in \eqref{double comm def 1} are therefore computed at this order by the following analytic continuations of $F^\rB$ or $F^\rF$:  
\begin{multline}
\label{LO bos double comm 1}
W(t_1,t_2) \equiv \lim_{\epsilon\rightarrow 0^+} \Bigl[ F \bigl( it_1, \tfrac{\beta}{2} + it_2; \tfrac{\beta}{2} + \epsilon, -\epsilon \bigr) \\
- F \bigl( it_1, \tfrac{\beta}{2} + it_2; \tfrac{\beta}{2} + \epsilon, \epsilon \bigr) + F \bigl( it_1, \tfrac{\beta}{2} + it_2; \tfrac{\beta}{2} - \epsilon, \epsilon \bigr) \\
- F \bigl( it_1, \tfrac{\beta}{2} + it_2; \tfrac{\beta}{2} - \epsilon, -\epsilon \bigr) \Bigr] \,.
\end{multline}
From here on, we will often omit the superscripts $\rB$ and $\rF$, since most of the steps are completely independent from them. We shall also consider another inequivalent double commutator where $\bar{A}$ and $B$ are swapped with respect to \eqref{double comm def 1}: 
\begin{align}
& \tfrac{1}{N^2} \Tr \bigl( e^{-\beta H} \, \bigl[ \overline{C}_a \bigl( \tfrac{\beta}{2} \bigr), \overline{A}_b \bigl( \tfrac{\beta}{2} + it_1 \bigr) \bigr] \, \bigl[ B_b(it_2), D_a(0) \bigr] \bigr) \nn \\
&= (-1)^{\xi_2[\bar A,B,\bar C]} \lim_{\epsilon\rightarrow 0^+} \Bigl\langle G_{\bar{A}B} G_{\bar{C}D} \bigl( \tfrac{\beta}{2} + it_1, it_2; \tfrac{\beta}{2} + \epsilon, -\epsilon \bigr) \nn \\
&\quad - G_{\bar{A}B} G_{\bar{C}D} \bigl( \tfrac{\beta}{2} + it_1, it_2; \tfrac{\beta}{2} + \epsilon, \epsilon \bigr) \nn \\
&\quad + G_{\bar{A}B} G_{\bar{C}D} \bigl( \tfrac{\beta}{2} + it_1, it_2; \tfrac{\beta}{2} - \epsilon, \epsilon \bigr) \nn \\
&\quad - G_{\bar{A}B} G_{\bar{C}D} \bigl( \tfrac{\beta}{2} + it_1, it_2; \tfrac{\beta}{2} - \epsilon, -\epsilon \bigr) \Bigr\rangle \,,
\label{double comm def 2}
\end{align}
where we defined $\xi_2[\bar A,B,\bar C] \equiv F[\bar C] \, F[\bar A B]$. Analogously, it is computed at $\cO(1/N)$ by different analytic continuations of $F^\rB$ or $F^\rF$:  
\begin{multline}
\label{LO bos double comm 2}
\wt W(t_1,t_2) \equiv
\lim_{\epsilon\rightarrow 0^+} \Bigl[ F \bigl( \tfrac{\beta}{2} + it_1, it_2; \tfrac{\beta}{2} + \epsilon, -\epsilon \bigr) \\
- F \bigl( \tfrac{\beta}{2} + it_1, it_2; \tfrac{\beta}{2} + \epsilon, \epsilon \bigr) + F \bigl( \tfrac{\beta}{2} + it_1, it_2; \tfrac{\beta}{2} - \epsilon, \epsilon \bigr) \\
- F \bigl( \tfrac{\beta}{2} + it_1, it_2; \tfrac{\beta}{2} - \epsilon, -\epsilon \bigr) \Bigr] \,.
\end{multline}

According to one measure of quantum chaos, the order $\cO(1/N)$ contributions to the double commutators of a chaotic theory at late times $t_1=t_2=t$ should behave as 
\be
W^{\rB,\rF}(t,t) \,\propto\, \exp \bigl[ \lambda_L^{\rB, \rF} \, t \bigr] \;,
\ee
where $0 < \lambda_L^{\rB,\rF} \leq \frac{2\pi}{\beta}$ are the so-called Lyapunov exponents \cite{Murugan:2017eto}. The goal in the following is to compute these exponents using the retarded kernel approach.

\subsection{The retarded kernel}

\begin{figure}
\resizebox{0.95\columnwidth}{!}{\begin{tikzpicture}
	\draw [thick] (-4.5,0) -- (-2.5,0) to [out = 0, in = 180, looseness = 0.5] (-2, 2) to [out = 0, in = 180, looseness = 0.5] (-1.5, 0) -- (1.5,0) to [out = 0, in = 180, looseness = 0.5] (2, 2) to [out = 0, in = 180, looseness = 0.5] (2.5, 0) -- (4.5, 0);
	\filldraw (-2.5,0) circle [radius = 0.07] node [above left] {$-\epsilon$};
	\filldraw [red!80!black] (-1.5,0) circle [radius = 0.07] node [above right, black] {$\epsilon$};
	\filldraw [red!80!black] (1.5,0) circle [radius = 0.07] node [above left, black] {$\tfrac\beta2 - \epsilon$};
	\filldraw (2.5,0) circle [radius = 0.07] node [above right] {$\tfrac\beta2 + \epsilon$};
	\filldraw (-2,2) circle [radius = 0.07] node [left, shift = {(-0.15,0)}] {$i t_1$};
	\filldraw (2,2) circle [radius = 0.07] node [left, shift = {(-0.15,0)}] {$\tfrac\beta2 + i t_2$};
	\draw (4, 2.2) -- (4, 1.8) -- (4.4, 1.8);
	\node at (4.2, 2) {$\tau$};
\end{tikzpicture}}
\caption{\label{dble comm contour}%
Contour in the complex $\tau$ plane used for the evaluation of (\ref{integral eqns F^BF}) and \eqref{double comm def 1}. The four terms in the double commutator correspond to the choices of placing an operator at $\pm\epsilon$, and another one at $\frac{\beta}{2}\pm\epsilon$.}
\end{figure}
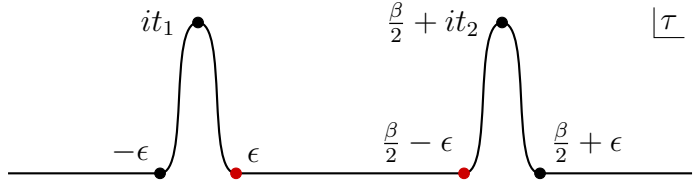

The strategy in the retarded-kernel approach is to use analytically continued versions of \eqref{integral eqns F^BF} to constrain $W$ and $\wt W$. Looking at the definitions \eqref{LO bos double comm 1} and \eqref{LO bos double comm 2}, it is apparent that one should sum four copies of \eqref{integral eqns F^BF} with $\tau_3$ and $\tau_4$ appropriately shifted by $\pm\epsilon$. We first consider $\tau_1=it_1$ and $\tau_2=\frac{\beta}{2}+it_2$, so that $W$ is obtained from the first term of \eqref{integral eqns F^BF}:
\begin{align}
& W(t_1,t_2) - \lim_{\epsilon\rightarrow 0^+} \int\! d\tau_3 \, d\tau_4 \; K \bigl( it_1, \tfrac{\beta}{2} + it_2; \tau_3, \tau_4 \bigr) \times {} \nn \\
\label{integral eqn W init}
& \times \Bigl[ F \bigl( \tau_3, \tau_4; \tfrac{\beta}{2} + \epsilon, -\epsilon \bigr) - F \bigl( \tau_3, \tau_4; \tfrac{\beta}{2} + \epsilon, \epsilon \bigr) \\
& + F \bigl( \tau_3, \tau_4; \tfrac{\beta}{2} - \epsilon, \epsilon \bigr) - F \bigl( \tau_3, \tau_4; \tfrac{\beta}{2} - \epsilon, -\epsilon \bigr) \Bigr] = W_0 (t_1,t_2) , \nn
\end{align}
where $W_0(t_1, t_2)$ is defined as in (\ref{LO bos double comm 1}) but using $F_0$ in place of $F$.
We should determine the integration contour for $\tau_3, \tau_4$ in the second term of \eqref{integral eqns F^BF}, and it is useful to note that $\tau_3, \tau_4$ are the locations of interaction vertices in the Feynman diagrams drawn in Fig.~\ref{schematic Feynman integral eqn}. In the path integral, Feynman vertices are derived from local interaction terms in the action, which have the form $\int\! d\tau \, O(\tau)$. Integrals over the positions of the vertices come directly from the time integral in the action, therefore specifying the complex time contour that is used to compute the path integral also specifies the contour for the vertex positions. Since we are computing the double commutator \eqref{double comm def 1}, the contour must pass through all operator insertions in the double commutator. We shall choose the same contour as in \cite{Murugan:2017eto}, shown in Fig.~\ref{dble comm contour}. The folds that pass through $\tau_1$ and $\tau_2$ are called the left and right rails; each rail has a left and a right side. The four terms contributing to \eqref{LO bos double comm 1} correspond to different positions for the operator insertions at the bottom of the rails.  

A priori the interaction vertices at $\tau_3$ and $\tau_4$ could be placed at any point on the contour in Fig.~\ref{dble comm contour}, and the integrals should be performed over the whole contour. However, it turns out that nonzero contributions to the integrals only occur when there is one vertex on each rail. This is because if the left rail is free of vertices, the 4-point functions computed by $F \bigl( \tau_3, \tau_4; \frac\beta2 + \epsilon, \pm\epsilon \bigr)$ are equal as $\epsilon\rightarrow 0$, since there is no difference in operator ordering. Consequently, the four terms under the integral in \eqref{integral eqn W init} cancel. Similarly, if the right rail is free of vertices, the 4-point functions computed by $F \bigl( \tau_3, \tau_4; \frac\beta2 \pm \epsilon, \epsilon \bigr)$ are equal and the same cancellation occurs. Moreover, $\tau_3$ must be on the right rail and $\tau_4$ on the left rail, but not the other way around. To see this, we first note that the time dependence in every component of the kernels $K^{\rB,\rF}$ is given by a product of three conformal 2-point functions:
\begin{align}
\label{kernel cpt t dep}
& K_{\bar A B,\bar C D}(\tau_1,\tau_2;\tau_3,\tau_4) \;\propto\; \\
&\quad G^\beta_{s_1}(\tau_{14};\Delta_1,\cE_1) \, G^\beta_{s_2}(\tau_{32};\Delta_2,\cE_2) \, G^\beta_{s_3}(\tau_{34};\Delta_3,\cE_3) \,. \nn
\end{align}
Suppose instead that $\tau_3$ is on the left rail. Then $\tau_3 = \pm \tilde\epsilon + it_3$ for some $\tilde\epsilon\in (0,\epsilon)$, where the plus/minus sign corresponds to $\tau_3$ being on the right/left side of the rail, respectively. Since $G^\beta_{s_2} (\tau_{32}; \Delta_2, \cE_2) = G_{s_2}^\beta( - \frac\beta2 \pm \epsilon + it_{32}; \Delta_2, \cE_2)$ has a smooth limit as $\epsilon\rightarrow 0$, the integrand in \eqref{integral eqn W init} is equal for the two sides of the rail. However, the contour in Fig.~\ref{dble comm contour} runs from left to right with increasing $\re\tau$, and there is a relative sign in the integration measure $d\tau_3 = \pm i \, dt_3$ between the sides of the rail. Hence the contributions from the two sides cancel. 

On the remaining contour where contributions are nonzero, $\tau_3 = \frac\beta2 + s_4 \tilde\epsilon + it_3$ and $\tau_4 = s_5 \tilde\epsilon + it_4$ with $\tilde\epsilon \in (0,\epsilon)$. The signs $s_{4,5}=\pm 1$ parametrize whether $\tau_{3,4}$ are on the right or left side of the rails, and they should be summed over. The contour orientation depends on which side we are on, so $d\tau_{3,4} = - i \, s_{4,5} \, dt_{3,4}$. Substituting this contour and \eqref{kernel cpt t dep} into the integral \eqref{integral eqn W init}, every component is a sum of terms proportional to
\begin{widetext}
\begin{align}
\label{integral eqn cpt ret}
& \lim_{\epsilon \rightarrow 0^+} \sum_{s_4, s_5 = \pm 1} (-s_4s_5) \int_0^{t_2} \! dt_3 \int_0^{t_1} \! dt_4 \; G_1^\beta \bigl( - s_5 \tilde\epsilon + it_{14} \bigr) \, G_2^\beta \bigl( s_4 \tilde\epsilon + it_{32} \bigr) \, G_3^\beta \bigl( \tfrac{\beta}2 + it_{34} \bigr) \times {} \\
&\quad \times \Bigl[ F_{\bar C D} \bigl( \tfrac\beta2 + s_4 \tilde\epsilon + it_3, s_5 \tilde\epsilon + it_4; \tfrac{\beta}2 + \epsilon, -\epsilon \bigr) - F_{\bar C D}\bigl( \tfrac{\beta}2 + s_4 \tilde\epsilon + it_3, s_5 \tilde\epsilon + it_4; \tfrac{\beta}2 + \epsilon, \epsilon \bigr) \nn \\
&\hspace{2.3em} + F_{\bar C D} \bigl( \tfrac{\beta}2 + s_4 \tilde\epsilon + it_3, s_5 \tilde\epsilon + it_4; \tfrac{\beta}2 - \epsilon, \epsilon \bigr) - F_{\bar C D} \bigl( \tfrac{\beta}2 + s_4 \tilde\epsilon + it_3, s_5 \tilde\epsilon + it_4; \tfrac{\beta}2 - \epsilon, -\epsilon \bigr) \Bigr] \;, \nn
\end{align}
\end{widetext}
where $F_{\bar C D}$ is a component of $F$, and appropriate dependencies on the conformal ansatz parameters are implied for $G^\beta$. Since $\tilde\epsilon \in (0,\epsilon)$, then $\sign( \pm \epsilon - s_{4,5} \tilde\epsilon ) = \pm 1$ and the operator ordering in the four $F_{\bar C D}$ is unaffected if we send $\tilde\epsilon \rightarrow 0$ at fixed $\epsilon$. This simplifies \eqref{integral eqn cpt ret} to
\begin{multline}
\int_0^{t_2} \! dt_3 \int_0^{t_1} \! dt_4 \; G^\rR_{s_1} (t_{14}; \Delta_1, \cE_1) \; \overline{G^\rR_{s_2} (t_{23}; \Delta_2, \cE_2)} \times {} \\
{} \times G^\text{lr}_{s_3} (t_{34}; \Delta_3, \cE_3) \; \wt W_{\bar C D}(t_3,t_4) \,,
\end{multline}
where we defined
\begin{align}
& G_s^\rR (t; \Delta, \cE) \equiv \lim_{\epsilon \rightarrow 0^+} \sum_{s'=\pm 1} s' G_s^\beta (s' \epsilon + it; \Delta, \cE)  \\
&\qquad = \frac{ 2\pi^{2\Delta} \, e^{- i \pi \frac{1+s}{4} - \frac{ 2\pi i\cE}{\beta} t } }{ \bigl[ \beta \sinh \bigl( \frac{\pi t}{\beta} \bigr) \bigr]{}^{2\Delta} } \, \sin \Bigl( \pi \bigl( \Delta + \tfrac{1-s}{4} + i \cE \bigr) \Bigr) \nn
\end{align}
and
\be
\label{lr retarded def}
\hspace{-0.1em} G^\text{lr} (t; \Delta, \cE) \equiv G_s^\beta \bigl( \tfrac{\beta}2 + it; \Delta, \cE \bigr) = \frac{\pi^{2\Delta} \, e^{ - \frac{ 2\pi i\cE }{ \beta} t} }{ \bigl[ \beta \cosh \bigl( \frac{\pi t}{\beta} \bigr) \bigr]{}^{2\Delta} } .
\ee
Note that $G^\rR_s (-t; \Delta, \cE) = \overline{ G^\rR_s (t; \Delta, \cE)}$. In the literature, $G^\rR_s$ is sometimes called the retarded propagator while $G^\text{lr}$ is called the ladder-rung propagator, due to how they arise from propagators connecting the interaction vertices, which run horizontally between the vertical rails in Fig.~\ref{dble comm contour}.

Therefore, we have shown that \eqref{integral eqn W init} can be written as
\begin{multline}
\label{integral eqn W}
W(t_1, t_2) - \ts\int_0^{t_2} \! dt_3 \ts\int_0^{t_1} \! dt_4 \; K_\rR^+ (t_1, t_2; t_3, t_4) \; \wt W(t_3, t_4) \\
= W_0(t_1, t_2) \,,
\end{multline}
where the components of $K_\rR^{+}$ are obtained from those of $K$ through the replacements $G^\beta_1 (\tau_{14}) \, G^\beta_2 (\tau_{32}) \, G^\beta_3 (\tau_{34}) \rightarrow G^\rR_1 (t_{14}) \, \overline{G^\rR_2 (t_{23})} \, G^\text{lr}_3 (t_{34})$. On the other hand, one could choose $\tau_1 = \frac{\beta}2 + it_1$, $\tau_2 = it_2$ and sum over four copies of \eqref{integral eqns F^BF} to construct an analog of \eqref{integral eqn W init} in which the first term is $\wt W(t_1,t_2)$. Following the same steps as the above, one obtains an equation as (\ref{integral eqn W}) but in which $W \leftrightarrow \wt W$, $K_\rR^+ \to K_\rR^-$ and $W_0 \to \wt W_0$, where $\wt W_0$ is defined as in \eqref{LO bos double comm 2} but with $F$ replaced by $F_0$,
while the components of $K_\rR^-$ are obtained from those of $K$ via the replacement $G^\beta_1 (\tau_{14}) \, G^\beta_2 (\tau_{32}) \, G^\beta_3 (\tau_{34}) \rightarrow s_3 \, G^\rR_1(t_{14}) \, \overline{G^\rR_2 (t_{23})} \, G^\text{lr}_3 (t_{34})$.
Note that the only difference between the replacements to use in $K_\rR^+$ and $K_\rR^-$ is the sign $s_3$. Due to the mixing between $W$ and $\wt W$, it is convenient to group \eqref{integral eqn W} and its analog for $\wt W$ into a single equation, in which $W$ is substituted by a vector $\bigl( \begin{smallmatrix} W \\ \wt W \end{smallmatrix}\bigr)$ and similarly for $W_0$, while $K_\text{R}^+$ is substituted by the matrix $K_\rR = \Bigl( \begin{smallmatrix} 0 & K_\rR^+ \\ K_\rR^- & 0 \end{smallmatrix} \Bigr)$.
Assuming that $W$ and $\wt W$ are exponentially increasing at late times, the right hand side of such an equation is  in comparison exponentially suppressed and can be neglected. Furthermore, the lower limits of integration can be modified to $-\infty$ at this order of approximation, since the dominant contribution to the integral comes from $t_{3,4}$ close to $t_{1,2}$ \cite{Murugan:2017eto}.
The equation we are left with shows that if $W$ and $\wt W$ are chaotic, their columns are eigenvectors of the retarded kernel $K_\rR$ with eigenvalue $1$. We find it convenient to change variables to 
\be
\label{ret ker coord redef}
z_{1,4} = \exp\bigl( - \tfrac{2 \pi}\beta \, t_{1,4} \bigr) \,,\quad z_{2,3} = - \exp\bigl( - \tfrac{2\pi}\beta \, t_{2,3} \bigr) \,,
\ee
so that $\int_{-\infty}^{t_2} dt_3 \int_{-\infty}^{t_1} dt_4 = \int_{-\infty}^{z_2} dz_3 \int_{z_1}^\infty dz_4 \,  \frac{(\beta/2\pi)^2 }{ |z_3||z_4| }$, after which the eigenvalue equation for $(W, \wt W)(z_1, z_2)$ becomes
\be
\label{eigenvalue eqn W z coord}
\bigl(\! \begin{smallmatrix} W \\ \wt W \end{smallmatrix} \!\bigr) \!=\!\! \int_{-\infty}^{z_2} \!\hspace{-0.5em} dz_3 \!\int_{z_1}^\infty \!\hspace{-0.5em} dz_4 \, K_\rR(z_1,z_2; z_3,z_4) \, \bigl(\! \begin{smallmatrix} W \\ \wt W \end{smallmatrix} \!\bigr) (z_3,z_4) .
\ee
Notice that, with some abuse of notation, here we defined $K_\rR(z_1,z_2; z_3,z_4) = \frac{\beta^2}{4\pi^2 |z_3z_4|} K_\rR \bigl( t(z_1), t(z_2); t(z_3), t(z_4) \bigr)$.

\subsection{Exponents from the fermionic and bosonic retarded kernels}

We obtained the retarded kernel $K_\rR$ in the $t$ coordinates. After the adequate change of variables, the components of $K_\rR$ in the $z$ coordinates are retrievable. 
The goal is now to find eigenvectors of $K_\rR^\rB$ and $K_\rR^\rF$ with eigenvalue $1$, which form the columns of $\bigl( W^\rB, \wt W{}^\rB \bigr){}^\sT$ and  $\bigl( W^\rF, \wt W{}^\rF \bigr){}^\sT$, and extract their exponents at late times.
Each component $K_{\bar A B,\bar C D}^\rR (z_1,z_2; z_3,z_4)$ has the form 
\begin{multline}
K_{\bar{A}B, \bar{C}D}^\rR = \frac{ g_{L(A)} \bigl( \frac{2\pi}{\beta} \bigr){}^{\Delta_A + \Delta_B} |z_1|^{ \Delta_A + i\cE_A} |z_2|^{\Delta_B - i\cE_B} }{ g_{L(C)} \bigl( \frac{2\pi}{\beta} \bigr){}^{\Delta_C + \Delta_D} |z_3|^{\Delta_C + i\cE_C} |z_4|^{\Delta_D - i\cE_D} } \times {} \\
\times \frac{K}{ |z_{14}|^{2\Delta_A} |z_{23}|^{2\Delta_B} |z_{34}|^{ 2 - \Delta_A - \Delta_B - \Delta_C - \Delta_D} } \,,
\end{multline}
where $K$ is some $z$-independent proportionality constant that does not depend on the coefficients $g_{\eta,\phi}$. Note that the components $K_{\bar{A}B, \bar{C}D}^\rR$ are labelled such that $\bar{A}B$ is a row index and $\bar{C}D$ is a column index in the matrices $K^{\rB,\pm}_\rR$, $K^{\rF,\pm}_\rR$. In a column of $\bigl( W^\rB, \wt W^\rB \bigr){}^\sT$ or $\bigl( W^\rF, \wt W^\rF \bigr){}^\sT$ on which $K^{\rB,\pm}_\rR$ and $K^{\rF,\pm}_\rR$ act respectively, we naturally label the component that is being contracted with $K_{\bar{A}B, \bar{C}D}^\rR$ as $W_{\bar{C}D}$ or $\wt W_{\bar{C}D}$. It is then straightforward to check that the eigenfunctions of $K_{\bar{A}B, \bar{C}D}^\rR$ under convolution are
\be
\label{ret eigenfunctions}
w_{\bar{C}D} \equiv g_{L(C)} \, \bigl( \tfrac{2\pi}{\beta} \bigr)^{\Delta_C + \Delta_D} \,  \frac{ |z_3|^{\Delta_C + i\cE_C} \, |z_4|^{\Delta_D - i\cE_D} }{ |z_{34}|^{\Delta_C + \Delta_D - h} }
\ee
where $h$ is a free parameter that will be determined below by requiring that (\ref{eigenvalue eqn W z coord}) be satisfied. Indeed we have
\begin{align}
& \int_{-\infty}^{z_2} \!\! dz_3 \int_{z_1}^\infty \!\! dz_4 \; K_{\bar{A}B, \bar{C}D}^\rR (z_1,z_2; z_3,z_4) \; w_{\bar{C}D} (z_3,z_4) = {} \nn\\
&\! = K \, \tfrac{ \Gamma(1-2\Delta_A) \, \Gamma(1-2\Delta_B) \, \Gamma(\Delta_A + \Delta_B - h) }{ \Gamma(2 - \Delta_A - \Delta_B - h) } \; w_{\bar{A}B} (z_1,z_2) .
\label{retarded ker eigenfn}
\end{align}
We therefore take $W_{\bar{A}B} (z_1,z_2)$ to be proportional to $w_{\bar{A}B}$. At $t_1=t_2=t$, the time dependence in $w_{\bar{A}B}$ is
\be
w_{\bar{A}B} \Bigl( e^{-\frac{2\pi t}{\beta}} , - e^{-\frac{2\pi t}{\beta}} \Bigr) \,\propto\, \exp\biggl[ \frac{2\pi t}{\beta} \, (-h + i\cE_B - i\cE_A ) \biggr] .
\ee
The Lyapunov exponent is thus given by $\lambda = - \frac{2\pi}{\beta} \re h$ for values of $h$ such that $K_\rR$ has eigenvalue $1$. In particular, we must have $-1\leq\re h < 0$ for consistency with the assumption of exponential growth that was used to simplify the computation previously, and with the maximal chaos bound $\lambda\leq\frac{2\pi}{\beta}$. 

When acting on \eqref{retarded ker eigenfn}, $K_\rR$ is represented by an ordinary matrix. We compute the chaos exponents by finding the values of $h$ at which $\det \big( K_\rR^{\rB,\rF} - \unit \bigr) = 0$. Remarkably, $\det \bigl[ K_\rR^\rF( h {=} -1/2 ) - \unit_{4\times 4} \bigr] = \det \bigl[ K_\rR^\rB( h {=} -1 ) - \unit_{8\times 8} \bigr]=0$ holds regardless of the dimensions $\Delta_{\eta,\phi}$, as long as they satisfy the supersymmetry constraint $p \Delta_\eta + q \Delta_\phi = \frac12$. This shows that all superconformal solutions exhibit the maximal chaos exponent and half of it in the double commutators computed by $\bigl( W^\rB, \wt W^\rB \bigr)$ and by $\bigl( W^\rF, \wt W^\rF \bigr)$, respectively.

\begin{figure*}[t]
\includegraphics[width=\textwidth]{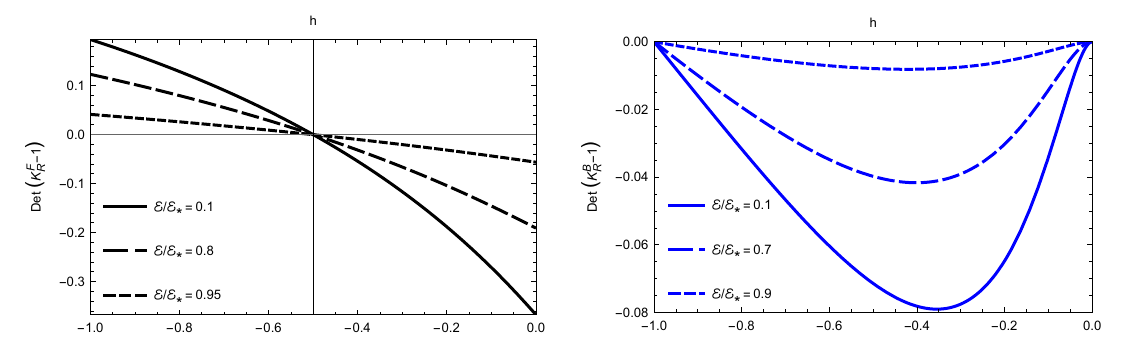}
\caption{\label{ret det(k-1) vs h}%
Plots of $\det \bigl(K_\rR^\rF - \unit_{4\times 4} \bigr)$ in black (left) and $\det \bigl( K_\rR^\rB - \unit_{8\times 8} \bigr)$ in blue (right) against $h$, for the solution \eqref{p1 q2 sol} at $\alpha = \frac{1}{2}$ and various values of $\cE$ as a fraction of the critical value $\cE_*=\frac{1}{\pi} \operatorname{arctanh} \bigl( 1/\sqrt{2} \bigr)$.}
\end{figure*}

In addition, for the analytic superconformal solutions found in Section~\ref{sec-existence} there are no other values of $h\in [-1,0)$ at which $\det \bigl( K_\rR^{\rB,\rF} - \unit \bigr) = 0$. For example, in Fig.~\ref{ret det(k-1) vs h} we plot $\det\bigl( K_\rR^\rF - \unit_{4\times 4} \bigr)$ in black on the left and $\det \bigl( K_\rR^\rB - \unit_{8\times 8} \bigr)$ in blue on the right against $h$, for the solution \eqref{p1 q2 sol} and various values of $\cE$. It is clear that there are no other zeros except at $h=-\frac{1}{2},-1$. The same qualitative behavior is found for other superconformal solutions. This suggests that for superconformal solutions there are no other chaos exponents except the maximal one and half of it.

\section{Discussion}
\label{sec: discussion}

For $p=1$, our model bears many similarities with that of \cite{Biggs:2023mfn}, although the abundance parameter and the coupling of the fermions make it irredeemably distinct. In the conformal phase of our model, the low-energy behavior is closer to the standard SYK and thus completely avoids the ``curious'' behavior of the model in \cite{Biggs:2023mfn}. Naively, even outside of the conformal phase, the models behave differently, with the dynamical bosonic correlator taking an exponential-like profile instead of a constant. However, it is possible that it realizes a \textit{charged} version of the phase identified in \cite{Biggs:2023mfn}. This is consistent with $\Delta_\phi\rightarrow 0$ at the transition point, and numerically we do not rule out a sub-Schwarzian mode which scales with $T^\gamma$ for non-integer $\gamma$. However, further studies would be required to check whether there is a meaningful connection.

Another interesting aspect of this model that could benefit from analytical understanding is the ``emergent'' supersymmetric behavior discussed in Section~\ref{sec: numerics}. It would be interesting, for example, to derive the relation \eqref{eq:nonsusy-epsilon} analytically. A perhaps more crucial question is to ascertain whether this is a physical phenomenon or simply an artefact of the Schwinger--Dyson equations. It is not clear, for example, if one can reconcile such an IR behavior and the arguments in Appendix~\ref{app: conformal ansatz with chem} which set $2\pi\cE = -\beta \mu$ by combining Ward identities and (anti-) periodicity. Could the chemical potentials be ``renormalized'' in the IR? Or should one consider these solutions to be simply outside of the physical Hilbert space? An actual implementation of the model could clear this up, as could some holographic argument.

The dependency of the Schwarzian coupling on the charge $\cQ$ is another point for which we found numerical results but with insufficient precision to be conclusive. Far from the critical value of $\mu$, the variation of the coupling with the charge seems to be very flat but probably not constant. This contrasts with some known black hole setups such as \cite{Boruch:2022tno} where the coupling varies with a rational power of the charge. Perhaps a specific gravitational dual to this model could enlighten this aspect, since in SYK-like models the Schwarzian coupling is notoriously difficult to fix from the quantum mechanical side alone. 

Another aspect about which the numerical analysis is not conclusive and would benefit from analytic control is whether a phase transition truly happens at the critical chemical potential. A possible alternative scenario would be a smooth behavior across the critical chemical potential, since at the level of solutions there is no obvious discontinuity. However, the fact that the solutions for large chemical potential approximate the exponentially-damped non-conformal ones suggests that, should there be a phase transition, it could correspond to the fields acquiring a mass. It would also be interesting to investigate whether this transition is related to replica-symmetry breaking: a breakdown of the conformal ansatz could instead signal a spin-glass phase which is not captured by our analysis.

Another property of this model that matches gravitational physics is maximal chaos. We find that with four identical operators, importantly the physical bosons and fermions, the OTOCs exhibit chaotic behavior with maximal Lyapunov exponent. Curiously, for some OTOCs constructed with a mix of bosonic and fermionic fields, we find instead an exponent equal to half of the chaos bound. We suspect that this is a consequence of supersymmetry and it would be interesting to derive it from the super-Schwarzian theory.

This leads to a final interesting challenge, which is whether there is a concrete and specific gravity dual to this model. In particular, both the inspiration from the quantum mechanical model of \cite{Benini:2022bwa} and the phases of solutions described in Section~\ref{sec-existence} suggest that the case $p=1$, $q=2$ is the most ``black-hole-like'', which is also the case in which we found most of the interesting numerical results.

Finally, it would be interesting to study the finite $N$ model numerically, before taking the approximation of the Schwinger--Dyson equation. This would require to truncate the bosonic Hamiltonian, or to consider a version with spin or hard bosons. This could confirm the absence of a spin-glass phase.

\begin{acknowledgments}
We are grateful to Ofer Aharony, Micha Berkooz, Ziming Ji, Yiyang Jia, Ohad Mamroud, Cheng Peng, Subir Sachdev, and Adrian Sanchez Garrido for very useful discussions. The authors are partially supported
by the ERC-COG grant NP-QFT No.~864583 ``Non-perturbative dynamics of quantum fields: from new deconfined phases of matter to quantum black holes'',
by the MUR-FARE grant EmGrav No.~R20E8NR3HX ``The Emergence of Quantum Gravity from Strong Coupling Dynamics'',
by the MUR-PRIN2022 grant No.~2022NY2MXY,
as well as by the INFN ``Iniziativa Specifica ST\&FI''.
ZZ is also supported by NSFC No.~12175237, the Fundamental Research Funds for the Central Universities, and funds from the Chinese Academy of Sciences.
Some of the numerical calculations were carried out with the Ulysses HPC cluster at SISSA.
\end{acknowledgments}

\appendix

\section{Superspace conventions}
\label{app: superspace}

In this appendix we collect our conventions about superspace and super-reparametrizations. In terms of superspace coordinates, the supercharges are
\be
\label{poincare supercharges}
Q = \partial_\theta - \tfrac{\overline{\theta}}{2} \, \partial_\tau \;,\qquad\qquad \overline{Q} = - \partial_{\overline{\theta}} + \tfrac{\theta}{2} \, \partial_\tau \;,
\ee
with $Q^2 = \wb Q{}^2 = 0$ and $\{Q, \wb Q\} = \partial_\tau$.
The chiral and anti-chiral superspace derivatives are
\be
\label{superspace derivatives}
D = \partial_\theta + \tfrac{\wb\theta}{2} \, \partial_\tau \,,\;\; \overline{D} = - \partial_{\overline{\theta}} - \tfrac{\theta}{2} \, \partial_\tau \,,\;\; \{D,\overline{D} \} = - \partial_\tau \,.
\ee
They satisfy $\{D,Q\} = \{D,\wb Q\} = \{\wb D,Q\} = \{\wb D, \wb Q\} = 0$.
Chiral superfields $\Phi$ and Fermi superfields $\cY$ (and they anti-chiral counterparts) have expansions 
\bea
\Phi &= \phi + \theta\psi + \tfrac{1}{2} \theta\wb{\theta} \, \partial_\tau\phi \,,\;& \wb{\Phi} &= \wb{\phi} - \wb{\theta} \,\wb{\psi} - \tfrac{1}{2} \theta \wb{\theta} \, \partial_\tau \wb{\phi} \,, \\
\cY &= \eta - \theta f + \tfrac{1}{2} \theta \wb\theta \, \partial_\tau\eta \,,\;& \wb{\cY} &= \wb{\eta} - \wb{\theta}\,\wb{f} - \tfrac{1}{2} \theta \wb{\theta} \, \partial_\tau \wb\eta \,.
\eea
They satisfy $\wb D \Phi = D \wb\Phi = \wb D \cY = D \wb\cY = 0$. In our conventions, conjugation reverses the order of fermions.
The integration measure over full superspace is written as $d^3T\equiv d\tau d\theta d\wb\theta$, while the measures over chiral and anti-chiral superspaces are $d^2T\equiv d\tau d\theta$ and $d^2\wb{T}\equiv d\tau d\wb\theta$, respectively. The measures over half superspace are fermionic: they anti-commute among themselves and with other fermionic operators like $D$, $\wb{D}$. The superspace delta functions
\begin{align}
\label{superspace delta def}
\delta^3(T_1-T_2) &\equiv (\wb\theta_1 - \wb\theta_2) \, (\theta_1-\theta_2)\, \delta(\tau_1-\tau_2) \;, \\ 
\delta^2(T_1,T_2) &\equiv \overline{D}_2 \, \delta^3(T_1-T_2) \nn\\
&= (\theta_1-\theta_2) \, \delta\bigl( \tau_1-\tau_2 + \tfrac12 \theta_1 \wb\theta_1 - \tfrac12 \theta_2 \wb\theta_2 \bigr) \;, \nn\\
\wb\delta{}^2(T_1,T_2) &\equiv D_2 \, \delta^3(T_1-T_2) \nn\\
&= ( \wb\theta_1 - \wb\theta_2 ) \, \delta\bigl( \tau_1-\tau_2 - \tfrac12 \theta_1 \wb\theta_1 + \tfrac12 \theta_2 \wb\theta_2 \bigr) \;, \nn
\end{align}
are defined to satisfy
\begin{align}
\ts\int\! d^3T_1 \; \delta^3(T_1-T_2) \, X(T_1) &= X(T_2) \,,& &\text{for $X$ generic}, \nn\\
\ts\int\! d^2T_1 \; \delta^2(T_1-T_2) \, \Phi(T_1) &= \Phi(T_2) \,,& &\text{for } \wb{D} \Phi = 0 \,, \nn \\
\ts\int\! d^2\wb T_1 \;\wb\delta{}^2(T_1-T_2) \, \overline{\Phi}(T_1) &= \overline{\Phi}(T_2) \,,& &\text{for } D\overline{\Phi} = 0 .
\end{align}
Note that $\delta^2$ is chiral in both arguments, and $\wb\delta{}^2$ is anti-chiral in both arguments. Besides, $\delta^3$ is bosonic and even in the two arguments, whilst $\delta^2$ and $\wb\delta{}^2$ are fermionic and anti-symmetric when swapping the two arguments. For bilocal fields $A,B(T_1,T_2)$ which are anti-chiral in the first argument and chiral in the second argument, we define the convolutions
\bea
\label{superspace convo def}
(A\star B)(T_1,T_2) &\equiv \ts\int\! d^2T_3 \; A(T_2,T_3) \, B(T_1,T_3) \;, \\
(A\;\overline{\star}\;B)(T_1,T_2) &\equiv \ts\int\! d^2\wb T_3 \; A(T_3,T_1) \, B(T_3,T_2) \;.
\eea
Here $A\star B$ is anti-chiral in both arguments, whilst $A\,\overline{\star}\,B$ is chiral in both arguments. Inverse operators under convolution shall be defined such that
\be
\label{delta sign in inverse}
A^{-1}\star A = \wb\delta{}^2 \;,\qquad\qquad A^{-1}\;\wb\star\;A = \delta^2 \;.
\ee
Notice that one could have alternatively defined the convolutions in (\ref{superspace convo def}) with a swap of $T_1$ and $T_2$ on the right-hand-side, which would have led to an opposite sign in the definition of the inverses (because $\delta^2$ and $\wb\delta{}^2$ are anti-symmetric). The convention we chose was dictated by comparing the equations of motion in \eqref{integral eom antichiral} with their superspace derivation in \eqref{superspace sigma deriv}. With this definition of $A^{-1}$, one can derive
\be
\label{derive inverse supermat}
\frac{\delta A^{-1}(T_1,T_2)}{\delta A(T_3,T_4)} = A^{-1}(T_1,T_4) \, A^{-1}(T_3,T_2) \;.
\ee

By starting with a general ansatz and imposing invariance under the supersymmetry transformation $\delta=\chi Q-\overline{\chi}\,\overline{Q}$, one can show that the most general supersymmetric completion of the translationally invariant quantity $\tau_1-\tau_2$ is $\tau_1 - \tau_2 + \tfrac12 \wb\theta_2 \theta_1 + \tfrac12 \theta_2 \wb\theta_1 + a (\theta_1-\theta_2)(\wb\theta_1 - \wb\theta_2 )$ with a constant $a$.
Imposing that the combination is anti-chiral in $T_1$ and chiral in $T_2$ further fixes $a=-\frac{1}{2}$. Therefore, a bilocal superfield $\cG(T_1,T_2)$ that is anti-chiral in $T_1$, chiral in $T_2$, translationally invariant, and supersymmetric, can only be a function of
\be
\label{transl chiral anti-chiral inv}
T_{12} \,\equiv\, \tau_1 - \tau_2 - \tfrac12 \bigl( \theta_1 \wb\theta_1 + \theta_2 \wb\theta_2 - 2 \theta_2 \wb\theta_1 \bigr)  \;.
\ee

\subsection{Super-reparametrizations}
\label{app: super-reparam}

The $\cN=2$ super-reparametrizations, denoted here as $\text{Diff}^+(S^{1|2})$, are general coordinate transformations $(\tau,\theta,\overline{\theta}) \rightarrow (\tau', \theta', \wb\theta{}')$ under which (anti-)chiral superfields remains (anti-)chiral in the new coordinates. This implies the conditions
\be
\label{super reparam def}
\wb{D} \theta' = D \wb\theta' = 0 \,,\quad \wb{D}\tau' = \tfrac{1}{2} \, \theta' \, \wb{D} \, \wb\theta' \,,\quad D\tau' = \tfrac{1}{2} \, \wb\theta' D\theta' \,.
\ee
They have general solutions parametrized by a real bosonic function $f(\tau)$, a complex fermionic function $\chi(\tau)$, and a real bosonic phase $\sigma(\tau)$ which implements $\rU(1)_R$ rotations on $\theta$:
\begin{align}
\label{general super reparam}
\tau' &= f + \tfrac{1}{2} \theta \, |\rho|^2 \overline{\chi} + \tfrac{1}{2} \wb\theta \, |\rho|^2 \chi + \tfrac{1}{4} \theta \overline{\theta} \, \partial_\tau(|\rho|^2 \, \wb\chi \chi) \,, \\
\theta' &= \rho\chi + \theta\rho + \tfrac{1}{2} \theta \overline{\theta} \, \partial_\tau(\rho\chi) \,,\qquad \rho=|\rho| \, e^{-i\sigma} , \nn \\
\wb\theta{}' &= \overline{\rho} \, \overline{\chi} + \overline{\theta} \, \overline{\rho} - \tfrac{1}{2} \theta \overline{\theta} \, \partial_\tau(\overline{\rho}\,\overline{\chi}) \,,\nn\\
|\rho|^2 &= \partial_\tau f \, \bigl( 1-\tfrac{1}{2}\chi(\partial_\tau+i\partial_\tau\sigma)\overline{\chi}-\tfrac{1}{2}\overline{\chi}(\partial_\tau-i\partial_\tau\sigma)\chi \bigr)^{\! -1} . \nn
\end{align}
The chiral measure transforms as
\begin{align}
\label{chiral measure reparam}
& d\tau' d\theta' = d \Bigl( \tau' {+} \tfrac{1}{2} \theta' \wb\theta{}' \Bigr) d\theta' \\
&\;\; = \Ber \begin{pmatrix} \partial_\tau \bigl( \tau' {+} \frac12 \theta' \wb\theta{}' \bigr) & \partial_\tau \theta' \\ \partial_\theta \bigl( \tau' {+} \frac12 \theta' \bar\theta{}' \bigr) & \partial_\theta\theta' \end{pmatrix} d\tau d\theta = - \overline{D} \, \wb\theta{}' \, d\tau d\theta \nn
\end{align}
where we defined the Berezinian as
\be
\label{berezinian}
\Ber \biggl( \begin{matrix} A & B \\ C & D \end{matrix} \biggr) = \det \bigl( A - B D^{-1} C \bigr) \, \det(D)^{-1} \,.
\ee
Here $A$ and $D$ are bosonic operators, while $B$ and $C$ are fermionic. In the last equality we used the identity $\Ber \smat{A & B \\ C & D} = \Ber \smat{ A & B \\ C+\eta A & D+\eta B }$ for any fermionic matrix $\eta$, the constraint (\ref{super reparam def}), and then in the derivation we took superspace derivatives of the constraints that imply $\partial_\tau \tau' = \frac12 \bigl( \partial_\tau \theta' \wb\theta{}' + \partial_\tau \wb\theta{}' \theta'\bigr) - D\theta' \wb{D} \, \wb\theta{}'$.
By analogous steps, one derives that the anti-chiral measure transforms as
\be
\label{antichiral measure reparam}
d\tau' d\wb\theta{}' = D\theta' \, d\tau d\theta \;.
\ee
Consequently, the chiral and anti-chiral delta functions transform as
\begin{align}
\delta^2(T_1,T_2) &= - \wb{D}_1 \wb\theta{}'_1 \; \delta^2(T_1',T_2') = - \wb{D}_2 \wb\theta{}'_2 \; \delta^2(T_1',T_2') \,, \nn \\
\wb\delta{}^2(T_1,T_2) &= D_1\theta'_1 \; \wb\delta{}^2(T'_1,T'_2) = D_2 \theta'_2 \; \wb\delta{}^2(T'_1,T'_2) \,.
\label{chiral delta reparam}
\end{align}
Chiral or anti-chiral primary operators of scaling dimension $\Delta$ transform as
\bea
\label{chiral primary trans}
\Phi(T) &\;\mapsto\; (-\wb{D} \, \wb\theta{}' )^{2\Delta} \, \Phi(T') \quad& &\text{if } \wb{D} \Phi = 0 \,, \\
\wb\Phi(T) &\;\mapsto\; (D\theta')^{2\Delta} \, \wb\Phi(T') \quad& &\text{if } D\wb\Phi = 0 \,.
\eea

\subsection{Superconformal algebra}

The $\cN=2$ superconformal group $\rSU(1,1|1)$ is a subgroup of the super-reparametrizations introduced above. Its bosonic subgroup is $\rSL(2,\bR)\times\rU(1)_R$ and there are $4$ fermionic generators. First, we want to derive generators of the algebra $\su(1,1|1)$ as differential operators on superspace in our conventions. This can be done by equating
\be
\label{superconf trans ito generators}
(-\overline{D}\,\wb\theta{}')^{2\Delta} \; \Phi(T') = \bigl( 1 + \epsilon \cQ + \cO(\epsilon^2) \bigr) \; \Phi(T)
\ee
with $\overline{D} \Phi = 0$ for various infinitesimal transformations, which allows us to identify the generators $\cQ$.

Firstly, supertranslations, $\rU(1)_R$ rotations, and dilations with $\tau'=\lambda^{-1}\tau$, $\theta'=\lambda^{-\frac{1}{2}}\theta$ directly satisfy \eqref{super reparam def}. From \eqref{superconf trans ito generators} one obtains:
\begin{align}
L_0 &= - \tau \partial_\tau - \tfrac{\theta}{2} \partial_\theta - \tfrac{\overline{\theta}}{2} \partial_{\overline{\theta}} - \Delta \,,\qquad L_1 = \partial_\tau \,, \nn\\
\label{supertrans dil R gen}
R_\text{sc} &= \overline{\theta} \partial_{\overline{\theta}} - \theta\partial_{\theta} + Q_R \,, \\
\cG_\frac{1}{2} &= Q = \partial_\theta - \tfrac{\overline{\theta}}{2} \partial_\tau \,,\qquad \overline{\cG}_\frac{1}{2} = \overline{Q} = - \partial_{\overline{\theta}} +\tfrac{\theta}{2} \partial_\tau \,. \nn
\end{align}
To be precise, by the above procedure in $R_\text{sc}$ we obtain $Q_R = 2\Delta$ due to the fact that $\Phi$ is chiral, but we wrote a generic $Q_R$ in order to represent the action on a generic superconformal primary. We denoted bosonic and fermionic generators by $L$ and $\cG$, respectively; their subscripts indicate their scaling dimension.

To get the bosonic generator of special conformal transformations, we consider the inversion $\tau'=-\frac{1}{\tau}$. In order to satisfy \eqref{super reparam def} and be an element of Diff$^+(S^{1|2})$, we also need $\theta' = \theta / |\tau|$. The Jacobian of the inversion is $-\overline{D}\,\wb\theta{}' = \frac{1}{\tau}-\frac{ \theta \overline{\theta} }{ 2\tau |\tau|}$. After composing an inversion, a translation by $\epsilon$ and another inversion, the result is:
\begin{multline}
\label{res inv trans inv}
\Phi(T,\theta,\overline{\theta}) \,\mapsto\, \biggl( \frac{1}{1-\epsilon\tau} + \frac{\theta\overline{\theta}}{2\tau(1-\epsilon\tau)^2} - \frac{\theta\overline{\theta}}{2\tau (1-\epsilon \tau )} \biggr)^{\! 2\Delta} \! \times \\
{} \times \Phi\biggl( \frac{\tau}{1-\epsilon\tau}, \frac{\theta}{1-\epsilon\tau}, \frac{\overline{\theta}}{1-\epsilon\tau} \biggr)
\end{multline}
for infinitesimal $\epsilon$. Expanding to linear order in $\epsilon$, we identify the generator of special conformal transformations:
\be
\label{spec conf gen}
L_{-1} = \tau^2\partial_\tau + \tau\theta\partial_\theta + \tau\overline{\theta}\partial_{\overline{\theta}} + 2\Delta\tau + \tfrac12 \, Q_R \, \theta\overline{\theta} \;.
\ee
Finally, the two remaining fermionic generators are obtained from the closure of the algebra:
\begin{align}
\label{conformal supercharges}
\cG_{-\frac{1}{2}} &= \bigl[ L_{-1},\cG_\frac{1}{2} \bigr] \\
&= \phantom{+} \tfrac{\tau\overline{\theta}}{2}\partial_\tau - \bigl( \tau + \tfrac12 \theta \overline{\theta} \bigr) \partial_\theta + \bigl( \Delta - \tfrac12 Q_R \bigr) \overline{\theta} \,, \nn \\
\overline{\cG}_{-\frac{1}{2}} &= \bigl[ L_{-1}, \overline{\cG}_\frac{1}{2} \bigr] \nn \\
&= - \tfrac{\tau\theta}{2}\partial_\tau + \bigl( \tau - \tfrac12 \theta \overline{\theta} \bigr) \partial_{\overline{\theta}} - \bigl( \Delta + \tfrac12 Q_R \bigr) \theta \,. \nn
\end{align}
We write the generators as $L_{m=-1,0,1}$, $\cG_{r=\pm\frac{1}{2}}$, $\overline{\cG}_{r=\pm\frac{1}{2}}$. Their commutation relations are:
\begin{align}
[L_m,L_n] &= (n-m) \, L_{m+n} \;,\quad& [R_\text{sc},\cG_r] &= \cG_r \;, \nn\\
[L_m,\overline{\cG}_r] &= \tfrac{2r-m}{2} \, \wb\cG_{m+r} \;, \quad& [R_\text{sc},L_m] &= 0 \;, \nn\\
[L_m,\cG_r] &= \tfrac{2r-m}{2} \, \cG_{m+r}  \;,\quad& [R_\text{sc},\overline{\cG}_r] &= - \overline{\cG}_r \;, \nn\\
\{\cG_r,\cG_s\} &= 0\;,\quad& \{ \wb\cG_r, \wb\cG_s\} &= 0 \;, \nn\\
\{\cG_r, \wb\cG_s\} &= L_{r+s} - \tfrac{r-s}{2} R_\text{sc} \;.
\label{superconf algebra}
\end{align}
In the presence of multiple coordinates, the generators acting on a particular coordinate $T_i$ will be denoted by $L_m^{(i)}$, $\cG_r^{(i)}$. We will also write $L_m^{(i)}(\Delta,Q_R)$ when necessary. The ``single-particle" quadratic Casimir which commutes with all the generators is
\begin{align}
\cC^{1\text{p}} &= L_0^2 - \tfrac12 \{ L_{-1}, L_1 \} - \tfrac14 R_\text{sc}^2 \\
&\quad + \tfrac12 \bigl[ \, \wb\cG_{-\frac12}, \cG_{\frac12} \bigr] + \tfrac12 \bigl[ \cG_{-\frac12}, \wb\cG_{\frac12} \bigr]
 = \Delta^2 - \tfrac14 \, Q_R^2 \,. \nn
\end{align}
On the right-hand-side we wrote the value of the Casimir for highest-weight representations, in terms of the dimension $\Delta$ and R-charge $Q_R$ of the superconformal primary, annihilated by $L_{-1}$, $\cG_{-\frac12}$, $\wb\cG_{-\frac12}$.
We also define ``two-particle" superconformal generators 
\begin{multline}
\label{2p sc gen def}
L^{2\text{p},\Delta}_m \; g(T_1,T_2)\equiv\\
\left[L^{(1)}_m(\Delta,-2\Delta)+L^{(2)}_m(\Delta,2\Delta)\right] \, g(T_1,T_2) \;,
\end{multline}
and similarly for $R_\text{sc}$, $\cG_r$, $\overline{\cG}_r$. They act on fields which are anti-chiral in $T_1$ and chiral in $T_2$. Analogously, the ``two-particle" quadratic Casimir is defined as
\begin{widetext}
\begin{align}
\label{2p casimir def}
\cC^{2\text{p},\Delta} &\equiv \bigl( L_0^{2\text{p},\Delta} \bigr){}^2 - \tfrac12 \bigl\{ L_{-1}^{2\text{p}, \Delta} , L_1^{2\text{p},\Delta} \bigr\} - \tfrac14 \bigl( R_\text{sc}^{2\text{p},\Delta} \bigr){}^2 + \tfrac12 \bigl[ \wb\cG{}_{-\frac12}^{\,2\text{p},\Delta} , \cG_{\frac12}^{2\text{p},\Delta} \bigr] + \tfrac12 \bigl[ \cG_{-\frac12}^{2\text{p},\Delta} , \wb\cG{}_{\frac12}^{\,2\text{p},\Delta} \bigr] \nn \\
&= - \bigl[ T_{12}^2 + T_{12}( \partial_{\wb\theta_1} T_{12}) (\partial_{\theta_2}T_{12}) \bigr] \partial_{\tau_1}\partial_{\tau_2} - T_{12}(\partial_{\wb\theta_1} T_{12})\partial_{\tau_1} \partial_{\theta_2} - T_{12}(\partial_{\theta_2}T_{12}) \partial_{\wb\theta_1} \partial_{\tau_2} \nn \\
&\quad\, +T_{12}\partial_{\wb\theta_1} \partial_{\theta_2} +2\Delta T_{12}(\partial_{\tau_1}-\partial_{\tau_2})+4\Delta^2 \;.
\end{align}
\end{widetext}

\subsubsection{Superconformal solutions in superspace}
\label{app: superconf sols in superspace}

The equations \eqref{consistency match coeff} and \eqref{eqn for coeffs susy} determining the superconformal solutions can be derived rather quickly in superspace. Writing the Fourier transforms of the 2-point functions as $\cG_{\cY,\Phi}(\omega,\theta_1,\theta_2)$ and $\Sigma_{\cY,\Phi}(\omega,\theta_1,\theta_2)$, the expression for $T_{12}$ in (\ref{transl chiral anti-chiral inv}) implies
\begin{align}
\label{susy sigma G fourier}
\cG_{\cY,\Phi}(\omega, \theta_1, \theta_2) &= \bigl[ 1 - \tfrac{i}2 \omega \bigl( \theta_1 \wb\theta_1 + \theta_2 \wb\theta_2 + 2 \wb\theta_1 \theta_2 \bigr) \\
&\qquad\qquad -\tfrac14 \omega^2 \theta_1 \wb\theta_1 \theta_2 \wb\theta_2 \bigr] \, G_{\bar\eta \eta, \, \bar\phi \phi} (\omega) \,, \nn\\
\Sigma_{\cY,\Phi} (\omega,\theta_1,\theta_2) &= \bigl[ 1 - \tfrac{i}2 \omega \bigl( \theta_1 \wb\theta_1 + \theta_2 \wb\theta_2 + 2 \wb\theta_1 \theta_2 \bigr) 
\nn\\
&\qquad\qquad -\tfrac14 \omega^2 \theta_1 \wb\theta_1 \theta_2 \wb\theta_2 \bigr] \, \Sigma_{\bar{f} f, \, \bar\psi \psi} (\omega) \,. \nn
\end{align}
Similarly, since the anti-chiral delta function can be written as
\be
\wb\delta{}^2(T_1,T_2) = \bigl( \wb\theta_1 - \wb\theta_2 + \tfrac12 \theta_1 \wb\theta_1 \, \wb\theta_2 \partial_{\tau_1} + \tfrac12 \wb\theta_1 \theta_2 \wb\theta_2 \partial_{\tau_1} \bigr) \, \delta(\tau_1-\tau_2) ,
\ee
its momentum-space expression is
\be
\wb\delta{}^2(\omega) = \wb\theta_1 - \wb\theta_2 + \tfrac{i}2 \omega \bigl( \theta_1 \wb\theta_1 \, \wb\theta_2 + \wb\theta_1 \theta_2 \wb\theta_2 \bigr) \,.
\ee
Plugging \eqref{susy sigma G fourier} into (\ref{integral eom antichiral}) (dropping the first term on the LHS), we get the equations
\be
\label{superspace int eq after int}
i \omega \, \Sigma_{\bar{f} f}(-\omega) \, G_{\bar\eta \eta}(\omega) = - i \omega \, \alpha^{-1} \, \Sigma_{\bar\psi \psi}(-\omega) \, G_{\bar\phi \phi}(\omega) = 1 \,.
\ee
With the ansatz (\ref{G conformal ansatz in omega}) for $G_{\bar\eta \eta}$ and $G_{\bar\phi \phi}$, the algebraic equations determine $\Sigma_{\bar{f}f}$ and $\Sigma_{\bar\psi \psi}$. Comparing with their general expression in \eqref{sigma G FT} implies the relations (\ref{susy constr dim eps}) among dimensions and spectral asymmetries, as well as the equations \eqref{consistency match coeff} and \eqref{eqn for coeffs susy}.

\section{Large \tps{\matht{N}}{N} action and EOMs in components}
\label{app: component expr}

We collect here actions and equations of motion in components, whose superspace expressions are presented in the main text.
The definitions of the bilocal fields $\cG_\cY$, $\Sigma_\cY$ and $\cG_\Phi$, $\Sigma_\Phi$ are:
\begin{align}
\cG_\cY(T_1, T_2) &= \Bigl[ G_{\bar\eta \eta} + \theta_2 \, G_{\bar\eta f} - \wb\theta_1 \, G_{\bar{f} \eta} + \wb\theta_1 \theta_2 \, G_{\bar{f}f} \Bigr] \nn\\
\label{bilocal definition component}
\Sigma_\cY(T_1, T_2) &= \Bigl[ \Sigma_{\bar{f} f} - \theta_2 \, \Sigma_{\bar\eta f} - \wb\theta_1 \, \Sigma_{\bar{f} \eta} + \wb\theta_1 \theta_2 \, \Sigma_{\bar\eta \eta}\Bigr] \\
\cG_\Phi(T_1, T_2) &= \Bigl[ G_{\bar\phi \phi} + \theta_2 \, G_{\bar\phi \psi} - \wb\theta_1 \, G_{\bar\psi \phi} + \wb\theta_1 \theta_2 \, G_{\bar\psi \psi} \Bigr] \nn\\
\Sigma_\Phi(T_1, T_2) &= \Bigl[ \Sigma_{\bar\psi \psi} - \theta_2 \, \Sigma_{\bar\phi \psi} - \wb\theta_1 \, \Sigma_{\bar\psi \phi} + \wb\theta_1 \theta_2 \, \Sigma_{\bar\phi \phi} \Bigr]  \,, \nn
\end{align}
where on the right-hand-side all functions are evaluated at $\bigl( \tau_1 - \tfrac12 \theta_1 \wb\theta_1 ,\, \tau_2 + \tfrac12 \theta_2 \wb\theta_2 \bigr)$.
Here the fields $G$ are the 2-point functions
\be
\label{def bilocal general}
G_{\bar{A} B} (\tau_1, \tau_2) = \tfrac1N \, \bigl\langle \, \bar{A}_a (\tau_1) \, B_a (\tau_2) \, \bigr\rangle \,,
\ee
where $A$ and $B$ stand for the species $\eta$, $f$, $\phi$, or $\psi$, and a sum over repeated indices $a$ is implied. The action for these bilocal fields is:
\begin{widetext}
\begin{align}
\label{averaged action component}
& \tfrac1N \, S[G, \Sigma] = -\log \Ber \begin{pmatrix} ( \partial_{\tau_1} + \mu_\eta) \, \delta(\tau_1 - \tau_2) - \Sigma_{\bar\eta \eta}(\tau_1, \tau_2) & \Sigma_{\bar{f} \eta}(\tau_1, \tau_2) \\ - \Sigma_{\bar\eta f} (\tau_1, \tau_2) & - \delta(\tau_1 - \tau_2) - \Sigma_{\bar{f} f} (\tau_1, \tau_2) \end{pmatrix} \\
&\qquad\qquad\;\; + \alpha \log \Ber \begin{pmatrix} ( \partial_{\tau_1} + \mu_\phi ) \, \delta(\tau_1 - \tau_2) - \Sigma_{\bar\phi \phi} (\tau_1, \tau_2) & -\Sigma_{\bar\psi \phi}(\tau_1,\tau_2) \\ \Sigma_{\bar\phi \psi}(\tau_1,\tau_2) & - \delta(\tau_1 - \tau_2) - \Sigma_{\bar\psi \psi}(\tau_1, \tau_2) \end{pmatrix} \nn\\
& + \!\int\! d\tau_1 d\tau_2 \biggl\{ \Sigma_{\bar\eta \eta} G_{\bar\eta \eta} + \Sigma_{\bar{f} \eta} G_{\bar\eta f} + \Sigma_{\bar\eta f} G_{\bar{f} \eta} + \Sigma_{\bar{f} f} G_{\bar{f} f} + \Sigma_{\bar\phi \phi} G_{\bar\phi \phi} + \Sigma_{\bar\psi \phi} G_{\bar\phi \psi} + \Sigma_{\bar\phi \psi} G_{\bar\psi \phi} + \Sigma_{\bar\psi \psi} G_{\bar\psi \psi} - \frac{J}{q} \Bigl[ G_{\bar{f} f} G_{\bar\eta \eta} \nn\\
& + (p-1) G_{\bar{f} \eta} G_{\bar\eta f} \Bigr] G_{\bar\eta \eta}^{p-2} G_{\bar\phi \phi}^q  - \frac{J}{p} \Bigl[ G_{\bar\psi \psi} G_{\bar\phi \phi} + (q-1) G_{\bar\psi \phi} G_{\bar\phi \psi} \Bigr] G_{\bar\eta \eta}^p G_{\bar\phi \phi}^{q-2} - J \Bigl[ G_{\bar{f} \eta} G_{\bar\phi \psi} - G_{\bar\eta f} G_{\bar\psi \phi} \Bigr] G_{\bar\eta \eta}^{p-1} G_{\bar\phi \phi}^{q-1} \biggr\} , \nn
\end{align}
where in the last integral all functions have arguments $(\tau_1,\tau_2)$. In the low-energy limit, dropping the kinetic terms, the action reduces to
\be
\label{low energy averaged action component}
\frac{S_{\text{IR}}[G, \Sigma]}{N} = - \log \Ber \Biggl( \begin{matrix} - \Sigma_{\bar\eta \eta}(\tau_1, \tau_2) \!\!\! & \Sigma_{\bar{f} \eta}(\tau_1, \tau_2) \\ -\Sigma_{\bar\eta f}(\tau_1, \tau_2) \!\!\! & - \Sigma_{\bar{f} f}(\tau_1, \tau_2) \end{matrix}\Biggr)
\!+ \alpha \log \Ber \Biggl( \begin{matrix} - \Sigma_{\bar\phi \phi}(\tau_1, \tau_2) \!\!\! & - \Sigma_{\bar\psi \phi}(\tau_1, \tau_2) \\ \Sigma_{\bar\phi \psi}(\tau_1, \tau_2) \!\!\! & - \Sigma_{\bar\psi \psi}(\tau_1, \tau_2) \end{matrix} \Biggr) + \int \dots \;,
\ee
where the integral part is the same as in \eqref{averaged action component}.
Varying the action \eqref{averaged action component} one obtains two sets of equations. Varying with respect to the $\Sigma$ fields leads to a first set of integro-differential equations of motion:
\begin{align}
- \bigl( \partial_{\tau_2} {+} \mu_\eta \bigr) G_{\bar\eta \eta}(\tau_1, \tau_2) + \ts\int d\tau_3 \, \bigl( \Sigma_{\bar\eta \eta}(\tau_2, \tau_3) \, G_{\bar\eta \eta}(\tau_1, \tau_3) + \Sigma_{\bar{f} \eta}(\tau_2, \tau_3) \, G_{\bar\eta f}(\tau_1, \tau_3) \bigr) &= \delta(\tau_1 {-} \tau_2) \,, \nn\\
G_{\bar\eta f}(\tau_1, \tau_2) + \ts\int d\tau_3 \, \bigl( \Sigma_{\bar{f}f}(\tau_2, \tau_3) \, G_{\bar\eta f}(\tau_1, \tau_3) - \Sigma_{\bar\eta f}(\tau_2, \tau_3) \, G_{\bar\eta \eta}(\tau_1, \tau_3) \bigr) &= 0 \,, \nn\\
- G_{\bar{f} f}(\tau_1, \tau_2) - \ts\int d\tau_3 \, \bigl( \Sigma_{\bar{f} f}(\tau_2, \tau_3) \, G_{\bar{f}f}(\tau_1, \tau_3) + \Sigma_{\bar\eta f}(\tau_2, \tau_3) \, G_{\bar{f} \eta}(\tau_1, \tau_3) \bigr) &= \delta(\tau_1 {-} \tau_2) \,, \nn\\
\bigl( \partial_{\tau_2} {+} \mu_\eta \bigr) G_{\bar{f} \eta}(\tau_1, \tau_2) + \ts\int d\tau_3 \, \bigl( \Sigma_{\bar{f} \eta}(\tau_2, \tau_3) \, G_{\bar{f}f}(\tau_1, \tau_3) - \Sigma_{\bar\eta \eta}(\tau_2, \tau_3) \, G_{\bar{f} \eta}(\tau_1, \tau_3) \bigr) &= 0 \, \nn\\
\bigl( \partial_{\tau_2} {+} \mu_\phi \bigr) G_{\bar\phi \phi}(\tau_1, \tau_2) - \ts\int d\tau_3 \, \bigl( \Sigma_{\bar\phi \phi}(\tau_2, \tau_3) \, G_{\bar\phi \phi}(\tau_1, \tau_3) + \Sigma_{\bar\psi \phi}(\tau_2, \tau_3) \, G_{\bar\phi \psi}(\tau_1, \tau_3) \bigr) &= \alpha \, \delta(\tau_1 {-} \tau_2) \,, \nn\\
G_{\bar\phi \psi}(\tau_1, \tau_2) + \ts\int d\tau_3 \, \bigl( \Sigma_{\bar\psi \psi}(\tau_2, \tau_3) \, G_{\bar\phi \psi}(\tau_1, \tau_3) - \Sigma_{\bar\phi \psi} (\tau_2, \tau_3) \, G_{\bar\phi \phi}(\tau_1, \tau_3) \bigr) &= 0 \,, \nn\\
G_{\bar\psi \psi}(\tau_1, \tau_2) + \ts\int d\tau_3 \, \bigl( \Sigma_{\bar\psi \psi}(\tau_2, \tau_3) \, G_{\bar\psi \psi}(\tau_1, \tau_3) + \Sigma_{\bar\phi \psi}(\tau_2, \tau_3) \, G_{\bar\psi \phi}(\tau_1, \tau_3) \bigr) &= \alpha \, \delta(\tau_1{-}\tau_2) \,, \nn\\
\! \bigl( \partial_{\tau_2} {+} \mu_\phi \bigr) G_{\bar\psi \phi}(\tau_1, \tau_2) + \ts\int d\tau_3 \, \bigl( \Sigma_{\bar\psi \phi}(\tau_2, \tau_3) \, G_{\bar\psi \psi}(\tau_1, \tau_3) - \Sigma_{\bar\phi \phi}(\tau_2, \tau_3) \, G_{\bar\psi \phi}(\tau_1, \tau_3) \bigr) & = 0 \,.
\label{integro-differential equations component}
\end{align}
Varying the action with respect to the $G$ fields, instead, we obtain a second set of algebraic equations of motion. All fields appearing below have argument $(\tau_1,\tau_2)$:
\begin{align*}
\Sigma_{\bar\eta \eta} &= J \, \Bigl[ \tfrac{p-1}{q} \Bigl( G_{\bar{f}f} G_{\bar\eta \eta} + (p-2) \, G_{\bar{f} \eta} G_{\bar\eta f} \Bigr) \, G_{\bar\phi \phi}^2 + \Bigl( G_{\bar\psi \psi} G_{\bar\phi \phi} + (q-1) \, G_{\bar\psi \phi} G_{\bar\phi \psi} \Bigr) \, G_{\bar\eta \eta}^2 \\
\stepcounter{equation}\tag{\theequation}\label{algebraic equations component}
& \qquad\;\; + (p-1) \Bigl( G_{\bar{f} \eta} G_{\bar\phi \psi} - G_{\bar\eta f} G_{\bar\psi \phi} \Bigr) \, G_{\bar\eta \eta} G_{\bar\phi \phi} \Bigr] \, G_{\bar\eta \eta}^{p-3} G_{\bar\phi \phi}^{q-2} \,, \\
\Sigma_{\bar\phi \phi} &= J \, \Bigl[ \Bigl( G_{\bar{f}f} G_{\bar\eta \eta} + (p-1) \, G_{\bar{f} \eta} G_{\bar\eta f} \Bigr) \, G_{\bar\phi \phi}^2 + \tfrac{q-1}{p} \Bigl( G_{\bar\psi \psi} G_{\bar\phi \phi} + (q-2) \, G_{\bar\psi \phi} G_{\bar\phi \psi} \Bigr) \, G_{\bar\eta \eta}^2 \\
& \qquad\;\; + (q-1) \Bigl( G_{\bar{f} \eta} G_{\bar\phi \psi} - G_{\bar\eta f} G_{\bar\psi \phi} \Bigr) \, G_{\bar\eta \eta} G_{\bar\phi \phi} \Bigr] \, G_{\bar\eta \eta}^{p-2} G_{\bar\phi \phi}^{q-3} \,, \\
\Sigma_{\bar{f} \eta} &= J \, \bigl( \tfrac{p-1}{q} G_{\bar{f} \eta} G_{\bar\phi \phi} + G_{\bar\psi \phi} G_{\bar\eta \eta} \bigr) \, G_{\bar\eta \eta}^{p-2} \, G_{\bar\phi \phi}^{q-1} \,, \qquad\;\;\;
\Sigma_{\bar{f} f} = \tfrac{J}{q} \, G_{\bar\eta \eta}^{p-1} G_{\bar\phi \phi}^q \,, \\
\Sigma_{\bar\eta f} &= - J \, \bigl( \tfrac{p-1}{q} G_{\bar\eta f} G_{\bar\phi \phi} + G_{\bar\phi \psi} G_{\bar\eta \eta} \bigr) \, G_{\bar\eta \eta}^{p-2} \, G_{\bar\phi \phi}^{q-1} \,, \qquad
\Sigma_{\bar\psi \psi} = \tfrac{J}{p} \, G_{\bar\eta \eta}^p G_{\bar\phi \phi}^{q-1} \,, \\
\Sigma_{\bar\psi \phi} &= J \, \bigl( \tfrac{q-1}{p} G_{\bar\psi \phi} G_{\bar\eta \eta} + G_{\bar{f} \eta} G_{\bar\phi \phi} \bigr) \, G_{\bar\eta \eta}^{p-1} \, G_{\bar\phi \phi}^{q-2} \,, \\
\Sigma_{\bar\phi \psi} &= -J \, \bigl( \tfrac{q-1}{p} G_{\bar\phi \psi} G_{\bar\eta \eta} + G_{\bar\eta f} G_{\bar\phi \phi} \bigr) \, G_{\bar\eta \eta}^{p-1} \, G_{\bar\phi \phi}^{q-2} \,.
\end{align*}
We are often interested in consistent truncations where the fermionic components are set to zero. In those cases, the action for the bilocal field components takes the simple form:
\begin{align}
\label{averaged bosonic action component}
\tfrac1N \, S[G, \Sigma] &= - \log \det \Bigl[ \bigl( \partial_{\tau_1} {+} \mu_\eta \bigr) \, \delta(\tau_1 {-} \tau_2) - \Sigma_{\bar\eta \eta}(\tau_1, \tau_2) \Bigr] + \log \det \Bigl[ - \delta(\tau_1 {-} \tau_2) - \Sigma_{\bar{f} f}(\tau_1, \tau_2) \Bigr] \\
&\quad + \alpha \log \det \Bigl[ \bigl( \partial_{\tau_1} {+} \mu_\phi \bigr) \, \delta(\tau_1 {-} \tau_2) - \Sigma_{\bar\phi \phi}(\tau_1, \tau_2) \Bigr] - \alpha \log \det \Bigl[ - \delta(\tau_1 {-} \tau_2) - \Sigma_{\bar\psi \psi}(\tau_1, \tau_2) \Bigr] \nn\\
&\quad + \!\int\! d\tau_1 d\tau_2 \Bigl( \Sigma_{\bar\eta \eta} G_{\bar\eta \eta} + \Sigma_{\bar{f} f} G_{\bar{f} f} + \Sigma_{\bar\phi \phi} G_{\bar\phi \phi} + \Sigma_{\bar\psi \psi} G_{\bar\psi \psi} - \tfrac{J}{q} G_{\bar{f} f} G_{\bar\eta \eta}^{p-1} G_{\bar\phi \phi}^q - \tfrac{J}{p} G_{\bar\eta \eta}^p G_{\bar\psi \psi} G_{\bar\phi \phi}^{q-1} \Bigr) \nn
\end{align}
and the corresponding equations of motion are:
\begin{align}
\label{integro-differential bosonic equations component}
- \bigl( \partial_{\tau_2} {+} \mu_\eta \bigr) G_{\bar\eta \eta}(\tau_1, \tau_2) + \ts\int d\tau_3 \, \Sigma_{\bar\eta \eta}(\tau_2, \tau_3) \, G_{\bar\eta \eta}(\tau_1, \tau_3) &= \delta(\tau_1 {-} \tau_2) \\
G_{\bar{f} f}(\tau_1, \tau_2) + \ts\int d\tau_3 \, \Sigma_{\bar{f} f}(\tau_2, \tau_3) \, G_{\bar{f} f}(\tau_1, \tau_3) &= - \delta(\tau_1 {-} \tau_2) \nn \\
- \bigl( \partial_{\tau_2} {+} \mu_\phi \bigr) G_{\bar\phi \phi}(\tau_1, \tau_2) + \ts\int d\tau_3 \, \Sigma_{\bar\phi \phi}(\tau_2, \tau_3) \, G_{\bar\phi \phi}(\tau_1, \tau_3) &= - \alpha \, \delta(\tau_1 {-} \tau_2) \nn \\
G_{\bar\psi \psi}(\tau_1, \tau_2) + \ts\int d\tau_3 \, \Sigma_{\bar\psi \psi}(\tau_2, \tau_3) \, G_{\bar\psi \psi}(\tau_1, \tau_3) &= \alpha \, \delta(\tau_1 {-} \tau_2) \nn \\
J \, \bigl[ \tfrac{p-1}{q} \, G_{\bar{f} f} G_{\bar\phi \phi} + G_{\bar\psi \psi} G_{\bar\eta \eta} \bigr] \, G_{\bar\eta \eta}^{p-2} G_{\bar\phi \phi}^{q-1} = \Sigma_{\bar\eta \eta} \,,\qquad\qquad & \tfrac{J}{q} \, G_{\bar\eta \eta}^{p-1} G_{\bar\phi \phi}^q = \Sigma_{\bar{f} f} \,, \nn \\
J \, \bigl[ G_{\bar{f} f} G_{\bar\phi \phi} + \tfrac{q-1}{p} \, G_{\bar\psi \psi} G_{\bar\eta \eta} \bigr] \, G_{\bar\eta \eta}^{p-1} G_{\bar\phi \phi}^{q-2} = \Sigma_{\bar\phi \phi} \,,\qquad\qquad & \; \tfrac{J}{p} \, G_{\bar\eta \eta}^p G_{\bar\phi \phi}^{q-1} = \Sigma_{\bar\psi \psi} \,. \nn
\end{align}
\end{widetext}

\section{IR conformal ansatz and UV limit}
\label{app: IR and UV limits}

In this appendix we gather some properties of the conformal ansatz used in the IR limit, and the behavior of the model in the UV limit.

\subsection{Reality of Euclidean 2-point functions}
\label{reality eucl 2pt}

Since the Hamiltonian is Hermitian, Euclidean operators in the Heisenberg picture satisfy:
\be
\cO(\tau)^\dagger = \bigl( e^{\tau H} \, \cO \, e^{-\tau H} \bigr)^\dag = e^{-\tau H} \, \cO^\dag \, e^{\tau H} = \cO^\dag(-\tau)
\ee
where $\cO \equiv \cO(0)$.
This is the quantum-mechanical version of reflection positivity. Therefore the Euclidean $2$\nobreakdash-point function between $\cO^\dagger$ and $\cO$ is real, since
\begin{align}
\langle\cO^\dag(\tau) \, \cO(0)\rangle_\beta^* &= \bigl[ \Tr e^{-\beta H}\cO^\dag(\tau) \, \cO(0) \bigr]^* \\
&= \langle \cO^\dag(\tau) \, \cO(0) \rangle_\beta \;. \nn
\end{align}
Under analytic continuation in $\tau$, one similarly shows that $\cO(\tau)^\dag = \cO^\dag( - \bar\tau)$ and that $\langle \cO^\dag(\tau)\, \cO(0) \rangle_\beta^* = \langle \cO^\dag(\bar\tau) \, \cO(0) \rangle_\beta$. Setting $\tau = it$, the Lorentzian $2$-point function satisfies
\begin{align}
& \braketmatrix{0}{\cT \, \cO^\dag(t) \, \cO(0)}{0}_\text{L}^* \\
&\qquad = \lim_{\substack{\epsilon \rightarrow 0^+ \\ \beta \rightarrow\infty}} \Bigl[ \Theta(t) \, \langle \cO^\dag(\epsilon + it)\, \cO(0) \rangle_\beta \nn\\[-1.2em]
&\qquad\qquad\qquad + \Theta(-t) \, \langle \cO^\dag(-\epsilon + it) \, \cO(0) \rangle_\beta \Bigr]^* \nn\\
&\qquad = \lim_{\substack{\epsilon\rightarrow 0^+ \\ \beta\rightarrow\infty}} \Bigl[ \Theta(t) \, \langle \cO^\dag(\epsilon - it) \, \cO(0) \rangle_\beta \nn\\[-1.2em]
&\qquad\qquad\qquad  + \Theta(-t)\, \langle \cO^\dag(-\epsilon - it) \, \cO(0) \rangle_\beta \Bigr] \nn\\
&\qquad = \braketmatrix{0}{\cT_- \, \cO^\dag(-t) \, \cO(0)}{0}_\text{L} . \nn
\end{align}
Here $\cT$ is the time ordering operator which arranges operators in order of increasing time from right to left, while $\cT_-$ is the anti-time ordering operator.

\subsection{Conformal ansatz with chemical potential}
\label{app: conformal ansatz with chem}

We review the discussion in Appendix~B of \cite{Davison:2016ngz}. Consider a reparametrization-invariant quantum mechanics with a global $\rU(1)$ symmetry, placed on a Euclidean $S^1$ with circumference $\beta$. To turn on a chemical potential, we couple the theory to an external gauge field $A_\tau = i\mu$ for the $\rU(1)$ symmetry. Under reparametrizations of $S^1$, the gauge field transforms as
\be
\label{reparam transf gf}
\tau \;\mapsto\; \tau'(\tau) \,,\qquad A_\tau \;\mapsto\; A_\tau' = (d\tau/d\tau') \, A_\tau \,.
\ee
Therefore we do not end up in a theory with the same chemical potential
\footnote{The integrated chemical potential $\protect\int d\tau\, A_\tau$ remains invariant.}.
In order to get back the same chemical potential, we need to perform a compensating (complexified) $\rU(1)$ gauge transformation with parameter
\be
\label{compen gauge transf}
\Lambda = i\mu \, \bigl(\tau'(\tau) - \tau \bigr) \;,
\ee
under conventions where the gauge field transforms as $A_\tau \,\mapsto\, A_\tau + \partial_\tau\Lambda$.

Now, suppose we have an operator $\cO$ with scaling dimension $\Delta$ and $\rU(1)$ charge $Q$. The conformal Ward identity for the thermal $2$-point function $\langle\cO^\dag \cO\rangle_\beta$ is
\begin{multline}
\label{conf WI with mu}
\bigl\langle \cO^\dag(\tau'(\tau_1)) \, \cO(\tau'(\tau_2)) \bigr\rangle_\beta = 
e^{Q\mu \left( \rule{0pt}{0.6em} \tau'(\tau_1) - \tau'(\tau_2) - \tau_1 + \tau_2 \right)} \times {} \\
\times \bigl[ \tfrac{d\tau'}{d\tau}(\tau_1) \, \tfrac{d\tau'}{d\tau}(\tau_2) \bigr]^{-\Delta} \langle \cO^\dag(\tau_1) \, \cO(\tau_2) \rangle_\beta \,.
\end{multline}
Notice the extra exponential factor due to the compensating gauge transformation. We seek the most general 2-point function that satisfies \eqref{conf WI with mu} for the subgroup $\rPSL(2,\bR)\subset\text{Diff}^+(S^1)$ which acts as
\be
\label{sl2r action s1}
\tan\biggl( \frac{\pi\tau'}{\beta} \biggr) = \frac{a \tan \bigl( \frac{\pi\tau}{\beta} \bigr) + b}{c \tan\bigl( \frac{\pi\tau}{\beta} \bigr) + d} \quad\text{with}\quad ad-bc = 1 \;.
\ee
This action is obtained by pulling back the fractional linear transformations on $\bR$ under the map $\tau \in \bigl( -\frac{\beta}{2}, \frac{\beta}{2} \bigr) \,\mapsto\, x = \tan \bigl( \frac{\pi\tau}{\beta} \bigr) \in \bR$. Due to translational invariance, the $2$-point function is only a function of $\tau \equiv \tau_1 - \tau_2$. Then one can check that 
\be
\label{gen 2pt conf ward}
\langle \cO^\dag(\tau) \, \cO(0) \rangle_\beta =
 e^{Q\mu\tau} \biggl\lvert \frac{\pi}{ \beta \sin \bigl( \frac{\pi\tau}{\beta} \bigr)} \biggr\rvert^{2\Delta} \! \bigl( A \,\Theta(\tau) + B \, \Theta(-\tau) \bigr)
\ee
with $A,B \in \bR$ indeed satisfies \eqref{conf WI with mu} under the transformations \eqref{sl2r action s1}, for all those values of $\tau_{1,2}$ such that $\sgn\bigl( \tau'(\tau_1) - \tau'(\tau_2) \bigr) = \sgn(\tau_1 - \tau_2)$.
The functional form is fixed by \eqref{conf WI with mu} and one allows for a discontinuity at $\tau=0$ where the operators coincide. The reality of the constants $A$, $B$ follows from the reality of the Euclidean $2$-point function.

Recall that there are two equivalent ways of introducing a chemical potential. One can either work with fields with boundary conditions twisted by a $\rU(1)$ transformation, or work with periodic fields but introduce a background gauge field in the Lagrangian. We chose the latter when deriving \eqref{conf WI with mu}, which means that the Hamiltonian is deformed to $\tilde H=H + \mu Q$, and Heisenberg operators are defined using $\tilde H$ instead of $H$. Now:
\bea
\label{KMS condition}
\langle \cO^\dag(\tau+\beta) \, \cO(0) \rangle_\beta &= \Tr e^{-\beta\tilde H} \, \cO^\dag(\tau+\beta) \, \cO(0) \\
&= s \, \langle\cO^\dag(\tau) \, \cO(0)\rangle_\beta \,,
\eea
where $s = +1$ if $\cO$ is bosonic and $s=-1$ if $\cO$ is fermionic. Notice that a cyclic permutation of fermionic operators in the trace does not introduce any sign (as it is clear by representing the trace as a sum of matrix elements, and inserting a complete basis of states in the middle), while the sign $s$ appears in the last equality when we swap the two operators.
Imposing the KMS relation \eqref{KMS condition} to \eqref{gen 2pt conf ward} implies $A \, e^{Q\mu\beta} = s \, B$.
The solution is $A = g \, e^{-\frac{Q\mu\beta}{2}}$ and $B = g \, s \, e^{\frac{Q\mu\beta}{2}}$ for some $g \in \bR$.
Defining $2\pi \cE = - Q \mu \beta$
\footnote{When we match $\cE$ with $\mu$ using the KMS relation, we implicitly assume that the conformal ansatz holds in the UV, which is not true in the actual model.},
the thermal 2-point function takes the form
\begin{multline}
\bigl\langle \cO^\dag(\tau) \, \cO(0) \bigr\rangle_\beta = g \, e^{- 2\pi\cE \frac{\tau}{\beta}} \, \biggl\lvert \frac{\pi}{ \beta \sin \bigl( \frac{\pi\tau}{\beta} \bigr)} \biggr\rvert^{2\Delta} \times {} \\
\times \Bigl( e^{\pi\cE} \, \Theta(\tau) + s \, e^{-\pi\cE} \, \Theta(-\tau) \Bigr) \,.
\end{multline}
Taking the zero-temperature limit $\beta\rightarrow\infty$ while keeping $\cE$ and $\tau$ finite, this becomes
\be
\label{2pt ansatz KMS}
\bigl\langle \cO^\dag(\tau) \, \cO(0) \bigr\rangle_{\beta=\infty} = \frac{g}{|\tau|^{2\Delta}} \, \Bigl( e^{\pi\cE} \, \Theta(\tau) + s \, e^{-\pi\cE} \, \Theta(-\tau) \Bigr) \;.
\ee

One can show that the coefficient $g$ must be non-negative in order to be consistent with the K\"all\'en-Lehmann spectral representation. Consider the Lorentzian $2$-point function
\bea
\label{advanced green fn O}
\braketmatrix{0}{\cO^\dag(t) \, \cO(0)}{0}_\text{L} &= \lim_{\epsilon\rightarrow 0^+} \langle \cO^\dag(\epsilon+it) \, \cO(0) \rangle_{\beta=\infty} \\
&= \frac{g \, e^{\pi\cE-i\pi\Delta}}{(t-i\epsilon)^{2\Delta}} \;,
\eea
that is obtained from analytically continuing \eqref{2pt ansatz KMS} for $\tau>0$. By inserting a complete set of energy eigenstates $\{\ket{n}\}$ with energies $E_n\geq E_0$, it can be rewritten as
\be
\braketmatrix{0}{\cO^\dag(t) \, \cO(0)}{0}_\text{L}
= \int_{-\infty}^\infty \frac{d\omega}{2\pi} \, e^{i\omega t} \, \rho(\omega) \,,
\ee
where
\be
\rho(\omega) = \sum\nolimits_n 2\pi \, \delta(\omega+E_n-E_0) \, \bigl\lvert \braketmatrix{n}{\cO(0)}{0} \bigr\rvert^2 > 0 \,.
\ee
Consequently the Fourier transform of the $2$-point function only has support on $\omega \leq 0$ and it must be non-negative. We shall impose this condition to the Fourier transform of \eqref{advanced green fn O}:
\bea
\rho(\omega) &= g \, e^{\pi\cE-i\pi\Delta} \lim_{\epsilon\rightarrow 0^+} \int_{-\infty}^\infty \!\! dt \, \frac{e^{-i\omega t}}{(t-i\epsilon)^{2\Delta}} \\
&= 2 \pi \, g \, \frac{e^{\pi\cE}}{\Gamma(2\Delta)} \, \Theta(-\omega) \, |\omega|^{2\Delta-1} \;.
\eea
In the last equality we used that the integrand has a branch point at $t=i\epsilon$, and the branch cut must be taken in the upper half plane in order not to intersect the real axis. For $\omega >0$, one can deform the contour so that it runs from $-i\infty$ to $0$ and back, giving zero. For $\omega <0$, we deform the contour so that it hugs the left and right side of the branch cut that we take along the positive imaginary axis, and the result follows from the discontinuity. Finally, requiring $\rho(\omega)>0$ implies $g>0$.

\subsection{UV limits}
\label{app: UV limits}

\paragraph{Chiral multiplet.}
Consider the free chiral multiplet $(\phi,\psi)$ with one-derivative kinetic term. Without chemical potential the action is topological
\footnote{The Lagrangian written as $\cL_\text{L} = i \phi^\dag \partial_t \phi + i \psi^\dag \gamma_t \psi$ is manifestly a 1-form. When the fermion Lagrangian is written simply as $\cL_\text{L} = \psi^\dag \psi$, the fermion $\psi$ takes values in a square root of the cotangent bundle.},
and it remains fully re\-pa\-ra\-me\-tri\-za\-tion invariant in the presence of a chemical potential $\mu$ if we think of the latter as a gauge connection and accompany the reparametrization (\ref{reparam transf gf}) with the gauge transformation (\ref{compen gauge transf}).
The fermion $\psi$ is auxiliary, has Lagrangian $\cL_\text{L} = \psi^\dag \psi$ and has Euclidean $2$-point function
\be
\label{free aux psi 2pf}
\langle \, \wb\psi(\tau) \, \psi(0) \, \rangle_\beta = \delta(\tau) \;.
\ee
The complex boson $\phi$ in Lorentzian signature has Lagrangian $\cL_\text{L} = i \phi^\dag \partial_t \phi - \mu \phi^\dag \phi$. We could decompose $\phi = \frac1{\sqrt{2}} (x+ip)$ and rewrite the Lagrangian as $\cL_\text{L} = p\dot x - \frac\mu2 (p^2 + x^2)$ up to total derivatives. Here $\mu$ is the chemical potential for the $\rU(1)$ global symmetry whose infinitesimal action is $\delta \phi = i \epsilon\phi$. The canonical momentum conjugate to $\phi$ is $\Pi_\phi = i \phi^\dag$, so that in the Hamiltonian formulation $[\phi, \phi^\dag] = 1$ and the Hamiltonian reads
\be
\label{free b H}
H = \mu \, Q_\phi = \frac\mu2 \, \{\phi^\dag , \phi \} \;,
\ee
where $Q_\phi = \frac12 \{\phi^\dag, \phi\} = \phi^\dag\phi + \frac12$ is the conserved charge for the $\rU(1)$ global symmetry, and we have fixed the ordering ambiguity for later convenience. As argued after (\ref{interacting Hamiltonian}), we should restrict to $\mu>0$. Defining the number operator $N = \phi^\dag \phi$, the Hilbert space is then a Fock space of eigenstates of $N$, \ie, $N |n\rangle = n |n\rangle$. They satisfy $\phi |0\rangle = 0$ and are generated by $\phi^\dag$, \ie, the normalized eigenstates are $|n\rangle = \frac1{\sqrt{n!}} (\phi^\dag)^n |0\rangle$. When going to Euclidean signature, we will use the notations $\wb\phi$ and $\phi^\dag$ interchangeably.

The partition function is
\be
\label{free b Z}
Z = \Tr e^{-\beta H} = \frac1{2 \sinh \bigl( \frac{ \mu\beta }2 \bigr) } \;.
\ee
The von Neumann entropy $S = (1 - \beta \partial_\beta) \log Z$ is
\be
\label{free b S entropy}
S = - \log\bigl[ 2 \sinh\bigl( \tfrac{\mu \beta}2 \bigr) \bigr] + \tfrac{\mu \beta}2 \coth\bigl( \tfrac{\mu \beta}2 \bigr) \;,
\ee
which diverges as $\mu \beta \to 0$.
The Euclidean thermal 2\nobreakdash-point function $\langle \wb\phi \phi \rangle_\beta$ is:
\begin{multline}
\label{fb eucl 2pt}
\langle \, \wb\phi(\tau) \, \phi(0) \, \rangle_\beta = Z^{-1} \Bigl[ \Theta(\tau) \Tr e^{-(\beta-\tau)H} \, \wb\phi \, e^{-\tau H} \, \phi \\
+ \Theta(-\tau) \Tr e^{-(\beta - |\tau|) H}\, \phi \, e^{- |\tau| H} \, \wb\phi \Bigr] \\
= \frac{e^{\mu\tau} }{ 2\sinh \bigl( \frac{\mu\beta}{2} \bigr)} \Bigl[ e^{ -\frac{\mu\beta}{2}} \, \Theta(\tau) + e^{ \frac{\mu\beta}{2}} \, \Theta(-\tau) \Bigr] .
\end{multline}
One can check invariance of the Euclidean action under the combination of \eqref{reparam transf gf} and \eqref{compen gauge transf}. The corresponding Ward identity for $\phi$, that has dimension zero, is
\be
\label{fb reparam ward}
\bigl\langle \, \wb\phi \bigl( \tau'(\tau) \bigr) \, \phi \bigl(\tau'(0) \bigr) \bigr\rangle_\beta = e^{\mu \left( \rule{0pt}{0.55em} \tau'(\tau) - \tau'(0) - \tau \right)} \, \bigl\langle \, \wb\phi(\tau) \, \phi(0) \, \bigr\rangle_\beta
\ee
for any reparametrization $\tau'(\tau)$, and it is satisfied by \eqref{fb eucl 2pt}.

In the zero-temperature limit $\beta\rightarrow\infty$, \eqref{fb eucl 2pt} becomes $\langle \, \wb\phi(\tau) \, \phi(0) \, \rangle_{\beta=\infty} = e^{\mu\tau} \, \Theta(-\tau)$. One can deduce the time-ordered $2$-point function in Lorentzian signature by performing an analytic continuation to $\tau = it$ with $\epsilon$-prescription:
\begin{multline}
\label{fb to 2pt}
\bigl\langle 0 \big| \cT \; \wb\phi(t) \, \phi(0) \big| 0 \bigr\rangle_\text{L} = \lim_{\epsilon \to 0^+} \Bigl[ \Theta(t) \, \langle\wb\phi(it + \epsilon) \phi(0) \rangle_{\beta=\infty} \\
+ \Theta(-t) \, \langle \wb\phi(it -\epsilon) \phi(0) \rangle_{\beta=\infty} \Bigr] = e^{i\mu t} \, \Theta(-t) \,.
\end{multline}
This matches the direct computation in Minkowski signature.
In order to define the Fourier transform of \eqref{fb to 2pt}, we use the regularization
\bea
\lim_{\epsilon \to 0^+} \int_{-\infty}^0 \! dt \, e^{ i (\omega + \mu - i\epsilon)t} &= \lim_{\epsilon \to 0^+} \frac{-i}{\omega + \mu - i\epsilon} \\
&= \cP \frac{-i}{\omega + \mu} + \pi \, \delta(\omega + \mu) \;,
\eea
where $\cP$ denotes the Cauchy principal value.

\paragraph{Fermi multiplet.}
Consider the theory of a free Fermi multiplet $(\eta,f)$
\footnote{The theory becomes topological if $\eta,f$ are an anticommuting scalar and a commuting spinor, respectively.}.
The auxiliary field $f$ has Euclidean 2-point function
\be
\label{ff aux 2pt}
\langle \, \wb{f}(\tau) \, f(0) \, \rangle_\beta = - \delta(\tau) \;.
\ee
The dynamical fermion $\eta$ has classical Lagrangian $\cL_\text{L} = i \eta^\dag \partial_t \eta - \mu \eta^\dag \eta$ in Lorentzian signature. The conjugate momentum to $\eta$ is $i\eta^\dag$ and the canonical commutation relation is $\{\eta, \eta^\dag\} = 1$. The Hamiltonian is
\be
H = \mu\, Q_\eta = \frac\mu2 \, [\eta^\dag, \eta] \;,
\ee
where $Q_\eta = \frac12 [\eta^\dag, \eta] = \eta^\dag \eta - \frac12$ is the conserved charge, and we have fixed the ordering ambiguity in a charge-conjugation invariant way. The Hilbert space consists of two states: the Fock vacuum $|0\rangle$ annihilated by $\eta$, and $\eta^\dag |0\rangle$. There are no restrictions on $\mu$.

For the fermion, the Euclidean partition function is
\be
\label{ff z}
Z = \Tr e^{-\beta H} = 2 \cosh \bigl( \tfrac{\mu\beta}{2} \bigr) \;.
\ee
The von Neumann entropy $S = (1 - \beta \partial_\beta) \log Z$ follows,
\be
\label{free f S entropy}
S = \log\bigl[ 2 \cosh\bigl( \tfrac{\mu \beta}2 \bigr) \bigr] - \tfrac{\mu \beta}2 \tanh\bigl( \tfrac{\mu \beta}2 \bigr) \;.
\ee
Lastly, the Euclidean thermal $2$-point function (using $\eta^\dag$ and $\wb\eta$ interchangeably) is
\begin{multline}
\label{ff eucl 2pt}
\langle \, \wb\eta(\tau) \, \eta(0) \, \rangle_\beta = Z^{-1} \Bigl[ \Theta(\tau) \Tr e^{-(\beta-\tau)H} \, \wb\eta \, e^{-\tau H} \, \eta \\
- \Theta(-\tau) \Tr e^{-(\beta - |\tau|) H}\, \eta \, e^{- |\tau| H} \, \wb\eta \Bigr] \\
= \frac{e^{\mu\tau} }{ 2\cosh \bigl( \frac{\mu\beta}{2} \bigr)} \Bigl[ e^{ -\frac{\mu\beta}{2}} \, \Theta(\tau) - e^{ \frac{\mu\beta}{2}} \, \Theta(-\tau) \Bigr] .
\end{multline}


\bibliography{BHEntropy_PRD}

\end{document}